\documentclass[11pt,final]{article}
\pdfoutput=1
\usepackage{MRnotes}

\newcommand{\Ts}{\delta {\sf T}}

\newcommand{\beq}{\begin{equation}}
\newcommand{\eeq}{\end{equation}}
 
\newcommand{\bea}{\begin{eqnarray}}
\newcommand{\ea}{\end{eqnarray}}

\newcommand{\ii}{\mathrm{i}}
\renewcommand{\d}{\dd}
\renewcommand{\i}{\ii}

\title{Looking at extremal black holes from very far away}

 \author{Maciej Kolanowski${}^a$, Donald Marolf${}^a$, Ilija Rakic${}^b$, Mukund Rangamani${}^b$, and Gustavo J. Turiaci${}^c$}
\affiliation{
	${}^a$ Department of Physics, University of California, Santa Barbara, CA 93106, USA}
 \affiliation{
	${}^b$ Center for Quantum Mathematics and Physics (QMAP),\\
	Department of Physics \& Astronomy, University of California, Davis, CA 95616 USA}
\affiliation{${}^c$ Physics Department, University of Washington, Seattle, WA USA}
\emailAdd{mkolanowski@ucsb.edu}
\emailAdd{marolf@ucsb.edu}
\emailAdd{irakic@ucdavis.edu}
\emailAdd{mukund@physics.ucdavis.edu}
\emailAdd{turiaci@uw.edu}

\abstract{ 
Near-extremal black holes are subject to large quantum effects, which modify their low-temperature thermodynamic behavior. Hitherto, these quantum effects were analyzed by separating the geometry into the near-horizon region and its exterior. It is desirable to understand and reproduce such corrections from the full higher-dimensional asymptotically flat or \AdS{} geometry's perspective. We address this question in this article and fill this gap. Specifically, we find off-shell eigenmodes of the quadratic fluctuation operator of the Euclidean gravitational dynamics, with eigenvalues that vanish linearly with temperature. We illustrate this for BTZ and neutral black holes with hyperbolic horizons in AdS in Einstein-Hilbert theory, and for the charged black holes in Einstein-Maxwell theory. The linear scaling with Matsubara frequency, which is a distinctive feature of the modes, together with the fact that their wavefunctions localize close to the horizon as we approach extremality, identifies them as responsible for the aforementioned quantum effects. We provide a contour prescription to deal with the sign indefiniteness of the Euclidean Einstein-Maxwell action, which we derive to aid our analysis.  We also resolve a technical puzzle regarding modes associated with rotational isometries in stationary black hole spacetimes.
}

\begin{document}\maketitle
\section{Introduction}\label{sec:intro}

Black holes are thermodynamic objects at a temperature set by the surface gravity, and an entropy given (at leading order) by the area of the horizon in Planck units (in Einstein-Hilbert gravity). Fluctuations of gravitons and matter around the geometry result in quantum corrections to the entropy~\cite{Sen:2012kpz,Sen:2012cj,Sen:2012dw}. These effects are of utmost importance for near-external black holes. We now understand that such black holes do not have macroscopic degeneracy, but rather behave as ordinary quantum systems~\cite{Ghosh:2019rcj,Iliesiu:2020qvm,Iliesiu:2022onk}.\footnote{Supersymmetric black holes are an exception where the vacuum degeneracy is protected and gapped from the rest of the states~\cite{Heydeman:2020hhw,Iliesiu:2022kny}.}

The key insight which led to this understanding was to  realize the presence of strongly coupled collective modes in the near-horizon region of near-extremal black holes. The fluctuation determinant of these modes suppresses the saddle point contribution in the gravitational path integral, thereby bringing down the degeneracy. This provides a resolution of some long-standing puzzles, outlined e.g., in~\cite{Preskill:1991tb,Michelson:1999kn,Page:2000dk}, in the case of non-supersymmetric near-extremal black holes.

Extremal black holes, by virtue of having a degenerate Killing horizon, have an enhanced $\mathrm{SL}(2,\mathbb{R})$ symmetry in their near-horizon region~\cite{Kunduri:2007vf}. The modes, alluded to above, correspond to large diffeomorphisms or gauge transformations of this near-horizon geometry. In particular, one always finds a set of large diffeomorphisms, referred to as the \emph{Schwarzian} modes, originating from the breaking of the near-horizon 1d conformal symmetry, at the locus where the geometry starts to transition to the asymptotic domain. In addition, there could be modes associated with large gauge symmetries, which may arise from either Killing isometries of the black hole, or from physical gauge fields in the background. 

Despite these remarkable developments, there are a few outstanding questions which we hope to clarify in this work. Our aims are twofold:
\begin{itemize}[wide,left=0pt]
\item First, we wish to provide a perspective on the quantum corrections without referring to the near-horizon geometry directly, but rather working in the full near-extremal black hole spacetime.\footnote{The original analysis of~\cite{Iliesiu:2020qvm} or \cite{Ghosh:2019rcj} was carried out in the full geometry, but relied on a split between the near-horizon, analyzed in a low-temperature expansion, and an external part. These two contributions were glued together to evaluate the functional integral. We will instead work directly in the near-extremal solution, and not make particular use of the near-horizon region.} Relatedly, we would like to reproduce the modes responsible for such quantum effects by working at finite temperature, and not strictly in a small temperature perturbation expansion. 

\item Second, we will clarify aspects of the large gauge transformations associated with abelian rotational isometries, clearing up a puzzle encountered in the analysis of~\cite{Rakic:2023vhv}.
\end{itemize}
In the rest of the introduction, we elaborate on these two points and highlight the salient results of our analysis. 

\paragraph{The Schwarzian mode in the black hole spacetime:} The infinite \AdS{2} throat  of extremal black holes is not truly decoupled from the asymptotic region. Indeed, it is this lack of decoupling that activates the near-horizon strongly coupled modes. As articulated in~\cite{Sen:2008yk, Sen:2008vm, Banerjee:2010qc, Sen:2012kpz} one may nevertheless use this \AdS{2} region to probe quantum effects. In particular, the $\mathrm{SL}(2,\mathbb{R})$ symmetry enables determination of explicit eigenspectrum of the quadratic fluctuation operators. Contained herein are the zero modes, elements in the kernel of the associated operator, generated by large gauge transformations of the near-horizon geometry. 

How do the fluctuations in the full geometry manifest the physical effect of these near-horizon zero modes? Generically, there cannot be zero modes in the entire black hole spacetime (away from extremality). In the case of near-extremal black holes, based on the near-horizon analysis, we should expect to see a set of nearly-gapless modes. One might guess that the eigenspectrum of the quadratic fluctuation operator contains modes whose eigenvalues scale linearly with the regulating temperature. Ideally, the support of such modes should also descend into the near-horizon throat as we approach extremality. We demonstrate this picture explicitly in several examples below, showing that the near-horizon zero modes of the extremal solution arise from very light off-shell modes in the full geometry.

The analysis of the eigenspectrum in the full black hole geometry is more involved than in the near-horizon. Nevertheless, we will explicitly verify this picture in a couple of examples: (a) analytically in the near-extremal rotating BTZ geometry, and (b) numerically in the hyperbolic \AdS{4} black hole, and the Reissner-Nordstr\"om-\AdS{4} black hole, respectively. Specifically, we will exhibit eigenmodes of the spin-$2$ fluctuation operator that exhibit this linear in $T$ scaling. Based on their properties, we will argue they correspond to the Schwarzian modes encountered in the near-horizon geometry. They contribute a factor of $T^{\frac{3}{2}}$ to the one-loop determinant universally. In addition, we will also find low-lying modes corresponding to gauge field, and rotational isometries. Using these results, we shall reproduce the near-extremal BTZ partition function originally obtained in~\cite{Ghosh:2019rcj}. Specifically, we will confirm that the result scales as $T^{\frac{3}{2}}\, e^{S_0}$, with $S_0$ being the (naive) extremal entropy, irrespective of the choice of ensemble (see below).

Before turning to other zero modes, we should highlight a technical result that we derive and use in our analysis. It is well known that the Euclidean Einstein-Hilbert action is sign indefinite, owing to the wrong sign kinetic term for the conformal mode. The standard prescription to deal with this issue is to integrate along purely imaginary conformal factors~\cite{Gibbons:1978ac}. A similar prescription for electrically charged black holes in Einstein-Maxwell theory is, to the best of our knowledge, not available in the literature. Taking inspiration from recent analyses~\cite{Marolf:2022ntb,Liu:2023jvm} we provide a prescription for linearized Einstein-Maxwell theory, and demonstrate its efficacy. Our prescription to deal with modes with non-positive definite action in short amounts to integrating the Maxwell fluctuations along an imaginary contour (along with rotating negative norm eigenmodes of the quadratic fluctuation operator).

\paragraph{Rotational and gauge zero modes:} Isometries of the black hole, as well as higher-dimensional gauge fields present in the theory, result in additional zero modes in the near-horizon~\cite{Sen:2012kpz}. For example, spherically symmetric solutions have non-abelian zero modes. These can be identified by dimensionally reducing the solution onto \AdS{2} whence one encounters non-abelian gauge fields, whose large gauge transformations constitute the aforesaid modes, see~\cite{Iliesiu:2020qvm} for an explicit analysis in Reissner-Nordstr\"om. Likewise, one expects axisymmetric solutions to have abelian rotational zero modes. 

Note that rotational and gauge zero modes are closely related. For one, should one uplift black holes to solutions in string or M-theory in 10 or 11 dimensions, then most gauge fields become geometric.\footnote{We thank Ashoke Sen for emphasizing this point and for extensive discussions regarding zero modes.} For isometries generated by vector fields with a constant norm, this identification is immediate. 
However, the dimensional reduction to \AdS{2} gravity to identify rotational zero modes is not always fully justified. 

A case in point is the Kerr solution, where the near-horizon geometry~\cite{Bardeen:1999px} is non-trivially warped over the polar angle of the $\mathbf{S}^2$. Therefore,~\cite{Rakic:2023vhv} attempted to directly construct the rotational zero mode in the four-dimensional near-horizon extremal Kerr geometry, and encountered a curious puzzle.\footnote{ The analysis of quantum effects in the near-extremal Kerr geometry was undertaken in two concurrent papers~\cite{Kapec:2023ruw} and \cite{Rakic:2023vhv} (cf.~\cite{Maulik:2024dwq} for a subsequent analysis in higher dimensions). The former focused solely on the Schwarzian modes, while the latter analyzed both the Schwarzian and the rotational zero modes, encountering the issue described in the text.} The putative rotational zero mode failed to satisfy the harmonic gauge condition chosen to carry out the calculation, which one would have intuitively expected it to do. Specifically, they demonstrated that the large diffeomorphism which would have led to such a zero mode, is not generated by a vector field that is smooth everywhere, should one demand the harmonic gauge be respected. 

We provide a prosaic resolution to this puzzle, by pointing out the breakdown of the harmonic gauge for any Ricci flat near-horizon geometry (which, in particular,  includes Kerr). In such cases, one encounters a gauge redundancy, by demonstrating the existence of normalizable diffeomorphisms that preserve the gauge choice. In particular, we show that there are indeed perturbations such as the rotational zero mode that cannot be brought into the harmonic gauge. Relatedly, the ghost one-loop determinant vanishes, confirming the failure of the gauge choice. We show that this problem does not exist for non-Ricci flat near-horizon geometries, such as Kerr-Newman and Kerr \AdS{} black holes. 

This simple resolution to the puzzle clarifies issues raised in~\cite{Rakic:2023vhv}, where the authors also pointed out that the absence of rotational zero modes would be a serious problem for supersymmetric black holes. In particular, the matching of the BMPV black hole entropy with the microscopic count carried out in~\cite{Sen:2012cj} is contingent on such modes being present. We demonstrate that the BMPV black hole near-horizon indeed supports the requisite number of bosonic zero modes. 

In the process of resolving this puzzle, we also check for additional zero-modes in the near-horizon of the extremal BTZ black hole. In line with the above argument, we find one associated with the rotational isometry. However, such a zero mode should not contribution to the low-temperature thermodynamics. Using the dual CFT partition function~\cite{Ghosh:2019rcj} argued for a low-temperature behavior $T^{\frac{3}{2}}\, e^{S_0}$, irrespective of the choice of ensemble; viz., this result holds whether we study the system at finite angular momentum (canonical) or angular velocity (grand canonical). Addressing how the gravitational computation remains consistent with the CFT analysis, requires a careful analysis of boundary conditions and ensemble choices, which we elaborate on next.

\paragraph{Boundary conditions and ensemble choices:} 
While we motivated the study of low-temperature corrections using the full geometry, we were incomplete in our specification of the problem. One has to be cognizant of the boundary conditions for the two scenarios. The most natural boundary conditions in the \AdS{2} throat involve fixing all the macroscopic conserved charges~\cite{Sen:2008yk}, including the energy. However, the subtle quantum effects captured by the Schwarzian modes and its gauge and isometry cousins are better understood in the grand canonical ensemble, where we fix temperature and all chemical potentials. (One can revert back to the fixed charge ensemble upon a Laplace transform.) This choice of boundary conditions is, in fact, more appropriate for the comparison with the full geometry. In higher dimensions the non-normalizable modes, which are more natural to fix, correspond to chemical potentials.\footnote{This statement largely holds for both  flat and AdS asymptotics. There are some exceptional cases, like \AdS{4} where both the fixing charge and chemical potentials are admissible~\cite{Witten:2003ya,Marolf:2006nd}, but even in these cases we imagine fixing the latter.} 

An additional wrinkle in making the comparison is that, while the near-horizon zero modes are derived in the fixed-charge ensemble (from the perspective of the throat), their contribution to the one-loop determinant has hitherto been estimated by passing to the grand canonical ensemble as an intermediate step. The essential idea facilitating this is that the temperature-regulated zero mode action reduces to the quantum mechanics of a particle moving on a group manifold (this perspective is natural if one dualizes the gauge field dynamics to a topological $BF$ theory). Nevertheless, this analysis does not always guarantee that these modes will (a) extend smoothly away from the throat, (b) be appropriately regulated by the finite temperature solution within the throat. These two questions are not unrelated, since the finite temperature correction to the metric in the throat is equivalent to the leading correction to the decoupling limit involved in gluing back to the exterior spacetime. We will analyze some examples where the modes do not extend outside the throat, making them unphysical. We will also argue in favor of the following criterion to identify this phenomenon. \emph{Whenever a gauge or rotational mode has a singular susceptibility\footnote{In some parts of the literature this has been referred to as the compressibility \cite{Sachdev:2019bjn}, but here we adhere to the more traditional thermodynamic definition, viz., variation of an extensive parameter with respect to its intensive conjugate. } (rate of change of charge against chemical potential) at zero temperature, the light mode associated to this symmetry will not extend to the full geometry.}

\paragraph{Outline of the paper:} We begin in~\cref{sec:schw} with an analysis of the Schwarzian mode in the full black hole spacetime. We present our analytic results for BTZ, and numerical results for the hyperbolic and Reissner-Nordstr\"om black holes in \AdS{4} in~\cref{sec:schw}. We also comment on the asymptotically flat Reissner-Nordstr\"om solution along the way. A detailed discussion of our contour prescription for dealing with the sign indefiniteness of Einstein-Maxwell theory can be found in~\cref{sec:contour}.  We then examine  the rotational modes in~\cref{sec:rot}, demonstrating their existence in Kerr-Newman and Kerr \AdS{} spacetimes, and explain the reason for the issues encountered in the case of Kerr. After explaining how to deduce analogous zero modes for BMPV and BTZ, we turn in~\cref{sec:BTZ_rot_mode} to the case of the low-temperature BTZ thermodynamics. We argue that in some cases the light modes (corresponding to rotation or gauge modes) found in the throat do not extend to the full geometry, and that one can identify when this happens from singularities in certain thermodynamic functions in the classical limit.

\section{Extending the nearly-gapless modes to the full geometry}\label{sec:schw}

We begin our discussion by demonstrating how the zero modes, localized in the near-horizon throat of the extremal black holes, uplift to the full geometry in the near-extremal solution. We will first discuss the rotating BTZ spacetime, where one has analytic control. We then describe other examples, such as the hyperbolic \AdS{4} solution. Finally, we find the Schwarzian and rotational modes in the full Reissner-Nordstr\"om solution in \AdS{4}.

\subsection{BTZ Black Hole}\label{Sec:BTZ_full}

In this section, we consider a theory of 3d gravity with a negative cosmological constant coupled to matter, with an action of the form
\beq
I = - \frac{1}{16\pi \,G_N}\, \int \d^3x\,\sqrt{g} \,\Big(R+\frac{2}{\lads^2}\Big) - \frac{1}{8\pi \,G_N}\,\oint \d^2 x\, \sqrt{\gamma}\, K + I_{\rm matter}\,.
\eeq
For simplicity, we work in units where $\Lambda=-1$, or equivalently, set the \AdS{3} radius to unity $\lads=1$. The matter sector can be arbitrary and will not be important for the discussion here, as we argue below. We are interested in studying quantum corrections around the asymptotically \AdS{3} rotating BTZ geometry, specifically in the near-extremal limit, and identifying the Schwarzian modes originating from metric fluctuations. 

The classical BTZ solution has the following line element~\cite{Banados:1992wn} in Euclidean signature
\begin{equation}\label{eq:btzmet}
\d s^2 = f(r)\, \d \tE^2 + \frac{\d r^2}{f(r)} + r^2 \,\Big( \d \varphi + \frac{\i\, r_+ \,r_-}{r^2} \, \d \tE\Big)^2\,,  
\end{equation}
where 
\begin{equation}\label{eq:fbtz} 
f(r) = \frac{(r^2-r_+^2)(r^2-r_-^2)}{r^2}\,.
\end{equation}
The solution is written in terms of $r_+$ and $r_-$, the outer and inner horizon radii, respectively. $\tE$ corresponds to Euclidean time and $\varphi \sim \varphi +2\pi$ the spatial coordinate of the conformal boundary with metric $\d s^2|_{\rm bdy} = \d \tE^2 + \d \varphi^2$. After imposing smoothness at the horizon, the parameters $r_+$ and $r_-$ are related to temperature and angular velocity by 
\beq
T = \frac{r_+^2 - r^2_-}{2\pi r_+} \,,\qquad \Omega= \frac{r_-}{r_+}\,,
\eeq
or equivalently the left-moving and right-moving temperatures
\beq
T_L = \frac{r_+-r_-}{2\pi}\,, \qquad T_R=\frac{r_++r_-}{2\pi}\,.
\eeq
In the grand canonical ensemble, these potentials enter the boundary geometry through the identifications $(\tE,\varphi) \sim (\tE+\beta, \varphi + \i \,\beta \Omega)$ together with $\varphi \sim \varphi+2\pi$. The charges of the solution are the mass and angular momentum
\beq
M = \frac{r_+^2 + r_-^2}{8 \,G_N}\,, \qquad J=\frac{r_+ r_-}{4 \,G_N}\,.
\eeq
Finally, the Brown-Henneaux derivation of the asymptotic symmetry implies a boundary central charge $c=\frac{3}{2\,G_N}$ \cite{Brown:1986nw}.  

The near-extremal limit we are interested in can be characterized in multiple ways, see \cite{Ghosh:2019rcj} for more details. In terms of left- and right-moving temperatures, we want to take $T_R$ fixed but with 
\beq
T_L \ll T_R \,.
\eeq
This limit also corresponds to low temperatures $T=2 \,\frac{T_L\, T_R}{T_L+T_R} \sim 2 \,T_L$ and fixed spin since $T_R \sim \sqrt{J/c}$ (or in terms of angular velocity $\Omega \sim 1$).

The classical on-shell action predicts a temperature-dependent mass and entropy in the near-extremal limit
\beq
M- J \sim 2\pi^2 \frac{T^2}{T_q}\,, \qquad S\sim 2\pi \sqrt{\frac{c\, J}{6}} + 4\pi^2 \frac{T}{T_q}\,.
\eeq
The temperature $T_q$ is the scale at which the classical approximation is argued to breakdown~\cite{Preskill:1991tb,Michelson:1999kn} and is given by
\beq\label{eq:TqBTZ}
T_q = \frac{24}{c} \,.
\eeq
The precise origin and consequences of the quantum corrections appearing at $T\lesssim T_q$ was explained in~\cite{Ghosh:2019rcj} for BTZ, and more generally in~\cite{Iliesiu:2020qvm}. We should emphasize that $T_q$ is a dimensionful scale, and the fact that we are working in units where the length of the spatial circle is $2\pi$ is implicit in \eqref{eq:TqBTZ}.

In fact, the quantum effects coming from the metric (as well as matter fluctuations) were computed in~\cite{Giombi:2008vd}. As already stated, the purpose of this section is not to reproduce the final answer, which is already known, but to identify the precise profile (in some convenient gauge) of the metric fluctuations in the full asymptotically \AdS{3} geometry corresponding to the so-called Schwarzian modes that live near the horizon. 

Let us begin by reviewing some generalities which will be useful in the rest of the article. We consider quantum fluctuations $h_{\mu\nu}$ around the BTZ metric $g_{\mu\nu}$ in~\eqref{eq:btzmet}
\begin{equation}
g_{\mu\nu} \to g_{\mu\nu} + h_{\mu\nu} \,.
\end{equation}
It will also prove useful to introduce the (partial) trace reversed fluctuation
\begin{equation}\label{eq:tracerevh}
\widetilde{h}_{\mu\nu} = h_{\mu\nu} - \frac{1}{2} g_{\mu\nu} h \,.
\end{equation}
We would like to integrate over all metric fluctuations. To do so, one needs to implement a gauge-fixing procedure. We do this by adding the following two terms to the action:
\beq
I_{\rm gf} = \underbrace{\frac{1}{32 \pi\, G_N}\,  
\int \d^3 x \, \sqrt{g} \; \nabla^\mu \widetilde{h}_{\mu\sigma} \,\nabla^\nu \widetilde{h}_{\nu}^{~\sigma}}_{\text{gauge fixing term}} + 
\underbrace{\frac{1}{32\pi \,G_N}\, \int \d^3 x \, \sqrt{g} \, \bar{\eta}_\mu\, (-g^{\mu\nu} \nabla^2 - R^{\mu\nu})\,\eta_\nu}_{\text{ghost action}}.
\eeq
The first one corresponds to our choice of gauge-fixing term, while the second corresponds to the action for the ghosts $\eta_\mu$ needed to introduce to carry out the Faddeev-Popov procedure. Since we aim to identify the Schwarzian modes, we focus on the quadratic action for $h_{\mu\nu}$ which is given by~\cite{Christensen:1978md,Christensen:1979iy}
\begin{equation}\label{eq:Lopmeasure}
I \supset \frac{1}{16\pi G_N} \int \d^3 x \sqrt{g} \, \widetilde{h}^{\mu\nu} (\Delta_L h)_{\mu\nu}
\end{equation}
The kernel appearing in this action is the spin-2 Lichnerowicz operator, which is
\begin{equation}\label{eq:Lichnerowicz_AdS}
\begin{split}
(\Delta_L h)_{\mu\nu} = -\frac{1}{4}\nabla_\rho \nabla^\rho h_{\mu\nu} + \frac{1}{2} \, R_{\rho(\mu}\,h\indices{_{\nu)}^\rho} -\frac{1}{2}\, R_{\mu\rho\nu\sigma}\, h^{\rho \sigma}  
- \left(R_{\sigma(\nu}-\frac{1}{4}\, g_{\sigma(\nu}\, R\right) h\indices{_{\mu)}^\sigma}  - \frac{\Lambda}{2} h_{\mu\nu}\,.
\end{split}
\end{equation}
This expression is actually valid in any dimension.
Let us recall the exact result for the one-loop quantum corrections originating from the metric in the grand canonical ensemble \cite{Giombi:2008vd} 
\bea
\log Z_{\text{graviton}}^{\text{1-loop}} &=& \log\left[ \prod_{n=2}^\infty \frac{1}{(1-e^{-4\pi^2 T_L n})(1-e^{-4\pi^2 T_R n})}\right]\nonumber\\
&\sim&  \underbrace{\frac{1}{24 \,T_L}}_{\text{Shift of extremal energy}}+\underbrace{\frac{3}{2} \log T_L}_{\text{Schwarzian mode}}\label{eq:GMY} 
\ea
Since the expression is naturally written in terms of left- and right-moving temperatures,\footnote{This expression for $Z_{\text{graviton}}^{\text{1-loop}}$ is nothing else than the product of left- and right-moving vacuum Virasoro characters. This was important in~\cite{Ghosh:2019rcj} but will not play a crucial role in our present discussion.} in the second line we took the near-extremal limit by keeping $T_R$ fixed and taking $T_L$ to be small. The first term in $\log Z_{\text{graviton}}^{\text{1-loop}}$ in the expression above corresponds to a renormalization of the extremal mass of the black hole, and will not be important for our purposes. What we are after is to identify the modes responsible for the second term in $\log Z_{\text{graviton}}^{\text{1-loop}}$ which, for obvious reasons, are referred to as the Schwarzian modes. In~\cite{Ghosh:2019rcj} it was explained how to reproduce such modes from analyzing quantum effects in the \AdS{2}$\times \mathbf{S}^1$ throat of the near-extremal BTZ black hole. Our goal here is to identify them in the full geometry, without restricting to the throat.\footnote{An attentive reader might be surprised that \eqref{eq:GMY} presents a $\frac{3}{2}\,\log T$ correction coming solely from the Schwarzian mode. The isometries of the throat are ${\rm SL}(2,\mathbb{R})\times {\rm U}(1)$ which suggest the presence of another contribution $\frac{1}{2}\, \log T$ from angular momentum fluctuations in the grand canonical ensemble. We will explain in~\cref{sec:BTZ_rot_mode} the reason for the rotational modes being absent in the full geometry.}

\medskip

We shall immediately present our result for the precise Schwarzian profiles, analyze their physical properties, and then mention how they were derived (and how the possibility of further zero modes for this system is excluded). In the harmonic gauge we are using to gauge-fix, the Schwarzian profiles are given by,
\begin{equation}\label{eq:hbtzb}
\begin{split}
h^{(n)}_{\mu\nu} \,\d x^\mu \,\d x^\nu 
&= 
     H_n(r) \,  
    \left[ \Big(r^2 \,f(r)^2 (\d \varphi-\i\,\d \tE)^2 + (r_+ + r_-)^2\, \d r^2\Big)\cos\Big( \frac{2\pi \,n\,\tE}{\beta}\Big) \right. \\
&\left.\qquad \qquad 
    - 2\, \i\, (r_+ + r_-) \,r\, f(r)\, \d r \,(\d \varphi - \i \,\d \tE)\sin \Big(  \frac{2\pi \,n\,\tE}{\beta}\Big) \right],
\end{split}
\end{equation}
where the function appearing in front of the metric is given by
\bea
H_n(r) = N_n\,2\, r^2 \,\Big( \frac{r^2-r_+^2}{r^2-r_-^2}\Big)^{\frac{\abs{n}}{2}-2} \, \Big( \frac{r_+^2 - r_-^2}{r^2-r_-^2} \Big)^{4-\frac{\abs{n}}{2} \frac{r_+ - r_-}{r_+}}\,.
\ea
The temporal dependence is fixed in terms of Matsubara frequencies given the identification $\tE \sim \tE + \beta$.
Looking at the full profile for $h^{(n)}_{\mu\nu}$ and how it behaves near the horizon shows that fluctuations with $\abs{n}<2$ are singular, while fluctuations with $\abs{n}\geq 2$ are smooth. The range for $n$ therefore excludes $n=0$ and $n=\pm 1$. $N_n$ is a normalization constant we will fix imminently. Let us first record some special properties of these fluctuation profiles:
\begin{itemize}
\item All these modes are both transverse and traceless
\beq
\nabla^\mu h_{\mu \nu} = h^\mu_{~\mu}=0,
\eeq
guaranteeing that it corresponds indeed to a spin-2 tensor. These two conditions imply that it also satisfies the harmonic gauge $\nabla^\mu \widetilde{h}_{\mu\nu} = 0$. Representations of the modes in other gauge-fixing implementations are related by diffeomorphisms. 

\item These modes are eigenmodes of the Lichnerowicz operator. Specifically, they satisfy\footnote{Note added in v2:  We have presented the eigenvalues for generic non-extremal BTZ black holes (in the sector displaying gapless modes at low temperatures). However, not all the modes in question are  in the physical spectrum. They fail to satisfy the boundary conditions  when $\lambda$ crosses $1/4$, at which point they merge with the continuum part of the spectrum. Thus, for finite $T_L$ we only have a finite number of modes in the eigenspectrum, whereas in the limit we are interested in $T_L \to 0$, the spectrum forms a discretuum. Curiously, despite this lack of normalizability, the ultralocal norm ends up being finite even when the modes are unphysical. Nevertheless, these subtleties do not affect the low-temperature expansion of the one-loop determinants. We thank Alejandra Castro for raising this issue.
}
\begin{equation}\label{eq:BTZLicheq}
(\Delta_L h)_{\mu\nu} = \underbrace{\frac{ \abs{n}\, T_L}{T_L+T_R} \Big( 1 - \abs{n} \frac{T_L}{T_L+T_R}\Big)}_{=\lambda_n} h_{\mu\nu} \,.
\end{equation}
The main property making these modes special is that the eigenvalue of the Lichnerowicz operator vanishes linearly as $T \sim T_L \to 0$, with $T_R$ fixed. Moreover, in the low-temperature limit the eigenvalues are proportional to the Matsubara frequency $n$\footnote{We remind the reader that the Schwarzian modes, when properly normalized by their natural symplectic measure, also have an action proportional to their Matsubara frequency, see for example \cite{Saad:2019lba} or \cite{Moitra:2021uiv}.}. These are the defining property of the Schwarzian modes, spin-2 fluctuations that are nearly zero-modes near extremality.

\item We argued earlier that, just like the Schwarzian modes, we should not include $n=0$ and $n=\pm 1$. In the approach that focuses on the throat, this restriction is due to the isometries of \AdS{2}. We can reach the same conclusion by recalling that the path integral measure over the metric fluctuations is defined through the ultralocal measure\footnote{In writing this formula we used the fact that we are working in units with $\lads=1$, otherwise some powers of $\lads$ should be included by dimensional analysis.}
\begin{equation}\label{eq:ulnorm}
\int {\rm D}h_{\mu\nu} \, e^{-\norm{h}^2} = 1\,, \qquad 
\norm{h}^2 = \int \d^3 x \sqrt{g}\,  \widetilde{h}^{\mu\nu} h_{\mu\nu}\,.
\end{equation}
We can expand metric fluctuations as 
\beq
h_{\mu\nu} = \sum_n \alpha_n h^{(n)}_{\mu\nu} + h^{\perp}_{\mu\nu},
\eeq
where $h^{\perp}_{\mu\nu}$ denotes metric fluctuations perpendicular to all $h^{(n)}_{\mu\nu}$ according to the inner product introduced above. Moreover, all $h^{(n)}_{\mu\nu}$ are perpendicular to each other since their eigenvalues are different. To determine the measure, we normalize the profiles $h^{(n)}_{\mu\nu}$ according to
\beq
\norm{h^{(n)}}^2 = N_n^2 \,\frac{16\,\pi^2 \,r_+ \,(r_++r_-)^8\,
    (r_+-r_-)^4 \,\Gamma\left(3 - \abs{n} +\abs{n}\frac{r_-}{r_+}\right)}{\Gamma\left(2+ \abs{n} \frac{r_-}{r_+}\right)}\, \Gamma(\abs{n}-1)\,.
\eeq
The final term on the right-hand side makes it clear that the norm diverges when $n=0$ or $\pm1$, indicating that the mode is non-normalizable and should not be included in the path integral. We can fix the normalization constant $\alpha_n$ for $n\neq 0, \pm 1$ such that $\norm{h^{(n)}}^2 =1$. The integration measure is then
\beq
\int {\rm D}h_{\mu\nu} = \int {\rm D}h^{\perp}_{\mu\nu} \int \prod_{\abs{n}\geq 2} \frac{\d \alpha_n}{\sqrt{\pi}},
\eeq
including solely normalizable modes.  A similar restriction on $n$ occurs in the evaluation of the one-loop determinant based on the ${\rm AdS}_2 \times \mathbf{S}^1$ throat. In that case, this restriction is due to the isometries of ${\rm AdS}_2$ making the $n=0,\pm1$ modes unphysical.

\item As we approach extremality, the modes are localized near the horizon.\footnote{We will give an explicit mapping to the Schwarzian modes of the near-horizon region in~\cref{sec:BTZ_rot_mode},} We can first show that the normalized modes vanish at a fixed value of $r>r_+$, in the limit $r_- \to r_+$. To see this, notice that $H_n (r) \sim N_n  (r_+ - r_-)^4$ while $N_n \sim (r_+-r_-)^{-2}$ implying that $H_n(r)\sim (r_+-r_-)^{2}\to 0$ as $r_+\to r_-$ at fixed $r$. This implies that either the modes disappear completely at extremality, or they become localized near the horizon. To show that the second option is correct, we consider the following observable, the regularized length of the throat
\beq
L = \int_{r_+}^\infty \,\d r\,\sqrt{g_{rr}+h_{rr}}  \approx \underbrace{\int_{r_+}^\infty\, \d r\, \sqrt{g_{rr}}}_{\text{BTZ length of the throat}} +\underbrace{ \int_{r_+}^\infty \,\d r\, \frac{h_{rr}}{2 \,\sqrt{g_{rr}}}}_{\text{fluctuation }\delta L} + \cdots\,.
\eeq
The BTZ answer is strictly divergent. We can, however, regulate the result by subtracting off this divergence and focusing instead on the change in length $\delta L$. Explicit integration leads, for one mode $\alpha_{n\geq 2}$, to
\beq
\delta L = \alpha_n \cos\Big(\frac{2\pi n \tE}{\beta}\Big)\,  \frac{N_n (r_+-r_-)^2\, (r_++r_-)^4\, \Gamma\left(\frac{n-1}{2}\right) \Gamma\left(2-\frac{n}{2}(1-\frac{r_-}{r_+})\right)}{4\,\Gamma\left(\frac{3}{2} + \frac{n}{2} \frac{r_-}{r_+}\right)} \,.
\eeq
Importantly, now the effect of the fluctuation mode on the length remains finite at zero temperature since $N_n\sim (r_+-r_-)^{-2}$, leading to $\delta L \sim \alpha_n \cos(\frac{2\pi n \tE}{\beta}) $. This shows that at extremality the zero-mode corresponds precisely to fluctuations in the length of the throat, which is the same geometric interpretation as the Schwarzian modes in the ${\rm AdS}_2$ description.

\begin{figure}[t!]
\centering
\includegraphics[width=0.5\textwidth]{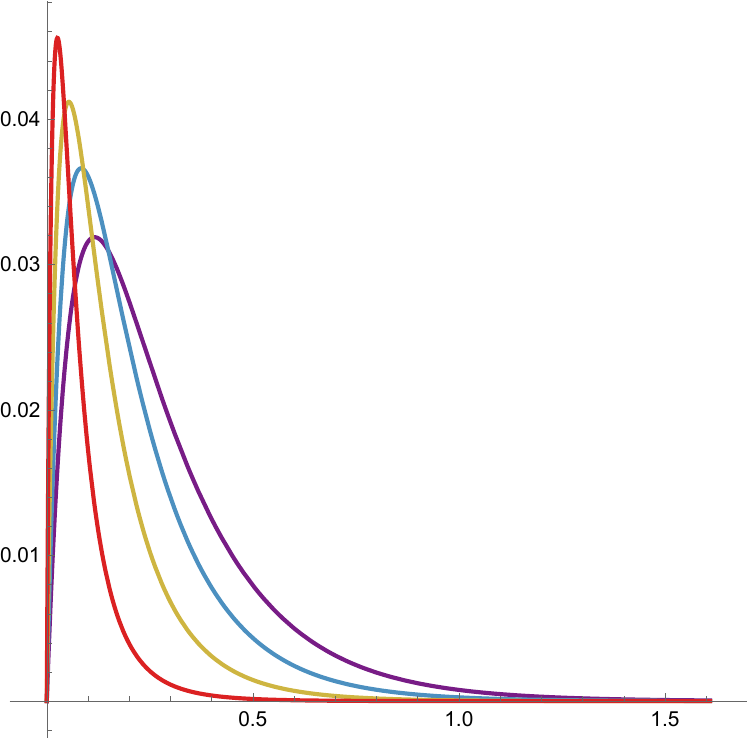}
\begin{picture}(0.3,0.4)(0,0)
\put(-230,220){\makebox(0,0){$\norm{h}^2_\text{den}$}}
\put(10,0){\makebox(0,0){$\log(r/r_+)$}}
\end{picture}
\vspace{0cm}
\caption{Plot of the norm density defined in~\eqref{eq:btznormden} as a function of the proper radial distance $\log(r/r_+)$ for the $n=3$. The figure shows the result for $r_-/r_+=0.6$ (purple), $0.7$ (blue), $0.8$ (yellow), $0.9$ (red). We see that, for a fixed $r$, the amplitude of the fluctuation becomes smaller as the black hole becomes colder. Instead, the maximum of $\sqrt{g} h^{\mu\nu} h_{\mu\nu}$ increases but progressively localizes closer to the horizon. Indeed, in the extremal limit the maximum amplitude takes place at a finite proper distance from the horizon, which is infinite proper distance down the throat from the boundary.}
\label{fig:h_squared_btz}
\end{figure}

\item We can also see that the modes are localized at a finite proper distance from the horizon. (We focus on $n\geq 2$ for simplicity; analogous formulas apply for $n\leq -2$.) To do so, let us change coordinates from $r$ to the proper distance from the horizon $z$:
\begin{equation}
    z = \textrm{arctanh} \left(
   \sqrt{ \frac{r^2-r_+^2}{r^2-r_-^2}}
    \right).
\end{equation}
Notice that for any $r>r_+$, in the extremal limit the proper distance blows up. We now may consider wavefunction density, the integrand of $\norm{h}^2$, which is 
\begin{equation}\label{eq:btznormden}
 \norm{h^2}_\text{den} =  \sqrt{g}\, h^{\mu \nu} h_{\mu \nu} \,.
\end{equation}
It is quite a lengthy function but easy to analyze. In particular, it vanishes both on the horizon ($z=0$) and at infinity ($z \to \infty$). Moreover, it has only one maximum at
\begin{equation}
    z = \textrm{arctanh} \left(\sqrt{\frac{(2\, n-3)\, r_+}{2 \,n \,r_- + 3\, r_+}}\right).
\end{equation}
In particular, in the extremal limit $r_- \to r_+$, it has a finite limit
\begin{equation}
  z=  \textrm{arctanh} \left(\sqrt{\frac{2\, n-3}{2\, n+3}}\right) < \infty.
\end{equation}
Moreover, if we evaluate the density,
\begin{equation}
    \int_0^\beta\, \d \tE \int_0^{2\pi}\, \d\varphi\, \sqrt{g}\, h^{\mu \nu}\, h_{\mu \nu},
\end{equation}
it is easy to check that it is finite at the maximum found above, even in the $r_- \to r_+$ limit. This is consistent with the solution vanishing for any $r>r_+$ because we need to `zoom' into the near-horizon region to see this result. At extremality all points with $r>r_+$ are located at an infinite proper distance from the horizon.
This strongly suggests that these modes are indeed localized deep in the \AdS{2} throat.
\item We can now add up the contributions from all these modes, after properly normalizing the eigenvalues of the Lichnerowicz operators according to the Einstein action.
\beq
\frac{1}{16\pi \,G_N} \, \int \d^3 x \, \sqrt{g} \,  \widetilde{h}^{\mu\nu}(\Delta_L h)_{\mu\nu} = 
\sum_{n\neq -1,0,1} \,\frac{1}{16\pi \,G_N}\, \lambda_n \, \textbf{}\alpha_n^2 + (\text{orthogonal modes})
\eeq
The orthogonal modes include $h^{\perp}_{\mu\nu}$. Assuming $\Delta_L$ is non-degenerate, there is no mixing between $h^\perp_{\mu\nu}$ and the $h^{(n)}_{\mu\nu}$.\footnote{This can be verified in BTZ but will need to be assumed in the higher-dimensional cases.} Since no matter field is turned on in the background apart from the metric, all other matter fluctuations are automatically orthogonal as well. The properly normalized contribution to the one-loop determinant from $h^{(n)}_{\mu\nu}$ is
\begin{equation}
\begin{split}
\log Z_{\rm graviton}^{\rm 1-loop} 
&\supset 
    -\frac{1}{2}\sum_{n\neq 0, \pm1} \log \Big(\frac{\lambda_n}{16 \pi\, G_N} \Big) \\ 
&=
    -\frac{1}{2}\sum_{n\neq 0, \pm1}\log \Big[\frac{1}{16 \pi\, G_N} \, \frac{ \abs{n} \,T_L}{T_L+T_R} \,\Big( 1 - \abs{n} \,\frac{T_L}{T_L+T_R}\Big) \Big]\\
&=
   - \sum_{n\geq 2} \log \Big(\frac{c}{24} \frac{ n\, T}{\sqrt{J/c}} \Big)\,\Big(1 +\cdots \Big) \\
&= \frac{3}{2} \log \Big( \frac{T}{T_q}\Big) + \cdots \,,
\end{split}
\end{equation}
where the final step follows via zeta-function regularization as described in~\cite{Iliesiu:2022kny,Iliesiu:2022onk}.
The ellipsis denotes terms that are finite in both the small temperature and in the large $c$ limits. This contribution reproduces the Schwarzian limit of the exact answer \eqref{eq:GMY}.

\end{itemize}

Therefore, \eqref{eq:hbtzb} is the desired result. For our choice of gauge-fixing, we have found the precise profile of the Schwarzian mode away from the throat region. In the rest of this section, we will extend this to black holes in four or higher-dimensions, although the analog of the mode functions such as $H_n(r)$ will have to be found numerically. For this reason, we discussed the BTZ case first.

Our strategy for solving for the modes was straightforward. We parameterized the metric fluctuation in terms of arbitrary axisymmetric functions. We use tracelessness and the divergence free condition to eliminate some functions, and then massage the eigenvalue equation for the Lichnerowicz operator $(\Delta_L h)_{\mu\nu} = \lambda\, h_{\mu\nu}$ to arrive at a second order ODE in the radial coordinate, which we then solved (for the explicit strategy see ~\cref{sec:HypSchw}).  The analysis was aided in part by the results of~\cite{Datta:2011za} and~\cite{Castro:2017mfj}. In fact, the result we seek is implicitly present in the latter work. We verified that our analysis agrees with theirs by direct computation.

Before continuing, two comments are in order. First, this is only one specific contribution to the full answer $Z_{\rm graviton}^{\rm 1-loop}$. Thanks to the solvability of the eigenvalue problem around the BTZ metric, cf.,~\cite{Datta:2011za,Castro:2017mfj}, one can verify these are the full set of modes with eigenvalue vanishing with the temperature. Any other mode contributes a temperature independent correction at extremality. Second, this should be surprising since one might expect zero-modes due to the rotational symmetry of BTZ. The exact result \eqref{eq:GMY} suggests these modes are not present, and we will analyze this problem further in~\cref{sec:BTZ_rot_mode}.

\subsection{Hyperbolic black hole}

Before we dive into the physically most interesting case of Reissner-Nordstr\"om black hole, let us consider a simpler pure gravity background also in four dimensions -- the Schwarzschild-AdS{4} solution with a hyperbolic horizon. Lack of both matter and rotation will allow us to identify relevant modes with greater ease. The geometry in question is a solution of Einstein-Hilbert theory with action
\beq
I = - \frac{1}{16 \pi\, G_N} \int \d^4 x \, \sqrt{g} \, (R - 2 \Lambda) + I_{\rm bdy} + I_{\rm matter}.
\eeq
The hyperbolic AdS black hole metric is  
\begin{equation} \label{eq:hyp_metric}
ds^2 = f(r)\,  \d \tE^2 + \frac{\d r^2}{f(r)} + r^2 \, \d \Sigma_H^2 \,, \qquad 
f(r)  = \frac{r^2}{\lads^2} - 1 - \frac{r_+}{r} \left(\frac{r_+^2}{\lads^2} - 1 \right) .
\end{equation}
and $\d\Sigma_H^2$ is a line element on any 2d surface with the Ricci curvature $-2$. Since we expect that Schwarzian modes should be in `s-wave' sector, the choice of the surface is not relevant.\footnote{A priori, we could take the 2d geometry to be a compact Riemann surface with genus $g >1$. However, holographic CFTs on compact hyperbolic spaces often have energies unbounded from below owing to the presence of conformally coupled scalars.  We will therefore work with the non-compact space $\mathbb{H}_2$ with an IR regulator. }  The horizon is located at $r=r_+$. It is easy to check that \eqref{eq:hyp_metric} satisfies Einstein equations with the cosmological constant $\Lambda = -\frac{3}{\lads^2}$. The surface gravity of the horizon, which is proportional to the black hole temperature, reads:
\begin{equation}
    \kappa := 2\pi\, T= \frac{1}{2} \, f'(r_+) = \frac{3\,r_+}{2\,\lads^2}-\frac{1}{2r_+}\,.
\end{equation}
We may proceed exactly as in the previous section with the expansion of the action and land on the same operator as in~\eqref{eq:Lichnerowicz_AdS}. It is our task to diagonalize that operator. 

\medskip 

This black hole has a near-extremal limit not too different from the one relevant for charged black holes. The reason is that the surface gravity can vanish at a finite value of $r_+$ given by
\beq
\kappa = 0 \;\;\Rightarrow\;\; r_+ = r_{\rm ext} = \frac{\lads}{\sqrt{3}}.
\eeq
The fact that the zero-temperature limit retains a macroscopic horizon suggests similar conceptual problems as the one identified in~\cite{Preskill:1991tb, Michelson:1999kn,Page:2000dk} and solved in~\cite{Ghosh:2019rcj,Iliesiu:2020qvm,Heydeman:2020hhw} by identifying the Schwarzian modes. Indeed, the near-extremal near-horizon geometry develops an \AdS{2} throat. The near-horizon version of that problem may be found in~\cite{Emparan:2023ypa}. The goal of this section is to identify the Schwarzian modes in the full asymptotically \AdS{4} higher-dimensional geometry.

At the classical level, we can verify the Schwarzian behavior by computing the low-temperature energy and entropy
\beq
S \sim  \frac{\lads^2 \,V_H}{12 \,G_N} + 4\pi^2 \frac{T}{T_q}\,,
\qquad 
E \sim\frac{\lads^3}{12\sqrt{3}\,\pi} \frac{V_H}{G_N} + 2\pi^2 \,\frac{T^2}{T_q},
\eeq
where $V_H$ is the (dimensionless) volume of the hyperbolic 2d surface $\Sigma_H$. This has precisely the form predicted by JT gravity and allows us to identify the temperature scale 
\beq
T_q= \frac{G_N}{V_H}\,\frac{12\sqrt{3}\pi}{\lads^3},
\eeq
at which the classical approximation breaks down and quantum effects are crucial.

\subsubsection{Schwarzian modes}\label{sec:HypSchw}

As already noted, we anticipate the Schwarzian modes should be homogeneous in the hyperbolic space. We can therefore preserve the 2d hyperbolic symmetry, and pick as our ansatz for the perturbation the following line-element: 
\begin{equation}
    h_{\mu \nu}\, \d x^\mu \, \d x^\nu = 
        e^{\frac{2\pi \i\, n}{\beta}\,\tE} \left(
f_1(r)\, f(r) \,\d \tE^2 + f_2(r)\, \frac{\d r^2}{f(r)} + 2\,\i\, f_3(r)\, \d \tE\, \d r + r^2\, f_4(r) \, \d \Sigma_H^2
    \right),
\end{equation}
where, as indicated, the functions $f_i(r)$ for $i\in\{1,\cdots, 4\}$  depend only on the radial coordinate. We have inserted factors of $f(r)$ and $r^2$ for convenience. They ensure that this perturbation will be in $L_2$ with respect to the ultralocal measure set by $\norm{h}^2$ as defined by
\begin{equation}\label{eq:ulnorm22}
\int {\rm D}h_{\mu\nu} \, e^{-\norm{h}^2} = 1\,, \qquad 
\norm{h}^2 = \frac{1}{\lads^4}\int \d^4 x \sqrt{g}\,  \widetilde{h}^{\mu\nu} h_{\mu\nu}\,,
\end{equation}
provided  $f_i$ are in $L^2$ (i.e., square-integrable) with respect to the uniform measure on $[r_+, \infty)$. We have also used the periodicity of the Euclidean time circle $\tE \sim \tE  + \beta$ to expand in Matsubara modes in time.\footnote{We explicitly only examine modes with $n\neq 0$. One can independently show that there are no nearly-gapless time-independent modes, but we will not explicitly demonstrate that in our analysis.\label{fn:nozerofreq} }  Since we are not restricting ourselves to the near-horizon, we expect Schwarzian modes to be physical (as opposed to pure gauge). Therefore, they should be traceless and transverse. It follows that we want to solve the following system, 
\begin{equation}\label{eq:HypLicheqns}
 \Delta_L h_{\mu \nu} = \frac{\lambda}{\lads^2}\,  h_{\mu \nu} \,, 
 \qquad 
 \nabla^\mu h_{\mu \nu} =0\,, \qquad  g^{\mu \nu} \, h_{\mu \nu}= 0  \,.
\end{equation}
The tracelessness condition allows us to eliminate $f_4(r)$ algebraically
\begin{equation}\label{eq:Hypf4sol}
     f_4 = -\frac{1}{2}\left(f_1 + f_2 \right),
\end{equation}
The transversality condition can be used to eliminate $f_1$, 
\begin{equation}
\begin{split}
f_1 
&= 
    \frac{1}{\beta  \left(\lads^2 (2 \,r- 3\, r_+)
        +3 \,r_+^3\right)} \bigg[\beta\, f_2 \left(\lads^2\, (5\, r_+ - 6\, r)+8\, r^3   -5 \,r_+^3\right) \\ 
&\qquad         
    +    2\, r \left(\beta \, f_2'\, (r_+ - r)  
    \left(\lads^2\,  - r^2 - r\, r_+ - r_+^2\right)- 2\pi \,  \
    \lads^2\, n\, r\, f_3\right) 
        \bigg]
\end{split}        
\end{equation}
Finally, $f_2$ and $f_3$ can themselves be expressed in terms of a scalar potential $u(r)$
\begin{equation}
f_2 = u - \frac{\beta}{4\pi\, \lads^2\, n\, r^2} \, 
     \left(\lads^2\,  (2\, r-3 \,r_+)+3\, r_+^3\right) f_3\,,
\end{equation}
and 
\begin{equation}
f_3 =  
    -\frac{4\pi \, \beta  \,\lads^2 \,n \,r^2 \left(
     \left(\lads^2 \,(6\, r-5 \,r_+)-8\, r^3+5\, r_+^3\right) u +
   2\, r \left(\lads^2 \,(r-r_+) - r^3 + r_+^3\right) u'\right)}{4\, \lads^2 \,r^4 ((2 \pi  \lads\, n)^2-\beta^2 )+\beta ^2 \,r_+\, (r_+^2-\lads^2) \left(\lads^2 \, (4\, r -3 \, r_+)  - 12\, r^3 + 3\, r_+^3\right)}.
\end{equation}

With these changes, the eigenvalue equation, the first equation in~\eqref{eq:HypLicheqns}, 
becomes a second order generalized eigenvalue problem for $u(r)$. Its explicit form is not very illuminating, so instead of writing it down, we only note a few salient properties. Before doing so, it is helpful to change variables from $r$ to $z$ by the map,
\begin{equation}\label{eq:rzmap}
  r = \frac{4\, r_+}{4-(1+z)^2}\,.  
\end{equation}
This makes the domain of integration compact, $z \in [-1,1]$. The two end-points $z= \pm 1$ end up being regular singular points of the eigenvalue equation.  
Let us examine the solutions near the boundaries. 

Near the asymptotic boundary, i.e., in the neighborhood of $z=1$, we find
\begin{equation}
    u(z) \sim (1-z)^{s_\pm}\,, \qquad s_\pm = \frac{1}{2} \left(7\pm\sqrt{9-16 \,\lambda}\right) .
\end{equation}
We require the solutions to preserve the asymptotic form of the metric, viz., obey Dirichlet boundary conditions. This is only achieved by the local solution that decays faster, viz., $(1-z)^{s_+}$. 
 
Near the horizon, viz., in the vicinity of $z=-1$, we find
\begin{equation}
    u(z) \sim (1+z)^{s'_\pm}\,, \qquad s'_\pm = -2 \pm n\,.
\end{equation}
The solution will be regular\footnote{To be precise, for regularity we demand that the solutions be sufficiently smooth to lie in $L^2$ with respect to the ultralocal measure. We will use this definition of regularity in the rest of this section. } at the horizon, only if we pick the faster decay; this requires the choice $s' = -2 + \abs{n}$. This, however, is not enough, we must also impose $\abs{n} \ge 2$. 

All told, from the local Frobenius analysis, we are led to a further substitution
\begin{equation}\label{eq:uwredef}
    u(z) =  (z+1)^{\abs{n}-2} \,(1-z)^\frac{5}{2}\, w(z)\,.
\end{equation}
Now the function $w(z)$ must be merely regular at the horizon and vanish at $z=1$. (Note that $w$ will be $C^2$ but not $C^3$ at that boundary). With this set of boundary conditions, our eigenvalue problem is well-posed as long as $\lambda<\frac{5}{16}$. For larger values, both solutions decay at  $z=1$. 

\begin{figure}[ht]
\begin{subfigure}[b]{\linewidth}
\centering
\subcaptionbox{The eigenvalue spectrum at low temperatures. \label{fig:hypschwzoom}}[\linewidth]{
\includegraphics[width=0.7\textwidth]{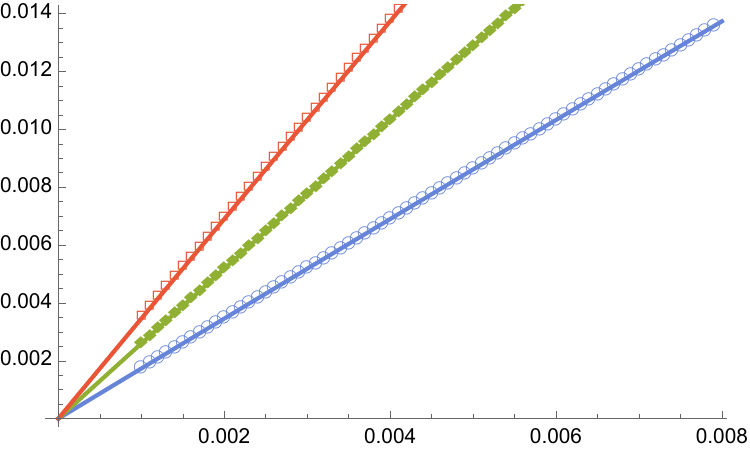}
\begin{picture}(0.3,0.4)(0,0)
\put(-320,175){\makebox(0,0){$\lambda_n$}}
\put(0,25){\makebox(0,0){$\lads\,\kappa$}}
\put(-210,135){\makebox(0,0){$n=4$}}
\put(-147,160){\makebox(0,0){$n=3$}}
\put(-130,125){\makebox(0,0){$n=2$}}
\end{picture}}
\end{subfigure}
\vspace{2cm}
\begin{subfigure}[b]{0.45\linewidth}
\centering
\subcaptionbox{The eigenvalue spectrum over an extended range.\label{fig:hypschwfull}}[\linewidth]{
\includegraphics[width=0.95\textwidth]{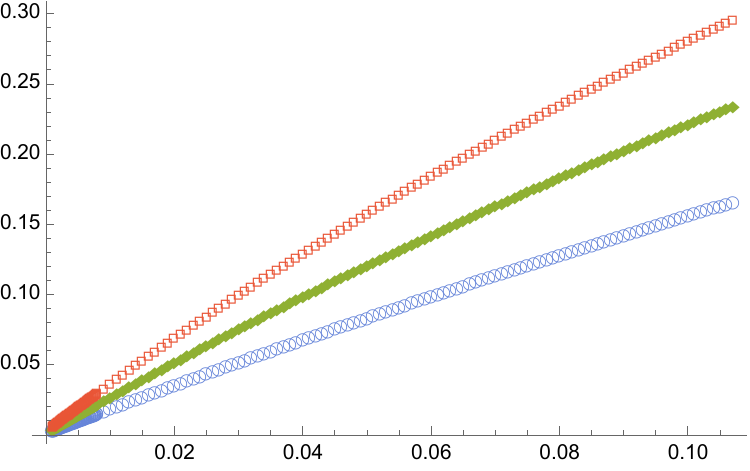}
\begin{picture}(0.3,0.4)(0,0)
\put(-200,100){\makebox(0,0){$\scriptstyle{\lambda_n}$}}
\put(0,15){\makebox(0,0){$\scriptstyle{\lads\,\kappa}$}}
\end{picture}}
\end{subfigure}
\hspace{1cm}
\begin{subfigure}[b]{0.45\linewidth}
\subcaptionbox{Rescaled eigenvalues demonstrating scaling.\label{fig:hypschwcollapse}}[\linewidth]{
\includegraphics[width=0.95\textwidth]{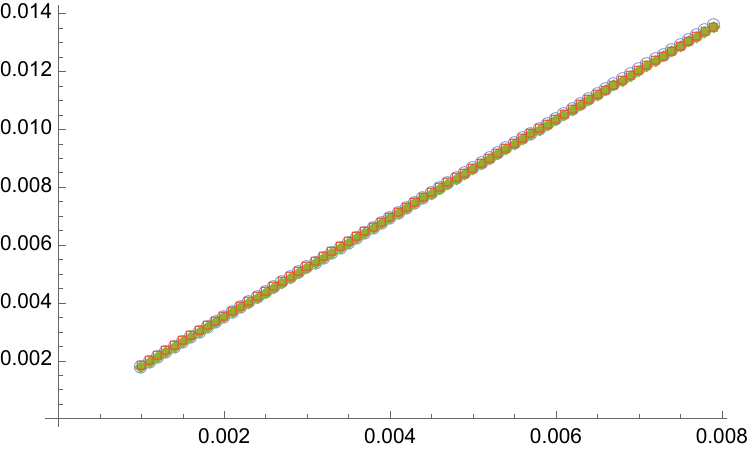}
\begin{picture}(0.3,0.4)(0,0)
\end{picture}}
\put(-210,100){\makebox(0,0){$\scriptstyle{\frac{2}{n}\lambda_n}$}}
\put(-20,15){\makebox(0,0){$\scriptstyle{\lads\,\kappa}$}}
\end{subfigure}
\vspace{-1.5cm}
\caption{The uplift of Schwarzian modes to the full hyperbolic \AdS{4} black hole geometry. We show the data for very low temperatures in the top panel,~\cref{fig:hypschwzoom}, demonstrating linear behavior with temperature. In the bottom left panel,~\cref{fig:hypschwfull}, we show the behavior over a greater range of temperatures, making the departures from linear growth manifest. Finally, in the bottom right panel~\cref{fig:hypschwcollapse}, we plot the eigenvalues rescaled by the overtone number $\frac{2}{n}\,\lambda_n$ to demonstrate that $\lambda_n \sim n\,T$ (notice the three curves are overlapping, which was achieved by working with extended precision numerics). }
\label{fig:hyperbolic_schwarzian}
\end{figure}

In the numerical analysis, we tracked the lowest set of eigenvalues of $\Delta_L$ as a function of temperature (for a fixed Matsubara frequency labeled by $n\in \mathbb{Z}$). The generalized eigenvalue problem for $u$ may be written using~\eqref{eq:uwredef} as the following ODE for $w(z)$:
\begin{equation}
    \mathcal{D} w = \lambda \,\mathcal{G} \, w,
\end{equation}
where
\begin{equation}
    \mathcal{D} = A(z) \,\partial_{z}^2 + B(z) \,\partial_z + C(z)
\end{equation}
and $\mathcal{G}(z)$ is a specific multiplicative factor in the source term proportional to the eigenvalue, fixed in terms of our choice of the Sturm-Liouville operator $\mathcal{D}$. We discretized the interval $[-1,1]$ into Gauss-Lobatto grid, $(z_i)$ with $i \in \lbrace 1,\cdots, N+1 \rbrace$, on which we used a spectral collocation method. Writing the unknown function $w$ as a vector of values on the grid $\vec{w} = \left(w(z_i) \right)$, we may approximate derivatives of $w$ by matrix multiplication by a certain matrix $\mathcal{D}_{\textrm{discretized}}$.\footnote{For more details on spectral methods and explicit implementations, see~\cite{trefethen2000spectral}. For the review of using them in gravity, see~\cite{Dias:2015nua}.} Then, the operators $\mathcal{D}$ and $\mathcal{G}$ takes the discretized form:
\begin{equation}
    \mathcal{D}_{\textrm{discretized}} = A(\vec{z})\, D^2 + B(\vec{z}) \, D + C(\vec{z})
\end{equation}
and
\begin{equation}
    \mathcal{G}_{\textrm{discretized}} = \mathcal{G}(\vec{z}),
\end{equation}
where $A(\vec{z})$ denotes a diagonal matrix with the diagonal entries given by values of $A(z)$ on the grid points and the same {\it mutatis mutandis} holds for $B(\vec{z})$, $C(\vec{z})$, and $\mathcal{G}(\vec{z})$. To implement the boundary conditions at infinity, we simply replace the row in $\mathcal{D}_{\textrm{discretized}}$  corresponding to $z=1$ with zeroes. The functions $A$ and $\mathcal{G}$ vanish at $z=-1$. The conditions at $B(-1)$ and $C(-1)$ ensure that the solution is regular (and not blowing up) at the horizon, so no further change is needed. Having done all that, we may just ask for the eigenvalues for this generalized eigenvalue problem using built-in functions in Mathematica. We found that already at $\kappa\, \ell_{AdS} = O(10^{-2})$ machine precision was not sufficient, and so we had to work with extended precision. The results of this numerical analysis for the lowest three positive frequency Matsubara modes ($n=2, 3,4$) is illustrated in~\cref{fig:hyperbolic_schwarzian}. 

The numerical data for the eigenvalues for $n=2,3,4$ fit well to a quadratic polynomial for low values of surface gravity, $\kappa\, \lads \in [0.001, 0.008]$. Letting, 
\begin{equation}
 \lambda_n = a_n + b_n \, \kappa\, \lads + c_n \, (\kappa\, \lads)^2   \,,   
\end{equation}
we find 
\begin{equation}
\begin{aligned}
b_2 &= 1.73195  \,, & \qquad c_2 &= - 1.97407 \,, & \qquad a_2 \sim \order{10^{-7}}\,, \\ 
b_3 &= 2.59772 \,, & \qquad c_3 &= -4.40843 \,, & \qquad a_3 \sim \order{10^{-7}} \,. \\ 
b_4 &= 3.46369\,, & \qquad c_4 &=  -7.82459 \,, & \qquad a_4 \sim \order{10^{-7}} \,.
\end{aligned}
\end{equation}
This is indeed consistent with these modes becoming zero modes as we attain extremality, i.e., at  $\kappa =0$. In particular, as expected, at very low temperatures, we have a set of modes whose eigenvalues scale linearly with temperature. Furthermore, 
\begin{equation}
    \frac{b_3}{b_2} = \frac{2.59772}{1.73195} \approx \frac{3}{2} - 0.000012\,,\qquad \frac{b_4}{b_3} = \frac{3.46369}{2.59772} \approx \frac{4}{3} - 0.000021\,.
\end{equation}
This illustrates that the eigenvalues have a slope that scales linearly with the overtone number $n$, consistent with them being the extension of the Schwarzian mode onto the full geometry.\footnote{The discrepancy of $\order{10^{-5}}$ is mainly because we truncated the Taylor expansion at quadratic order.} The upshot of this numerical results is that one has\footnote{A factor of $\lads^{-2}$ arises from our normalization of the Lichnerowicz operator eigenvalue, while another factor of $\lads^4$ arises from the normalization of the inner product $\norm{h}^2$.} 
\begin{equation}
    \lim_{T\to 0} \frac{\lambda_n\, \lads^2}{16\pi G_N } \approx \textbf{} \alpha\, n\, \frac{T}{T_q} + \order{T^2}\,,\qquad \abs{n} \geq 2\,,
\end{equation}
in excellent agreement with our expectation based on the near-horizon analysis  
(for explicit investigation in the case of the hyperbolic black hole, cf.~\cite{Emparan:2023ypa}). This result implies that the graviton one-loop determinant around the near-extremal hyperbolic black hole is
\beq
\log Z_{\rm graviton}^{\rm 1-loop} \supset - \sum_{n\geq 2} \log \Big(\frac{\lambda_n \lads^2}{16\pi G_N}\Big)  = \frac{3}{2} \log \frac{T}{T_q} + \mathcal{O}(1),
\eeq
consistent with the result obtained by gluing the near-horizon throat to the far-away region, and applying 2d gravity techniques on the throat.

\subsubsection{Probe gauge fields}\label{sec:GaugeHyp}

The analysis in~\cref{sec:HypSchw} focused on the fluctuation of gravitons. It is also interesting to analyze the one-loop determinant of matter around this geometry. One class of matter that is interesting to examine is a probe gauge field. Owing to the associated gauge invariance, one expects to see a set of large gauge transformations being manifested as zero modes in the near-horizon of the extremal solution.\footnote{ Per se, we do not expect the Klein-Gordon operator associated to scalar fields to have low-lying eigenvalues. We have  verified this numerically for some reasonably small values of $\kappa$. }
 
In fact, the analysis of a probe Maxwell field, which we now undertake, will provide a useful guide for the analysis around the Reissner-Nordstr\"om geometry later in the sequel. We will examine the dynamics in a $\mathrm{U}(1)$ theory, with the standard Maxwell action supplemented by the Lorenz gauge fixing term. To wit, 
\begin{equation}
\begin{split}
I &= 
    I_\text{Max} + I_\text{gf} =  
    \int \d^4 x \,\sqrt{g} \left[ F_{\mu\nu} \,F^{\mu\nu} +  
   2 \left(\nabla^\mu A_\mu \right)^2 \right] \\
&=
   2 \int\, \d^4 x\, \sqrt{g} \, A^\mu\left( -\delta\indices{_\mu^\nu}\, \nabla^2 + R\indices{_\mu^\nu} \right) A_\nu
\end{split}
\end{equation}
In the second line we have integrated by parts, discarding the boundary term, and thereby extracted the operator whose fluctuation determinant we wish to compute, viz., 
\begin{equation}
    \det \left(
- \delta\indices{^\nu_\mu} \,\nabla^2+ R^\nu_{\ \mu}
    \right).
\end{equation}
The gauge field vanishes in the background geometry~\eqref{eq:hyp_metric}, and the measure of integration for fluctuations $A_\mu$ are normalized with respect to the ultralocal measure on the space of fluctuations defined through the norm, i.e.,
\begin{equation}
   \norm{A}^2 = \int \d^4 x \,\sqrt{g}\, A^\mu A_\mu\,.
\end{equation}
As for the Schwarzian modes, potential light fluctuations should be physical (not gauge) modes in the full geometry. As such, we may require that they are divergence-free. It follows that we want to solve the following system of equations
\begin{equation}
    -\nabla^2 A_\mu + R^\nu_{\ \mu} A_\nu = \frac{\lambda}{\lads^2}\, A_\mu \,, \qquad  \nabla^\mu A_\mu = 0\,.
\end{equation}
Based on the intuition from the near-horizon analysis, we  expect to find the low-lying eigenmodes to be homogeneous on $\mathbb{H}_2$. Therefore, we pick the ansatz 
\begin{equation}\label{eq:gauge_hyp}
A_\mu\, \d x^\mu = e^{\frac{2 \pi\, \i\, n }{\beta} \,\tE}\, \sqrt{f}      \left(\i\, a_t(r)\,  \d\tE + a_r(r) \, \frac{\d r}{f}
    \right),
\end{equation}
parameterized by functions $a_t(r)$ and $a_r(r)$. The factors of $\sqrt{f}$ are inserted for convenience alone. They ensure that $A$ is in $L^2$ with respect to the ultralocal measure if $a_t, a_r$ are in $L^2$ with respect to the $r^2 \,\d r$ measure on $[r_+, \infty)$. 

The condition that fluctuations are divergence-free allows us to eliminate $a_t$:
\begin{equation}
    a_t = \beta \frac{\left(\lads^2 \,(4 \,r - 3\, r_+)-6 \,r^3 + 3\, r_+^3\right) a_r + 2\, r\left(\lads^2 \,(r-r_+) + r_+^3 -r^3 \right)  a_r'  }{4 \pi \, \lads^2 \,n \,r^2}\,.
\end{equation}
We are then left with a single radial ODE to solve
\begin{align}\label{eq:arevaleq}
\begin{split}
0 &= \left(\lads^2 \, (r_+ -r) + r^3 - r_+^4 \right) a_r''(r) - 2\, (\lads^2 - 3\, r^2)\, a_r'(r) + V_A(r)\, a_r(r) \,, \\
V_A(r) 
&\equiv 
   -\frac{\lads^4 \left(16 \pi^2 \,n^2\, r^4 +\beta^2\,  
    \left[4 \,\lambda\,  r^4 - 4 \,\lambda \, r^3 \,r_+ +8\, r^2 
    -24 \,r \,r_+ + 15 \,r_+ ^2\right]\right)}{ 4 \, \beta^2 \,r^2\, (r-r_+) }  \\ 
&\qquad    
    - \frac{2\,\lads^2 \left(-2\, \lambda\,  r^6 + 6\, r^4 + 2 \,\lambda\,  r^3 \,r_+^3 + 12\, r\, r_+^3 - 15\, r_+^4\right)
    +3 \,\beta^2 \left(5 \,r_+^6 - 8\, r^6\right)}{ 4 \, r^2\, (r-r_+) }\,.
    \end{split}
\end{align}

\begin{figure}[ht]
\begin{subfigure}[b]{\linewidth}
\centering
\subcaptionbox{The eigenvalue spectrum at low temperatures. \label{fig:hypu1zoom}}[\linewidth]{
\includegraphics[width=0.7\textwidth]{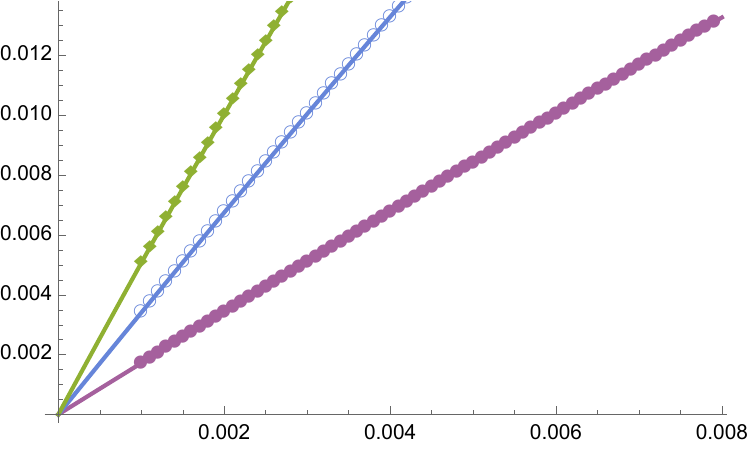}
\begin{picture}(0.3,0.4)(0,0)
\put(-320,175){\makebox(0,0){$\lambda_n$}}
\put(0,25){\makebox(0,0){$\lads\,\kappa$}}
\put(-240,135){\makebox(0,0){$n=3$}}
\put(-150,160){\makebox(0,0){$n=2$}}
\put(-130,125){\makebox(0,0){$n=1$}}
\end{picture}}
\end{subfigure}
\vspace{2cm}
\begin{subfigure}[b]{0.45\linewidth}
\centering
\subcaptionbox{The eigenvalue spectrum over an extended range.\label{fig:hypsu1full}}[\linewidth]{
\includegraphics[width=0.95\textwidth]{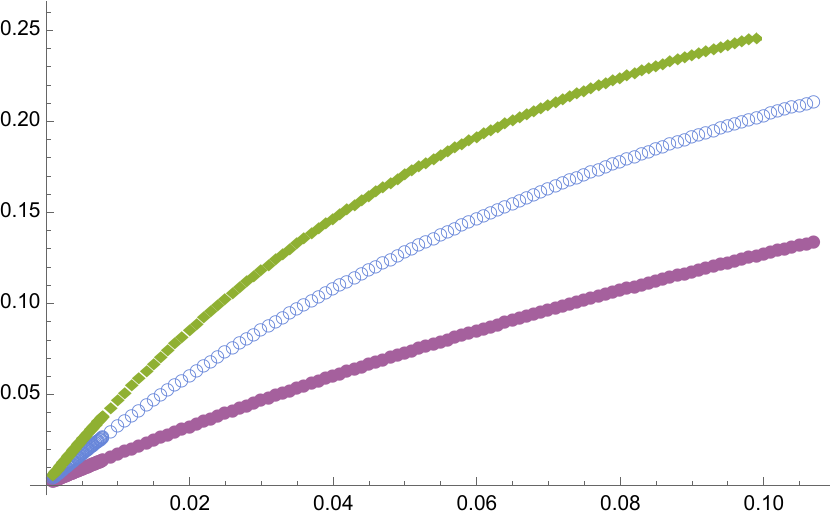}
\begin{picture}(0.3,0.4)(0,0)
\put(-200,100){\makebox(0,0){$\scriptstyle{\lambda_n}$}}
\put(0,15){\color{white}\circle*{0.25}}
\put(0,15){\makebox(0,0){$\scriptstyle{\lads\,\kappa}$}}
\end{picture}}
\end{subfigure}
\hspace{1cm}
\begin{subfigure}[b]{0.45\linewidth}
\subcaptionbox{Rescaled eigenvalues demonstrating scaling.\label{fig:hypu1collapse}}[\linewidth]{
\includegraphics[width=0.95\textwidth]{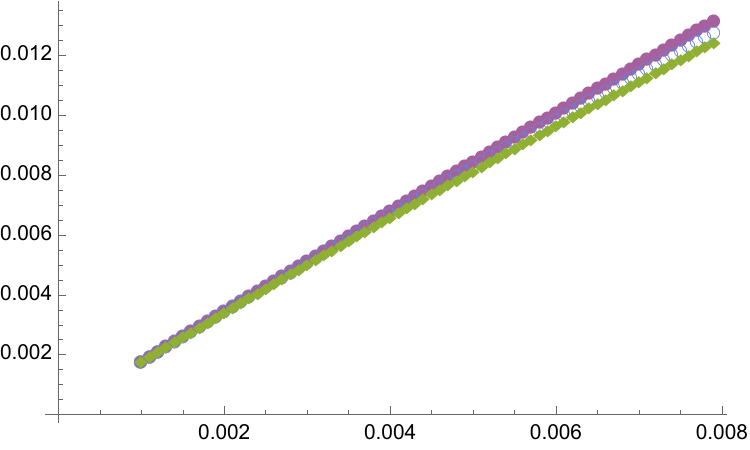}
\begin{picture}(0.3,0.4)(0,0)
\end{picture}}
\put(-205,100){\makebox(0,0){$\scriptstyle{\frac{\lambda_n}{n}}$}}
\put(-20,15){\makebox(0,0){$\scriptstyle{\lads\,\kappa}$}}
\end{subfigure}
\vspace{-1.5cm}
\caption{The uplift of probe Maxwell field zero modes to the full hyperbolic \AdS{4} black hole geometry. We show the data for very low temperatures in the top panel~\cref{fig:hypu1zoom}, demonstrating linear behavior with temperature. In the bottom left panel~\cref{fig:hypsu1full}, we show the behavior over a greater range of temperatures, making the departures from linear growth manifest. Finally, in the bottom right panel~\cref{fig:hypu1collapse}, we plot the eigenvalues rescaled by the overtone number $\frac{1}{n}\,\lambda_n$ to demonstrate that $\lambda_n \sim n\,T$.  }
\label{fig:hyperbolic_gauge}
\end{figure}

To analyze the equation, we again pass to the $z$-coordinate introduced in~\eqref{eq:rzmap} to compactify the domain of integration. As before, both $z =\pm 1$ are regular singular points of~\eqref{eq:arevaleq}. Analyzing the behavior near the horizon and asymptopia, we learn 
\begin{equation}
a_r(z) \sim 
\begin{dcases} 
     (1-z)^{s_\pm} \,,\qquad s_\pm = \frac{5}{2} \pm \sqrt{\frac{1}{4}-\lambda} \,, & 
        \qquad z\lesssim 1 \,,  \\
    (1+z)^{s'_\pm} \,, \qquad s'_\pm = \pm n -1\,,  &\qquad z 
        \gtrsim -1\,.
\end{dcases}
\end{equation}
We learn that the $s_+$ solution is normalizable and preserves Dirichlet boundary conditions at $z=1$, while at the horizon we need $s' = \abs{n}-1$. As for the Schwarzian modes earlier, this is not enough and one must impose, in this case, $\abs{n} \geq 1$. Redefining, $a_r(z)$ in favor of a new function $w(z) $ through, 
\begin{equation}
    a_r(z) = (1+z)^{\abs{n}-1}\, (1-z)^\frac{5}{2} \,w(z)\,,
\end{equation}
we conclude that $w(z)$ must be regular at $z=-1$ (horizon) and vanish at $z=1$ (infinity). The function $w(z)$ will only be $C^0$ at the latter boundary. We now have a well-posed eigenvalue problem. We solved it numerically in the same way as for the Schwarzian modes. The only difference being that we found machine precision to be sufficient for our purposes. This explains why the data presented below matches the near-horizon analysis with less precision. That may be best seen by comparing~\cref{fig:hypschwcollapse} and ~\cref{fig:hypu1collapse} for temperatures of order $\kappa \,\lads \sim 10^{-2}$.

This analysis illustrates the presence how the gauge field zero modes in the near-horizon are manifest in the full geometry at low temperatures. The mechanics of the computation is similar in the spirit to the Schwarzian discussion, and much simplified in the present case, by the decoupling between the gauge field and metric perturbations.

The contribution to the one-loop determinant from this set of gauge (near) zero modes would be
\begin{equation}\label{eq:gauge1loop}
    \log Z_{\rm gauge}^{\rm 1-loop} \supset - \sum_{n\geq 1} \log \Big(\frac{\lambda_n\, \lads^2}{16\pi G_N }\Big) = \frac{1}{2} \log \frac{T}{T_q} + \order{1}\,.
\end{equation}
Unlike the Schwarzian modes, this answer is not by itself the complete story. These zero modes exist with fixed asymptotic chemical potential $\mu$, viz., in the grand canonical ensemble. Owing to charge quantization, we should sum over boundary conditions with shifted chemical potentials,  $\beta\mu \to \beta\mu + 2\pi \,\i\,\mathbb{Z}$.  The full partition function is given by the sum over these shifts, weighted by the classical action, which will itself receive corrections from the backreaction of the gauge fields. Every such configuration can be viewed as a new saddle, and as such, if the above backreaction is ignored each of them will have a one-loop contribution given by~\eqref{eq:gauge1loop}. A simple application of Poisson resummation (see, for example,~\cite[section 2]{Mertens:2019tcm} or~\cite[section 3]{Iliesiu:2020qvm}) suggests that the fixed charge partition function is temperature independent near-extremality. To give a complete argument for this feature, without using the throat, would require accounting for the backreaction of the gauge field in the presence of these integer fluxes (although one can already notice the backreaction will be small at low temperatures since the shift in the chemical potential vanishes linearly with temperature).

\subsubsection{Rotations -- a formal detour}

For asymptotically flat, or asymptotically globally AdS black holes, we also expect another family of gravitational zero modes. These arise from the background isometries. For spherically symmetric black holes such as Reissner-Nordstr\"om they originate from the rotations of the sphere (or, more geometrically, from the Killing fields of the sphere). Similar statements apply for the case of axisymmetric solutions like Kerr. 

It is interesting to ask if the black holes with hyperbolic horizons, which we are currently examining, allow for an analog of such modes. If so, they should be easier to analyze. The answer to this question depends on the nature of the horizon spatial cross-section. If the spatial cross-section was a compact hyperbolic space, then there would generically be no Killing vectors, and thus we won't encounter an analog of the rotational zero modes. On the other hand, on $\mathbb{H}_2$, which is a maximally symmetric space,  we do have Killing vector fields. These are, however, not normalizable. Consequently, the associated modes in the full spacetime will also not be normalizable as well. 

For the present, we will study this $\mathbb{H}_2$ model as an illustrative example. To analyze the modes, we will simply ignore the issue of normalizablity by working with an IR regulator.  One should keep in mind that these solutions will disappear from our functional space if we remove the regulator. Hence, one should treat this part as a rather formal (but useful) exercise.

Let $\mathfrak{v}$ be a Killing (co)-vector in $\mathbb{H}_2$. We pick an ansatz for the metric perturbations around~\eqref{eq:hyp_metric}
\begin{equation}\label{eq:rot_hyp}
h_{\mu \nu}\, \d x^\mu\, \d x^\nu =
    2\, e^{\frac{2 \pi \i \,n}{\beta}\, \tE} \, \mathfrak{v}\, r\, \sqrt{f}\left(\i\,
    g_1(r)\, \d\tE + g_2(r) \,\frac{\d r}{f} 
    \right) .
\end{equation}
 The inspiration for this is similar to the rationale behind the choice for the gauge field analysis, cf.~the discussion around~\eqref{eq:gauge_hyp}. The sprinkling of factors of $\sqrt{f}$ ensure that as long as $\{f_1, f_2\}$ are in $L^2$ with respect to the uniform measure on $[r_+,\infty)$, the metric perturbation has a finite ultralocal norm. We now aim to solve~\eqref{eq:HypLicheqns}.

We have ensured  that the perturbation is trace free, so only have to impose the divergenceless condition, which helps eliminate $g_1$
\begin{equation}
g_1 = -\beta \,\frac{\left(\lads^2\, (6 \,r-5\, r_+)-8 \, r^3+ 5\, r_+^3\right) g_2  + 2 \,r \left(\lads^2\, (r-r_+)-r^3 + r_+^3\right) g_2' }{4 \pi \, \lads^2\, n \,r^2} \,.
\end{equation}
The eigenvalue equation itself is a second order ODE for $g_2$, which we report here for completeness
\begin{align}
\begin{split}
0 
&= 
   \left(\lads^2\, (r-r_+) - r^3 + r_+^3\right) g_2''(r) 
   + 2 \left(\lads^2-3 \,r^2\right) g_2'(r) + \lads^2 \, V_{f_2}(r) \, g_2(r) \,, \\
V_{g_2} 
&\equiv
   \frac{\lads^2 \left(\beta ^2 \left(16\, \lambda \, r^3\, (r_+-r)
    -24 \,r^2 + 60 \,r \,r_+ - 35\, r_+^2\right)-16\, \pi^2\, n^2\, r^4\right)}{4\, \beta^2\, r^2 \left(\lads^2 (r-r_+)- r^3 + r_+^3\right)} \\ 
&\;    
    + \frac{\left(8 \,\lambda\,  r^6
        +6 \, r^4 -2\,  r^3\,  r_+ \left(4\, \lambda\, r_+^2+ 7\right) -
        30 \,r \,r_+^3 + 35\, r_+^4\right) +
        \beta^2 \left(16\, r^6 + 28\, r^3\, r_+^3 - 35\, r_+^6\right)}{2\, r^2 \left(\lads^2 (r-r_+)- r^3 + r_+^3\right)} \,,
\end{split}
\end{align}
where $\lambda$ is the eigenvalue we seek.
Passing, as before, to the compactified domain of the $z$-coordinate~\eqref{eq:rzmap}, we learn of the asymptotic behavior
\begin{equation}
g_2(z) \sim  
\begin{dcases} 
     (1-z)^{s_\pm} \,,\quad s_\pm = \frac{5}{2} \pm \sqrt{\frac{9}{4}-4\,\lads^2 \lambda} \,, & 
        \qquad z\lesssim 1 \,,  \\
    (1+z)^{s'_\pm} \,, \quad s'_\pm = \pm n -1\,,  &\qquad z 
        \gtrsim -1\,.
\end{dcases}
\end{equation}
This suggests trading $g_2(z)$ in favor of $w(z)$, which is related by
\begin{equation}
    g_2(z) = (1+z)^{\abs{n}-1}\, (1-z)^\frac{3}{2} \, w(z)\,,
\end{equation}
which again requires us to consider only $\abs{n}\geq 1$ just like the gauge modes. The function $w(z)$ is required to be regular at $z=-1$ (horizon) and vanish at $z=1$ (asymptopia). At the latter boundary $w(z)$ is $C^2$ but not $C^3$.  We now have a well-posed eigenvalue problem, which we can solve numerically in the same way as for the Schwarzian modes. The main difference from that case is that we found machine precision to be sufficient for our purposes. This explains why the data presented below matches the near-horizon analysis with less precision. That may be best seen by comparing  ~\cref{fig:hypschwcollapse} and ~\cref{fig:hypu1collapse} for temperatures of order $\kappa \lads \sim 10^{-2}$.

\begin{figure}[ht]
\begin{subfigure}[b]{\linewidth}
\centering
\subcaptionbox{The eigenvalue spectrum at low temperatures. \label{fig:hyprotzoom}}[\linewidth]{
\includegraphics[width=0.7\textwidth]{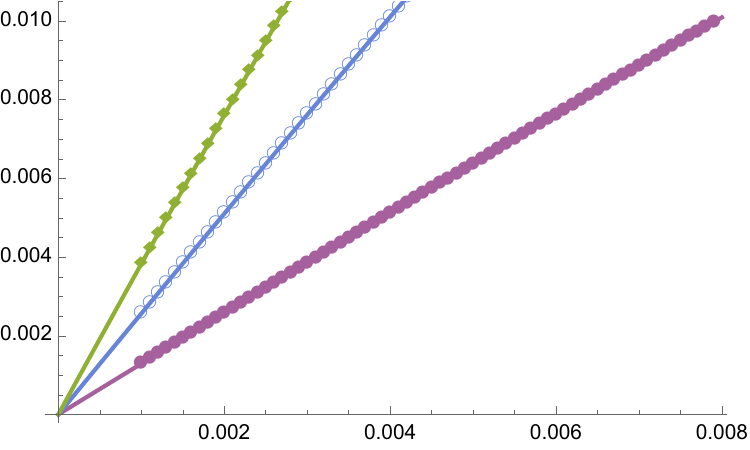}
\begin{picture}(0.3,0.4)(0,0)
\put(-320,175){\makebox(0,0){$\lambda_n$}}
\put(0,25){\makebox(0,0){$\lads\,\kappa$}}
\put(-240,135){\makebox(0,0){$n=3$}}
\put(-150,160){\makebox(0,0){$n=2$}}
\put(-130,125){\makebox(0,0){$n=1$}}
\end{picture}}
\end{subfigure}
\vspace{2cm}
\begin{subfigure}[b]{0.45\linewidth}
\centering
\subcaptionbox{The eigenvalue spectrum over an extended range.\label{fig:hyprotfull}}[\linewidth]{
\includegraphics[width=0.95\textwidth]{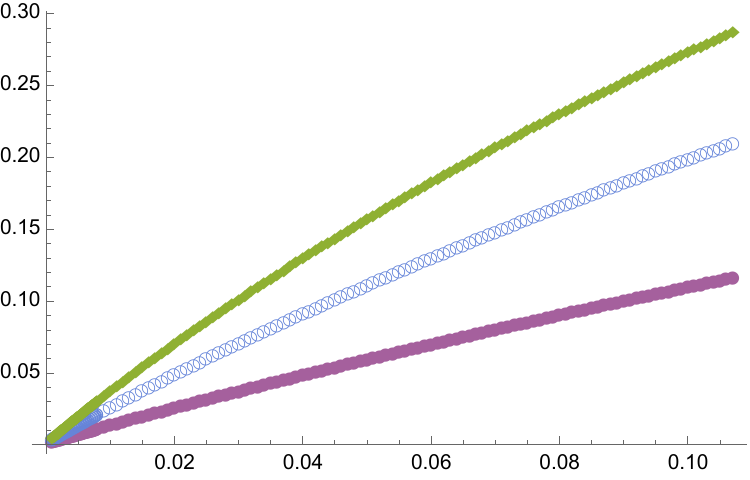}
\begin{picture}(0.3,0.4)(0,0)
\put(-200,100){\makebox(0,0){$\scriptstyle{\lambda_n}$}}
\put(0,15){\makebox(0,0){$\scriptstyle{\lads\,\kappa}$}}
\end{picture}}
\end{subfigure}
\hspace{1cm}
\begin{subfigure}[b]{0.45\linewidth}
\subcaptionbox{Rescaled eigenvalues demonstrating scaling.\label{fig:hyprotcollapse}}[\linewidth]{
\includegraphics[width=0.95\textwidth]{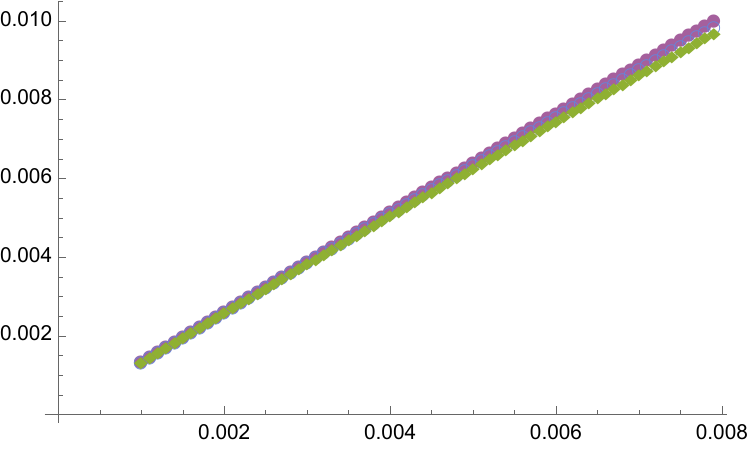}
\begin{picture}(0.3,0.4)(0,0)
\end{picture}}
\put(-205,100){\makebox(0,0){$\scriptstyle{\frac{\lambda_n}{n}}$}}
\put(-20,15){\makebox(0,0){$\scriptstyle{\lads\,\kappa}$}}
\end{subfigure}
\vspace{-1.5cm}
\caption{The uplift of the (formal) rotational field zero modes to the full hyperbolic \AdS{4} black hole geometry. We show the data for very low temperatures in the top panel,~\cref{fig:hyprotzoom}, demonstrating linear behavior with temperature. In the bottom left panel,~\cref{fig:hyprotfull}, we show the behavior over a greater range of temperatures, making the departures from linear growth manifest. Finally, in the bottom right panel,~\cref{fig:hyprotcollapse}, we plot the eigenvalues rescaled by the overtone number $\frac{1}{n}\,\lambda_n$ to demonstrate that $\lambda_n \sim n\,T$. While the plot very closely resembles that of the probe Maxwell field~\cref{fig:hyperbolic_gauge}, there are  subtle differences in the numerical data.} 
\label{fig:hyperbolic_rotation}
\end{figure}

These rotational zero modes will also give rise to a factor of $\frac{1}{2}\, \log T$ to $\log Z_{\rm graviton}^{\rm 1-loop} $. This is, as in the discussion at the end of~\cref{sec:GaugeHyp}, the contribution from a single saddle. We still have to sum over the configurations related by the shifts of the rotational chemical potential $\beta \Omega \to \beta \Omega + 2\pi\, \i\,n$ (or $4\pi\,\i\,n$ in theories with fermionic degrees of freedom), which will additionally entail computing the shifted value of the classical on-shell action. The expectation, based on a calculation done at the level of the throat, is that Poisson resummation will again remove the temperature dependence of the quantum correction near extremality. This might be harder to verify since the backreacted solutions involve a rotating version of the hyperbolic black hole.  

\subsection{Reissner-Nordstr\"om (AdS) black holes}\label{sec:RNads}

We are now ready to attack the most interesting problem -- the one-loop determinant for the Reissner-Nordstr\"om solution. For technical reasons, it is easier to work with AdS asymptotics, which simplifies imposition of boundary conditions. So we will initially work with a finite cosmological scale $\lads$, and subsequently comment on the limit $\lads \to \infty$ in~\cref{sec:flatlimit}.

The background metric for the spherically symmetric static solution reads
\begin{equation} \label{eq:RNAdS_metric}
\d s^2 = f(r)\,  \d \tE^2 + \frac{\d r^2}{f(r)} + r^2 \, \d \Omega_2^2 \,, \qquad 
f(r)  = \frac{r^2}{\lads^2} + 1 - \frac{2m}{r} + \frac{Q^2}{r^2}.
\end{equation}
We consider an electrically charged black hole and so the Maxwell potential is
\begin{equation} \label{eq:RNAdS_potential}
    A = \i \left(\frac{Q}{r}-\frac{Q}{r_+} \right) \d \tE\,.
\end{equation}

It will prove useful to parameterize the solution in terms of the outer and inner horizon radii, which we achieve via 
\begin{equation}\label{eq:rnQM}
\begin{split}
Q^2 
= 
    r_+\, r_- \left(1 + \frac{r_+^2 + r_-^2 + r_+ \, r_-}{\lads^2} \right) \,, \qquad 
2\,m  
= 
    r_+ \left(\frac{r_+^2}{\lads^2}\right) + \frac{Q^2}{r_+} \,.
\end{split}
\end{equation}
The black hole solution is a saddle point for the gravitational action   with fixed temperature and chemical potential, which are in turn
\begin{equation}\label{eq:rnTmu}
\begin{split}
\kappa 
&:= 
    \frac{T}{2\pi}    
    = \frac{r_+ - r_-}{2\, r_+^2} \left(1 + \frac{r_-^2 + 2\, r_+ \, r_- + 3\, r_+^2}{\lads^2} \right) ,  \\
\mu
&= 
    \sqrt{\frac{r_-}{r_+}} \; \sqrt{1 + \frac{r_+^2 + r_-^2 + r_+ \, r_-}{\lads^2} } \,.
\end{split}   
\end{equation}

 We will be interested in the near-extremal limit as before. If we take the limit $r_-\simeq r_+$, keeping $\lads$ fixed, then we obtain a solution with \AdS{4} asymptotics. On the other hand, taking the limit with $r_+ \simeq r_- \ll \lads$ allows us to make contact with black holes in flat spacetime. For later reference, note that setting  $r_+ = r_0 + \Ts$ and $r_- = r_0 - \Ts$, we have 
\begin{equation}\label{eq:rnTmusmall}
\kappa = \frac{\Ts}{r_0^2} \left(1+6\,\frac{r_0^2}{\lads^2}\right) + \order{\Ts^2} \,, \qquad \mu = \sqrt{1+3\,\frac{r_0^2}{\lads^2}} + \order{\Ts} \,,
\end{equation}
which implies
\begin{equation}
    \kappa \sim \frac{3 \,\Ts}{\lads^2}\, \frac{2\,\mu^2 -1}{\mu^2-1}\,.
\end{equation}
From the near-horizon analysis, we expect three families of low-lying modes. These modes were identified using the heat kernel in \cite{Sen:2012kpz}, or rewriting the theory in the throat as a theory of 2d gravity coupled to matter in \cite{Nayak:2018qej,Iliesiu:2020qvm}. These should correspond to the  Schwarzian diffeomorphisms, the $\mathrm{SO}(3)$ rotations, and the $\mathrm{U}(1)$ gauge transformations, respectively.  While the Schwarzian and the $\mathrm{U}(1)$ modes are decoupled in the near-horizon region, there is a priori no reason why they should do so in the full spacetime. Both would be expected to behave as scalars under $\mathrm{SO}(3)$ rotations, and therefore can mix. We will see that the modes indeed mix, and thus will treat them on the same footing in our analysis.

\subsubsection{A contour prescription for linearized Einstein-Maxwell dynamics}\label{sec:contour}

The Reissner-Nordst\"rom \AdS{4} solution is a solution to the Einstein-Maxwell theory, with Euclidean action 
\begin{equation}
\begin{split}
I 
&=
    - \frac{1}{16\pi\, G_N}\, \int \d^4 x\, \sqrt{g} \left(R + \frac{6}{\lads^2} - F_{\mu \nu} F^{\mu \nu} \right) -\frac{1}{8\pi\, G_N}\,\oint\, \d^3x\,\sqrt{\gamma}\, K + I_\text{bdy} \,, \\
I_\text{bdy} 
&=
    \frac{1}{8\pi\,G_N}\,\oint \, \sqrt{\gamma}\, \left[\frac{2}{\lads} +\lads\, \tensor[^\gamma]{R}{} \right].
\end{split} 
\end{equation}
We will discuss gauge fixing terms in the sequel. Due to the background Maxwell field, contrary to the hyperbolic example discussed above, we cannot decouple the conformal and transverse traceless modes. As a result, after expanding the action to the quadratic order, we obtain a non-positive definite quadratic form $I^{(2)}$ (or, to put it simply, the Euclidean action is not bounded from below).\footnote{For Einstein-Hilbert theory, this is the well-known statement that the conformal mode has the wrong sign kinetic term. In Einstein-Maxwell theory, the conformal mode mixes with the photons (and gravitons), but there is nevertheless a direction in field space where the action is unbounded below.} The general strategy of dealing with such problems was presented in~\cite{Marolf:2022ntb, Liu:2023jvm}. For the reader's convenience, we will briefly recapitulate the salient points. 

On the space of fluctuations, parameterized by elements $X$ and $Y$, we need to introduce an auxiliary inner product
\begin{equation}
  \bra{X}\ket{Y} = \left(X_i \right)^\star \,\mathbb{G}^{ij} \,Y_j
\end{equation}
that is not positive-definite (but non-degenerate). As already mentioned, the quadratic action is a quadratic form:
\begin{equation}
    I^{(2)} = X_i\, \mathbb{S}^{ij}\, X_j.
\end{equation}
Using $\mathbb{G}$ and $\mathbb{S}$, we may now introduce an operator $\hat{L}$ such that
\begin{equation}
    \hat{L} = \mathbb{G}^{-1} \cdot \mathbb{S}.
\end{equation}
By construction, $\hat{L}$ will be self-adjoint with respect to $\bra{\cdot} \ket{\cdot}$, and then we may write\footnote{This expression presumes that the complex conjugations $\star$ does not change $X$. As we will see presently, it is a slightly subtle assumption.}
\begin{equation}
    I^{(2)}(X) = \bra{X}\hat{L}\ket{X}.
\end{equation}
Hence, it is an operator $\hat{L}$ that we would like to study, in particular to find its spectrum. The rule of thumb of \cite{Marolf:2022ntb,Liu:2023jvm} is that whenever we encounter an eigenvector of negative norm, we Wick rotate the inner product (and thus, also the contour of integration in the path integral) to make it positive.\footnote{For Einstein-Hilbert dynamics this is the original prescription of~\cite{Gibbons:1978ac}.} Then, it is $\det\left(\hat{L} \right)$ that computes the one-loop corrections to the partition function. Due to the gauge and diffeomorphism invariance of the Einstein-Maxwell action, $\hat{L}$ is going to have a large kernel. To deal with it, one needs to introduce gauge fixing terms that we should introduce promptly. In the following we will denote by $\hat{L}_0$ the operator without gauge-fixing terms and by $\hat{L}$ the full operator.

Before that, let us try to apply the prescribed procedure to the problem at hand. The simplest choice of the inner product is given by the ultralocal measure on the graviton and photon fluctuations, 
\begin{equation}
\norm{X}^2 = \lads^{-4}\int \d^4 x\, \sqrt{g} \,\left( \tilde{h}^{\mu \nu} h_{\mu \nu} + a^\mu a_\mu \right),
\end{equation}
where $\tilde{h}_{\mu\nu}$ is defined in~\eqref{eq:tracerevh}. The metric and Maxwell fluctuations are 
$h_{\mu \nu}$ and $a_\mu$, respectively. We will also use $h = g^{\mu \nu} h_{\mu \nu}$ (as before) and $f = \d a$.
Then, we should expand the action to the second order in $h_{\mu \nu}$ and $a_\mu$. In this expansion, we encounter the first problem. The terms describing mixing between gravitons and photons are schematically of the form
\begin{equation}\label{eq:interactions}
   \int \d^4 x \,\sqrt{g}\, h\, F\, f \,,
\end{equation}
with appropriate index contractions. However, since $F$ is clearly imaginary, these terms will evaluate to something imaginary as well for real $h$ and $a$. It follows, that the quadratic form $\mathbb{S}$ is not real-valued and so $\hat{L}_0$ will not be self-adjoint with respect to $\mathbb{G}$. 
 
There are a few ways out of this impasse. The most common one is to analytically continue the black hole charge $Q$ to imaginary values. Then, all the background fields are real, and we do not have to worry about these problems. However, it is easy to check that the Reissner-Nordstr\"om black hole with imaginary charge at very low temperatures must have complex $r_+$. Thus, we could not be able to probe the regime we are most interested in. Another possibility is to work with magnetically charged black holes (and perhaps use the electromagnetic duality at the end of the day). Then, the background Maxwell field is real. This approach was successfully applied in the previous calculations of certain one-loop quantities (see for example~\cite{Brill:1992ydn, Dowker:1994up,Monteiro:2008wr}). However, this is a choice that can only be made in four dimensions. It does have an additional unsavory feature of needing multiple coordinate patches to describe the gauge potential. Therefore, we will not pursue this strategy either.

Instead, we proceed by simply declaring that we integrate the Maxwell fluctuations along the imaginary axis and write\footnote{Note, that this is different from just applying Wick rotations to real Lorentzian fluctuations. In this case, only $a_{\tE}$ would be imaginary.}
\begin{equation}
    a = \i\, \overline{a}\,.
\end{equation}
Then, $I^{(2)}$ will be real for $h, \overline{a} \in \mathbb{R}$. There is a price to pay for this choice of contour. Kinetic terms for $\overline{a}$ have wrong signs. As a result, even the purely Maxwell part of $I^{(2)}$ is not bounded from below. Since we are already using a formalism that is supposed to take care of these sorts of problems, we need not be troubled by this particular development. Indeed, let us notice that the inner product now reads
\begin{equation}
\norm{X}^2 =\lads^{-4} \int \d^4 x\, \sqrt{g} \left( \tilde{h}^{\mu \nu} h_{\mu \nu} - \overline{a}^\mu\, \overline{a}_\mu \right).
\end{equation}
Thus, we do not expect that this choice of contour should introduce (or remove) any additional negative modes in  $\hat{L}_0$.

As explained at the beginning of this section, this choice of inner product and contour (after fixing the gauge invariance of the theory, which we do next) leads to an operator $\hat{L}$ that is non-negative and self-adjoint. This is enough for our purpose of identifying the nearly zero-modes characteristic of an \AdS{2} throat. To complete the calculation of the partition function, one would need to evaluate the norm of the eigenfunctions of $\hat{L}$ using the inner product and further rotate the contour of any mode with negative norm (which would otherwise lead to a negative action). We will not attempt to give a complete characterization of this last step, although it is straightforward.

We are now ready to discuss the thorny issue of the gauge invariance. We need to break it explicitly in the action to remove the degeneracy of $\hat{L}_0$. In general, all choices are allowed. However, as we will see, only some of them are convenient. Let us start by naively repeating the analysis of the previous sections and trying to impose Lorenz and harmonic gauge. The relevant terms read
\begin{equation}
    I_{\textrm{gf}}' = \frac{1}{16\pi\, G_N}\,\int \d^4 x \,\sqrt{g} \left(
    2\, (\nabla^\mu a_\mu)^2 + \frac{1}{2}\, \nabla^\mu \tilde{h}_{\mu \nu}\,  \nabla_{\rho} \tilde{h}^{\rho \nu}
    \right).
\end{equation}
Treating $I_{\textrm{gf}}'$ as a quadratic form, we obtain a modified operator $\hat{L'}$ from $\hat{L}_0$. One can explicitly check that the eigenvalue equation
\begin{equation}
    \hat{L'} X = \lambda X
\end{equation}
for $X = (h,\overline{a})$ is inconsistent with the gauge conditions
\begin{equation}
    \nabla_\mu \tilde{h}^{\mu \nu} = 0 \,, \qquad  
    \nabla_\mu a^\mu =0 \,.
\end{equation}
Thus, if we wanted to continue along this line, we would have to deal with pure gauge degrees of freedom, which would make our life unnecessary hard (for one, we would have to then also compute the eigenspectrum of the ghost operators). 
Instead, we use the method of~\cite{Marolf:2022ntb} to find a more convenient choice of gauge. 

The idea is to enforce a gauge where all physical fluctuations (viz., the ones satisfying gauge conditions) are orthogonal to pure gauge modes. The space of gauge parameters in our case is parametrized by a vector field $\xi$ and a scalar field $\chi$. Pure gauge configurations are described by a map $\mathcal{P}$ 
\begin{equation}
    \mathcal{P}(\xi, \chi) = \left(
\mathcal{L}_\xi g_{\mu \nu},\  \mathcal{L}_{\xi} A_\mu + \nabla_\mu \chi
    \right) = (h_{\mu \nu}, a_{\mu})
\end{equation}
from the gauge parameters to the space of fluctuations. Hence, the physical fluctuations must be orthogonal to the image of $\mathcal{P}$. If we equip the gauge parameter space (which is automatically a vector space) with the inner product to make it a Hilbert space, we will have a well-defined notion of $\mathcal{P}^\dagger$:
\begin{equation}
    \bra{X} \ket{\mathcal{P}(\xi, \chi)} = \bra{\mathcal{P}^\dagger X} \ket{(\xi, \chi)} ,
\end{equation}
where the first inner product is in the space of fluctuations, and the other one in the space of gauge parameters. Hence, our gauge condition reads\footnote{There are certain subtleties connected to this procedure, which are discussed in~\cite{Liu:2023jvm}.}
\begin{equation}
    \mathcal{P}^\dagger X = 0.
\end{equation}
The gauge-fixing term can then be taken to be (up to a constant)
\begin{equation}
  I_{\textrm{gf}} \sim  \bra{X} \mathcal{P} \mathcal{P}^\dagger \ket{X}\,.
\end{equation}
It follows that the eigenvalue equation
\begin{equation}
    \hat{L} X = \frac{\lambda}{\lads^2} X\,,
\end{equation}
is consistent with the gauge condition. This choice will thus simplify the analysis by allowing us to focus on physical fluctuations (and therefore simplify the number of degrees of freedom in the equations). 

Note that the choice of $\mathcal{P}^\dagger$ will depend on the choice of the auxiliary inner product on the space of gauge parameters. The simplest choice is
\begin{equation}
    \norm{(\xi, \chi)}^2 = \lads^{-4}\int \d^4x\, \sqrt{g} \left( \,\xi_\mu \xi^\mu + \chi^2 \right).
\end{equation}
Then, our gauge conditions now read\footnote{In deriving the form of $\mathcal{P}^\dagger$ one should keep in mind that it is $\overline{a}$ that is invariant under complex conjugation, not $a$. Nevertheless, if we considered real background fields and real fluctuations, we would obtain the same form of the gauge conditions.}
\begin{subequations} \label{gauge_cond_RN_AdS}
    \begin{equation}
    \nabla^\mu \tilde{h}_{\mu \nu} -\frac{1}{2} \,F_{\nu \mu} \,a^\mu +  A_\nu\, \frac{1}{2} \nabla^\mu a_\mu = 0\,,
\end{equation}
\begin{equation}
    \nabla^\mu a_\mu = 0 \,.
\end{equation}
\end{subequations}
One can explicitly check that they are indeed consistent with the eigenvalue problem. 

We could now proceed by expanding the action and finding operator $\hat{L}$ explicitly. However, it is easy to make a mistake while trying to establish if various terms with abstract indices correspond to Hermitian operators.\footnote{As an example, one can check that even the Lichnerowicz operator as given in \eqref{eq:Lichnerowicz_AdS} is not Hermitian unless the background metric is Einstein.} As an alternative, we will decompose the perturbations with respect to different representations of time-translations and $\mathrm{SO}(3)$. In this way, we reduce the problem to one dimension in which it is much easier to explicitly find the form of the operator  $\hat{L}$. Before dealing with the details of this procedure, let us point out that the background fields $(g, F)$ are even and odd under time-reversal,\footnote{Or, more precisely, what a quantum field theorist would call ${\sf CT}$.} respectively. It follows that we may decompose the perturbations with respect to their time-reversal parity. We will have two sectors: in which the metric fluctuations are even and the Maxwell field is odd or vice versa.

\subsubsection{The Schwarzian and U(1) gauge modes}
 
As noted above, the Schwarzian and the $\mathrm{U}(1)$ modes should preserve the background $\mathrm{SO}(3)$ rotation isometry. So we can focus on a spherically symmetric ansatz, and use time-translational invariance to decompose them into Matsubara modes. We may additionally decompose metric and Maxwell fluctuations into even and odd parity modes, respectively, under time-reversal. All in all, we can motivate the following general ansatz to parameterize the perturbation:
\begin{equation}
\begin{split}
h_{\mu \nu}\d x^\mu \d x^\nu 
&= 
    \frac{1}{\sqrt{\pi \beta} r} \left[ \cos( \frac{2\pi n\, \tE}{\beta}) \left(f_1 \,f\, \d \tE^2 + f_2\, \frac{\d r^2}{f} + \frac{r^2\, f_4}{4} \, \d\Omega_2^2
        \right) + \sin( \frac{2\pi n\,\tE}{\beta}) f_3\, \d \tE \d r
        \right], \\ 
a_\mu\, \d x^\mu 
&= 
    \frac{\i}{\sqrt{2\pi \beta}\; r} \left(
    \cos( \frac{2\pi n\,\tE}{\beta} ) \, \sqrt{f}\, a_t\, \d\tE + \sin \left( \frac{2\pi n\,\tE}{\beta} \right) a_r\, \frac{\d r}{\sqrt{f}}
        \right),
\end{split}
\end{equation}
where $f_i$ with $i=1,\cdots 4$, $a_t$, and $a_r$ are functions of $r$ alone. The inner product written in terms of these read:
\begin{equation}\label{eq:schnorm}
\int \d^4 x \sqrt{g} \left(
    \tilde{h}^{\mu \nu} \,h_{\mu \nu} + a_\mu \,a^\mu \right) 
= \int_{r_+}^{\infty} \d r \left( (f_1 - f_2)^2 + f_3^2 - (f_1 +f_2) \,f_4 -  
         a_t^2 - a_r^2 \right).
\end{equation}
It follows that for perturbations to land in our Hilbert space, they should be sufficiently regular around $r=r_+$ and vanish sufficiently fast around $r=\infty$.\footnote{ Note that if we act on the fluctuations with the isometry $\tE \mapsto \tE + \frac{\beta}{2 n}$, we obtain a perturbation with reversed parity properties. Hence, we do not have to consider them separately. We also remind the reader that we only focus on modes with $n\neq 0$, cf.~\cref{fn:nozerofreq}.}

After some integration by parts, we arrive at the desired quadratic fluctuation operator $\hat{L}$. Its form is too long to write down here, instead we mention a few properties. 
\begin{itemize}[wide,left=0pt]
\item As advertised, we may decouple gauge degrees of freedom. Imposing \eqref{gauge_cond_RN_AdS} allows us to eliminate $a_t$, $f_1$ and $f_2$. Specifically, 
\begin{equation}
\begin{split}
a_t 
&= 
    \frac{\beta}{4 \pi \, n\, r}  \left(\left(r\, f'+2 f\right)a_r + 2 \,r \,f \,a_r'  \right), \\
f_1 
&= 
    \frac{\sqrt{2} \,Q \,\beta\,  a_r \,\sqrt{f} + 2\, r \left(\beta \, f_3\, \dv{r}(rf) +\beta\,  r\, f\, f_3'+2 \pi\,  n \,r \,f_2+
    \pi  n\, r\, f_4\right)}{4 \pi\,  n\, r^2} \,,\\
f_2 
&= 
    \frac{1}{8 \pi  n\,\beta\, r^2 f} \left[
     \beta ^2 \,r \left(r^2\, f_3\, f'^2 + r\,f \left(r\, f_3\, f'' + f'\left(3 \,r f_3'+f_3\right)\right) +f^2 \left(r^2 \,f_3''-2 \, f_3\right)\right)\right.  \\ 
&\left.   \qquad   
    +\; \pi n \,  \beta\,r^3 \left(f_4 \,f'+2 \,f \,f_4'\right)-4 \,\pi ^2 n^2 \,r^3\, f_3 +
    \sqrt{2}\, Q\, \beta ^2\, \sqrt{f} \left(r\, f\, a_r' + a_r \left(r\, f'-f\right)\right)\right] .
\end{split}
\end{equation}
\item Eliminating the above functions, the eigenvalue problem becomes a system of three second-order ODEs for $f_3$, $f_4$, and $a_r$.
\item To solve this, we pass over to the coordinate $z$ introduced in~\eqref{eq:rzmap}, and deduce that $z = \pm 1$ are regular singular points of this eigenvalue problem.
\item Near the AdS boundary ($z =1$) we find the behavior of $f_3$ and $f_4$ to be correlated, and a set of six decays in addition. Specifically, parameterizing
\begin{equation}
  f_3 \sim (1-z)^{s_3}\,, \qquad f_4 \sim (1-z)^{s_4}\,, \qquad a_r \sim (1-z)^{s_r}\,,
\end{equation}
we find $s_4 = s_3-1$ and 
\begin{equation}\label{eq:rnadsSchp1}
\begin{aligned}
s_3 &= 
    \frac{3}{2} \pm \sqrt{\frac{9}{4} - 4 \,\lambda } \,, \qquad 
s_3 &= 
    \frac{3}{2}  \pm \sqrt{\frac{33}{4} - 4 \,\lambda}\,, \qquad 
s_3 &= 
    \frac{7}{2} \pm \sqrt{\frac{1}{4}-\frac{1}{2}\, \lambda}  \,, \\  
s_r &= 
    \frac{7}{2} \pm \sqrt{\frac{9}{4} - 4 \,\lambda}\,, \qquad 
s_r &= 
    \frac{7}{2} \pm \sqrt{\frac{33}{4} - 4\, \lambda} \,,\qquad
s_r &= 
    \frac{3}{2} \pm \sqrt{\frac{1}{4}-\frac{1}{2}\, \lambda}\,.  
\end{aligned}
\end{equation}
These fall-offs correspond to the decays of massive gravitons, photons, and scalars (conformal mode) near the AdS boundary. Since we impose Dirichlet boundary conditions at infinity, we should choose positive signs everywhere in~\eqref{eq:rnadsSchp1}.
\item At the horizon (near $z =-1$), we require the perturbations to be smooth. This leads to demanding 
\begin{equation}
\begin{split}
f_3 \sim (1+z)^{\abs{n}-2} \,, \qquad  
f_4  \sim (1+z)^{\abs{n}-1}\,, \qquad  
a_r \sim (1+z)^{\abs{n}-2}\,.
\end{split}    
\end{equation}
Just as at the boundary, this behavior can be identified with that of a mixture of decays of gravitons and photons for neutral black holes.\footnote{There are also additional asymptotics near $z=-1$ corresponding to the behavior of scalars. However, they decay faster at the horizon, and therefore are automatically included as subleading terms in the provided expressions.}
\end{itemize}

Since there is no a priori reason why the different modes should decouple, we must take all of these fall-offs into account. We do so by redefining the perturbations after stripping off some leading factors at the horizon and boundary with the ansatz 
\begin{equation}\label{eq:Schwarzian_subst}
\begin{split}
f_3(z) & = (1-z)^\frac{1}{2}\,  (1+z)^{\abs{n}-2} \, U_3(z) \,, \\ 
f_4(z) &= (1-z)^{-\frac{1}{2}} \, (1+z)^{\abs{n}-1}\,  U_4(z) \,, \\ 
a_r(z) &= (1-z)^{\frac{3}{2}}\, (1+z)^{\abs{n}-2} \, U_r(z)\,.
\end{split}
\end{equation}
The Dirichlet boundary conditions at the AdS asymptopia translate to $U_3(1) = U_4(1) = U_r(1) =0$, as long as
\begin{equation}\label{eq:lambdabd}
    \lambda < \frac{5}{16}\,.
\end{equation}
We can now use this eigensystem to solve for the uplift of the Schwarzian and the $\mathrm{U}(1)$ gauge modes. We will begin with the former, and then tackle the latter.

\paragraph{The Schwarzian modes:} Let us first  record a useful observation from~\eqref{eq:Schwarzian_subst}. Owing to the factors of $(1-z)^{\abs{n}-2}$, the metric and gauge field perturbations will fail to be in $L^2$ with respect to the ultralocal measure and thus will not have finite action for $\abs{n} \le 1$. Hence, for now, we shall consider only $n \ge 2$ (without loss of generality, we may take $n$ to be positive). This is already suggestive of the Schwarzian modes from our preceding analyses.

\begin{figure}[ht]
\begin{subfigure}[b]{\linewidth}
\centering
\subcaptionbox{The eigenvalue spectrum at low temperatures. \label{fig:rnschwzoom}}[\linewidth]{
\includegraphics[width=0.7\textwidth]{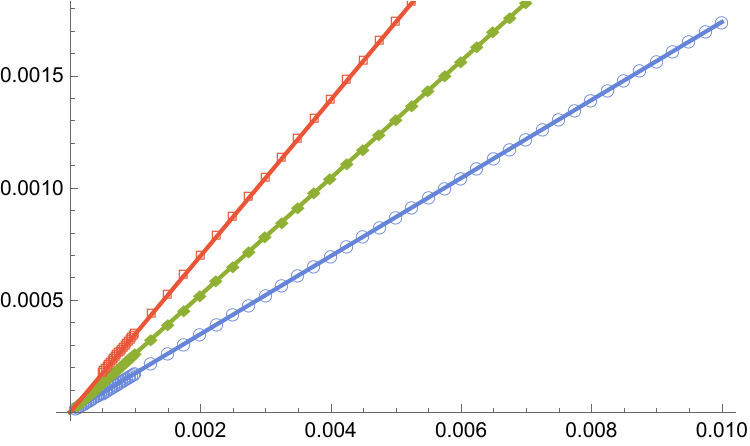}
\begin{picture}(0.3,0.4)(0,0)
\put(-320,175){\makebox(0,0){$\lambda_n$}}
\put(0,25){\makebox(0,0){$\lads\,\kappa$}}
\put(-210,135){\makebox(0,0){$n=4$}}
\put(-147,160){\makebox(0,0){$n=3$}}
\put(-130,125){\makebox(0,0){$n=2$}}
\end{picture}}
\end{subfigure}
\vspace{2cm}
\begin{subfigure}[b]{0.45\linewidth}
\centering
\subcaptionbox{The eigenvalue spectrum over an extended range.\label{fig:rnschwfull}}[\linewidth]{
\includegraphics[width=0.95\textwidth]{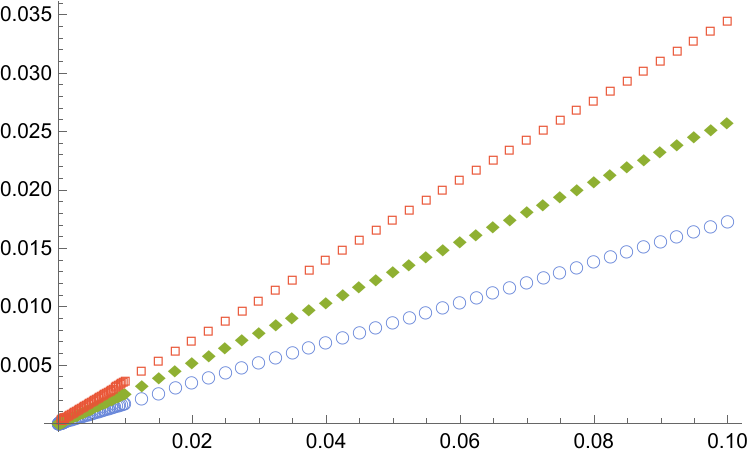}
\begin{picture}(0.3,0.4)(0,0)
\put(-200,100){\makebox(0,0){$\scriptstyle{\lambda_n}$}}
\put(0,15){\makebox(0,0){$\scriptstyle{\lads\,\kappa}$}}
\end{picture}}
\end{subfigure}
\hspace{1cm}
\begin{subfigure}[b]{0.45\linewidth}
\subcaptionbox{Rescaled eigenvalues demonstrating scaling.\label{fig:rnschwcollapse}}[\linewidth]{
\includegraphics[width=0.95\textwidth]{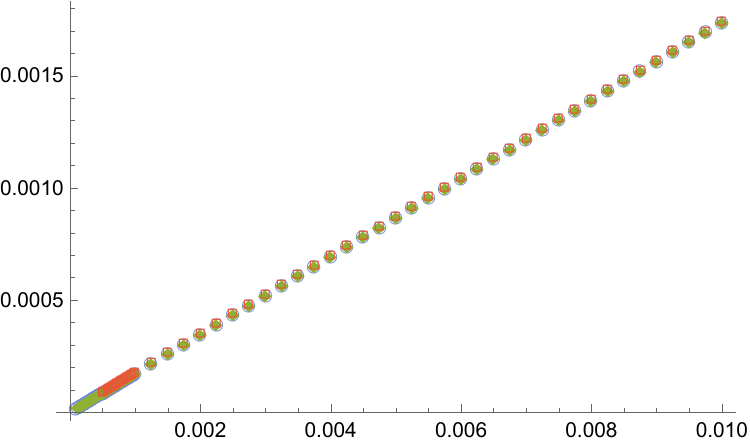}
\begin{picture}(0.3,0.4)(0,0)
\end{picture}}
\put(-210,100){\makebox(0,0){$\scriptstyle{\frac{2}{n}\lambda_n}$}}
\put(-20,15){\makebox(0,0){$\scriptstyle{\lads\,\kappa}$}}
\end{subfigure}
\vspace{-1.5cm}
\caption{The uplift of Schwarzian modes to the full Reissner-Nordstr\"om-\AdS{4} black hole geometry with $\mu=10$. We show the data for very low temperatures in the top panel,~\cref{fig:rnschwzoom}, demonstrating linear behavior with temperature. In the bottom left panel,~\cref{fig:rnschwfull}, we show the behavior over a greater range of temperatures, making the (mild) departures from linear growth manifest. Finally, in the bottom right panel~\cref{fig:rnschwcollapse}, we plot the eigenvalues rescaled by the overtone number $\frac{2}{n}\,\lambda_n$ to demonstrate that $\lambda_n \sim n\,T$ (notice the three curves are overlapping, which was achieved by working with extended precision numerics). }
\label{fig:rn_schwarzian}
\end{figure}
Using the numerical method described in~\cref{sec:HypSchw}, we were able to track the lowest-lying eigenvalue for different $n$. For example, working with $\mu = 10$ we examine the behavior within the interval $\kappa\, \lads \in (0.000975,0.1)$.\footnote{The choice of $\mu$ was dictated by convergence of the numerics for low temperatures. We  find that one needs to go to even lower temperatures at as we lower $\mu$.} Sample results of this numerical analysis for the lowest three overtones are illustrated in~\cref{fig:rn_schwarzian}. The numerical data for the eigenvalues for $n=2,3,4$ fit well to a quadratic polynomial for low values of surface gravity. Parameterizing the eigenvalues in the low-temperature regime $\kappa\,\lads \in [0.000975,0.001]$  as 
\begin{equation} \label{eq:fit_Sch}
 \lambda_n = a_n + b_n \, \kappa\, \lads + c_n \, (\kappa\, \lads)^2   \,,   
\end{equation}
we find 
\begin{equation}
\begin{aligned}
b_2 &= 0.17408  \,, & \qquad c_2 &= - 0.01512 \,, & \qquad a_2 \sim \order{10^{-10}}\,, \\ 
b_3 &= 0.26112 \,, & \qquad c_3 &= - 0.03021 \,, & \qquad a_3 \sim \order{10^{-9}} \,. \\ 
b_4 &= 0.34815\,, & \qquad c_4 &=  - 0.05022 \,, & \qquad a_4 \sim \order{10^{-9}} \,.
\end{aligned}
\end{equation}
This is indeed consistent with these modes becoming zero modes as we attain extremality, i.e., at  $\kappa =0$. In particular, as expected, at very low temperatures, we have a set of modes whose eigenvalues scale linearly with temperature. Furthermore, 
\begin{equation}
    \frac{b_3}{b_2} = \frac{0.26112}{0.17408} \approx \frac{3}{2} - 9.9 \times 10^{-7}\,,\qquad \frac{b_4}{b_3} = \frac{0.34815}{0.26112} \approx \frac{4}{3} - 3.715 \times 10^{-6}\,.
\end{equation}
This illustrates that the eigenvalues have a slope that scales linearly with the overtone number $n$, consistent with them being the extension of the Schwarzian mode onto the full geometry.\footnote{As before, the discrepancy of $\order{10^{-6}}$ is because we truncated the Taylor expansion at the quadratic order.} The upshot of our numerical results is that one has for a near-extremal Reissner-Nordstr\"om \AdS{4} black hole, a set of eigenmodes with eigenvalues 
\begin{equation}
    \lim_{T\to 0} \frac{\lambda_n \lads^2}{16\pi G_N} \approx \textbf{} \alpha\, n\, \frac{T}{T_q} + \order{T^2}\,,\qquad \abs{n} \geq 2\,,
\end{equation}
in excellent agreement with the expectation based on the near-horizon analysis.

\begin{figure}[ht]
\begin{subfigure}{0.45\linewidth}
\centering
\subcaptionbox{Norm against the proper distance from horizon.\label{fig:rnschhnorm}}[\linewidth]{
\includegraphics[width=0.9\textwidth]{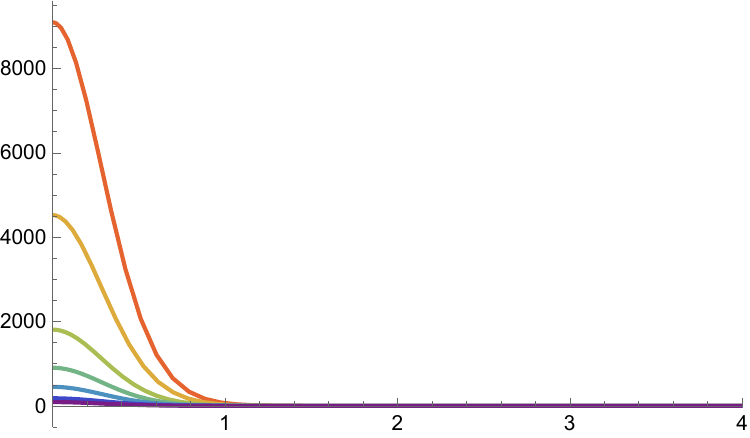}
\begin{picture}(0.3,0.4)(0,0)
\put(-190,100){\makebox(0,0){$\scriptstyle{\norm{X}^2}$}}
\put(0,15){\makebox(0,0){$\scriptstyle{\frac{\rho}{\lads}}$}}
\end{picture}}
\end{subfigure}
\hspace{0.6cm}
\begin{subfigure}{0.45\linewidth}
\subcaptionbox{Norm density versus proper distance from horizon\label{fig:rnschhnormdensity}}
[\linewidth]{
\includegraphics[width=0.9\textwidth]{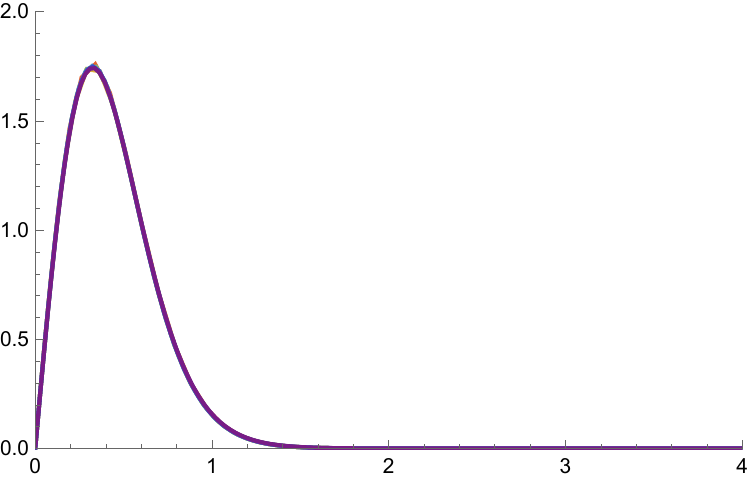}
\begin{picture}(0.3,0.4)(0,0)
\put(-200,100){\makebox(0,0){$\scriptstyle{\sqrt{f}\,\norm{X}^2}$}}
\put(0,15){\makebox(0,0){$\scriptstyle{\frac{\rho}{\lads}}$}}
\end{picture}}
\end{subfigure}
\caption{
The behavior of the norm of the perturbations, with $n=2$, as a function of proper distance $\rho \sim \lads \, \log(r/r_+)$ from the horizon. As with earlier results, this is for a Reissner-Nordstr\"om-\AdS{4} black hole geometry with  $\mu=10$. 
The norm $\norm{X}^2$ defined as the integrand of~\eqref{eq:schnorm}, viz., the specific combination of the functions $f_i$, $a_t$ and $a_r$. 
The data is shown for $\kappa \lads \in \{0.01, 0.005,0.002,0.001,0.0005,0.0002, 0.0001\}$.  On the left panel~\cref{fig:rnschhnorm} we show the bare norm which increases in amplitude as we go to lower temperature, similar to the feature seen in the BTZ case~\cref{fig:h_squared_btz}. However, the norm density, which is normalized to integrate to unity, displays data collapse as depicted in~\cref{fig:rnschhnormdensity}.
} 
\label{fig:rn-sch-hsq}
\end{figure}

We can also find eigenfunctions from our numerical analysis.\footnote{Obtaining them, however, requires a larger grid than the eigenvalue computation.}  To get a sense of how the modes behave, we examine the norm of the perturbations, which is the integrand of~\eqref{eq:schnorm}. We depict the result in~\cref{fig:rn-sch-hsq}. It is clear that for the range of temperatures considered, these fluctuations are localized near the horizon, thus confirming the fact that they are indeed the uplift of the  Schwarzian modes onto the full geometry. Moreover, one can check that the metric contributions to the norm are a few orders of magnitude larger than the Maxwell ones, and so we may see these modes as being ``mainly metric" (as one would expect from the near-horizon analysis). However, at large $r$, the norm density becomes negative (albeit very small). This is connected with the fact that $a_r$ decays slower at infinity than $f_3, f_4$ and does not seem to have any physical significance.  

\paragraph{The $\mathrm{U}(1)$ modes:} Since we already have the spherically symmetric fluctuations of the Maxwell field, we also ought to find the modes associated with the $\mathrm{U}(1)$ gauge symmetry. These should be part of the same eigensystem, and we ought to have all the necessary ingredients at hand to find them. 

\begin{figure}[ht]
\begin{subfigure}[b]{\linewidth}
\centering
\subcaptionbox{The eigenvalue spectrum at low temperatures. \label{fig:rnu1zoom}}[\linewidth]{
\includegraphics[width=0.7\textwidth]{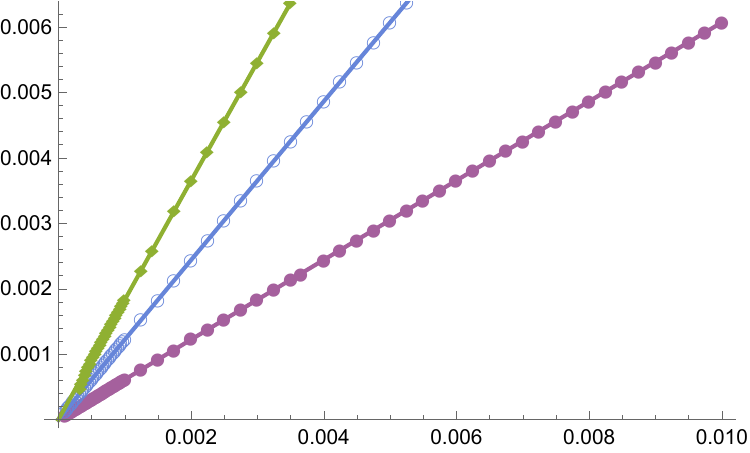}
\begin{picture}(0.3,0.4)(0,0)
\put(-320,175){\makebox(0,0){$\lambda_n$}}
\put(0,25){\makebox(0,0){$\lads\,\kappa$}}
\put(-240,135){\makebox(0,0){$n=3$}}
\put(-150,160){\makebox(0,0){$n=2$}}
\put(-130,125){\makebox(0,0){$n=1$}}
\end{picture}}
\end{subfigure}
\vspace{2cm}
\begin{subfigure}[b]{0.45\linewidth}
\centering
\subcaptionbox{The eigenvalue spectrum over an extended range.\label{fig:rnsu1full}}[\linewidth]{
\includegraphics[width=0.95\textwidth]{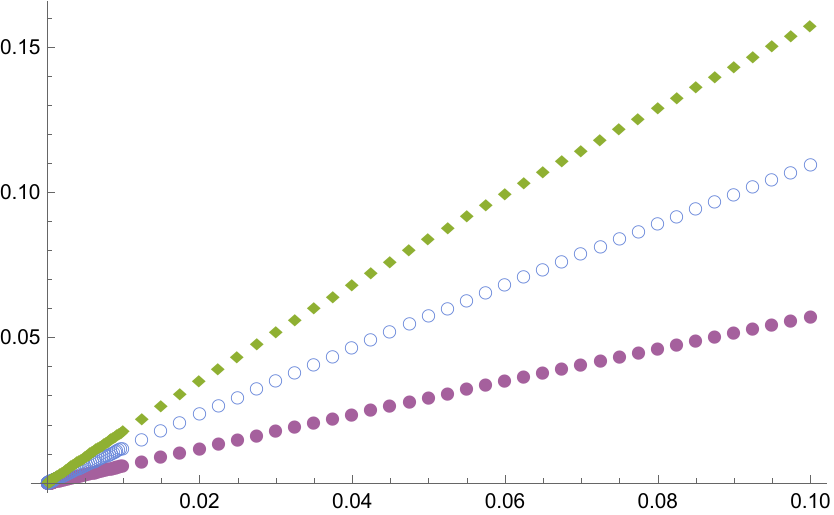}
\begin{picture}(0.3,0.4)(0,0)
\put(-200,100){\makebox(0,0){$\scriptstyle{\lambda_n}$}}
\put(0,15){\makebox(0,0){$\scriptstyle{\lads\,\kappa}$}}
\end{picture}}
\end{subfigure}
\hspace{1cm}
\begin{subfigure}[b]{0.45\linewidth}
\subcaptionbox{Rescaled eigenvalues demonstrating scaling.\label{fig:rnu1collapse}}[\linewidth]{
\includegraphics[width=0.95\textwidth]{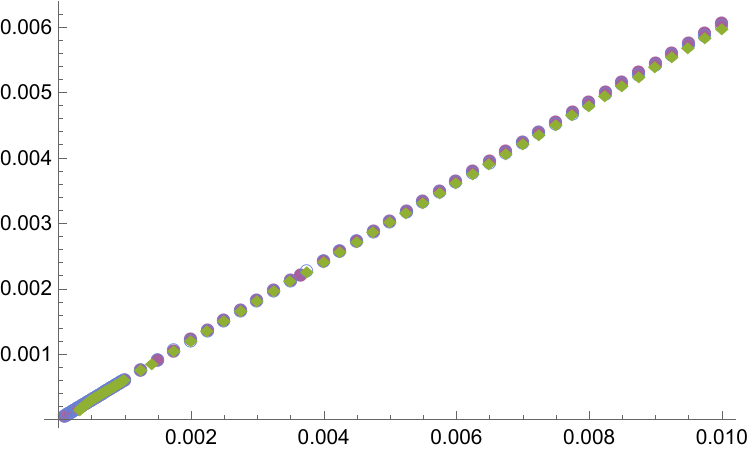}
\begin{picture}(0.3,0.4)(0,0)
\end{picture}}
\put(-205,100){\makebox(0,0){$\scriptstyle{\frac{\lambda_n}{n}}$}}
\put(-20,15){\makebox(0,0){$\scriptstyle{\lads\,\kappa}$}}
\end{subfigure}
\vspace{-1.5cm}
\caption{The uplift of $\mathrm{U}(1)$ gauge zero modes to the full Reissner-Nordstr\"om-\AdS{4} black hole geometry with $\mu=10$. We show the data for very low temperatures in the top panel~\cref{fig:rnu1zoom}, demonstrating linear behavior with temperature. In the bottom left panel~\cref{fig:rnsu1full}, we show the behavior over a greater range of temperatures, making the departures from linear growth manifest. Finally, in the bottom right panel~\cref{fig:rnu1collapse}, we plot the eigenvalues rescaled by the overtone number $\frac{1}{n}\,\lambda_n$ to demonstrate that $\lambda_n \sim n\,T$.  }
\label{fig:rn_gauge}
\end{figure}

Thus far, we tracked the lowest-eigenvalue and detected the Schwarzian modes. Now, we instead can track the second lowest-lying eigenvalue\footnote{Here the operator has an additional  factor of two relative to the operator that was considered for gauge modes in the hyperbolic \AdS{} black hole. The eigenvalues are therefore a factor of two larger.  } for a given Matsubara frequency. Unfortunately, one  needs to go to significantly lower temperatures to see the linear scaling properly. Since at lower temperatures, gradients near the horizon are larger, this makes the numerics more challenging. Sample results of this numerical analysis for the lowest three overtones are illustrated in~\cref{fig:rn_gauge}.
The numerical data for the eigenvalues for $n=1,2,3$ (see below) fit well to a quadratic polynomial for low values of surface gravity. For example, for $\mu = 10$ we may look at $\kappa\, \lads \in [9.75\times 10^{-4},10^{-3}]$. Letting, 
\begin{equation}
 \lambda_n = a_n + b_n \, \kappa\, \lads + c_n \, (\kappa\, \lads)^2   \,,   
\end{equation}
we find
\begin{equation}
\begin{aligned}
b_1 &= 0.61238\,, & \qquad c_1 &= - 0.56216 \,, & \qquad a_1 \sim \order{10^{-8}}\,, \\ 
b_2 &= 1.22216 \,, & \qquad c_2 &= - 1.75035 \,, & \qquad a_2 \sim \order{10^{-7}} \,. \\ 
b_3 &= 1.18378\,, & \qquad c_3 &= -4.19197 \,, & \qquad a_3 \sim \order{10^{-5}} \,.
\end{aligned}
\end{equation}
This is indeed consistent with these modes becoming zero modes as we attain extremality, i.e., at  $\kappa =0$. In particular, as expected, at very low temperatures, we have a set of modes whose eigenvalues scale linearly with the Matsubara frequency. Furthermore, 
\begin{equation}
    \frac{b_2}{b_1} = \frac{1.22216}{0.612376} \approx 2 - 0.0042\,,\qquad \frac{b_3}{b_1} = \frac{1.85748}{0.612376} \approx 3+ 0.001\,.
\end{equation}
This illustrates that the eigenvalues have a slope that scales linearly with the overtone number $n$, consistent with them being the extension of the $\mathrm{U}(1)$ mode onto the full geometry.\footnote{The numerical discrepancies in this case are slightly higher, but for similar reason as before.} The upshot of these numerical results is that one has 
\begin{equation}
    \lim_{T\to 0} \frac{\lambda_n\lads^2}{16\pi G_N} \approx \textbf{} \alpha\, n\, \frac{T}{T_q} + \order{T^2}\,,\qquad \abs{n} \geq 1\,,
\end{equation}
in excellent agreement with our expectation based on the near-horizon analysis. 

\begin{figure}[ht]
\centering
\includegraphics[width=0.7\textwidth]{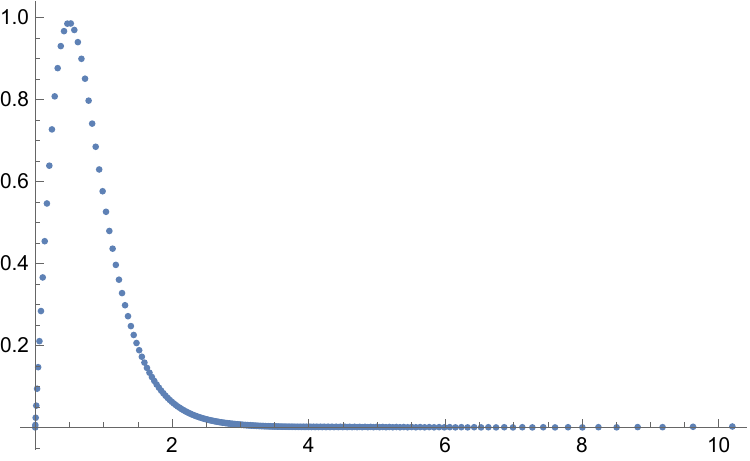}
\begin{picture}(0.3,0.4)(0,0)
\put(-340,180){\makebox(0,0){$\sqrt{f}\,\norm{X}^2$}}
\put(10,15){\makebox(0,0){$\dfrac{\rho}{\lads}$}}
\end{picture}
\caption{
The behavior of the norm density of the $n=1$ mode for $\kappa \,\lads = 0.001$ to illustrate that the $\mathrm{U}(1)$ gauge modes remain well-behaved. The conventions are as in~\cref{fig:rnschhnormdensity}.
} 
\label{fig:rn-u1-hsqn1}
\end{figure}

There is one issue that we need to explain, for the astute reader ought to be surprised by the inclusion of $\abs{n}=1$ mode. As argued around~\eqref{eq:Schwarzian_subst} such a mode should lead to singular behavior at the horizon. However, just as for Schwarzian modes, we can find the form of the eigenfunctions. In particular, we learn that $U_3(1) = U_r(1) = 0$ for $n=1$, and thus, $f_3, f_4, a_r$ are smooth, when we consider the second lowest-eigenvalue (but not for the lowest one). We illustrate this explicitly by plotting the norm for the $n=1$ mode for $\kappa\,\lads = 0.01$ in~\cref{fig:rn-u1-hsqn1}.

From these plots, it may seem surprising that $f_3$ and $a_r$ are of the same magnitude. One might have naively hoped that the $\mathrm{U}(1)$ modes have dominant support on gauge field fluctuations. However, this is not the case, unlike for the Schwarzian which was supported dominantly on graviton perturbations. The reason for this mixing turns out to be connected with our choice of gauge. For divergence free vectors we have
\begin{equation}
    \nabla^\mu \tilde{h}_{\mu \nu} = \frac{1}{2} F_{\nu \mu}a^\mu.
\end{equation}
Hence, as long as $F_{\nu \mu}a^\mu \neq 0$ (which here is equivalent to $a^\mu \neq 0$), $h_{\mu \nu}$ cannot vanish as well. However, let us emphasize that fluctuations $(h_{\mu \nu}, a_{\mu})$ and $(h_{\mu \nu} + \mathcal{L}_\xi g_{\mu \nu}, a_{\mu} + \mathcal{L}_\xi A_\mu)$ give the same contribution to the quadratic action (ignoring gauge fixing terms) and we can always choose $\xi$ in such a way $\nabla^\mu \tilde{h}_{\mu \nu} =0$.

\subsubsection{The rotational modes}

The final piece of our analysis is to unearth the modes corresponding to the $\mathrm{SO}(3)$ rotations, which we know exist in the near-horizon geometry. The incarnation of these modes in the full spacetime should lie in the vector sector of the theory. In particular, they should be built out of the Killing vectors of the $\mathbf{S}^2$. Exploiting this intuition, and accounting for the time-reversal properties, we can motivate the following ansatz for these modes:
\begin{equation}\label{eq:rnadsrotansatz}
\begin{split}
h_{\mu \nu} \,\d x^\mu \,\d x^\nu 
&= 
    \sqrt{\frac{3}{2\pi \,\beta}}\,  \mathfrak{v} \, 
    \left(\cos(\frac{2\pi n \,\tE}{\beta}) \,g_2 \, \frac{\d r}{\sqrt{f}} 
    - \sin(\frac{2\pi n\, \tE}{\beta}) \, g_1\, \sqrt{f} \,\d\tE
        \right) \,,\\ 
a_\mu \,\d x^\mu 
&= 
    \i \sqrt{\frac{3}{4\pi\, \beta}} \,\mathfrak{v} \sin(\frac{2\pi n \tE}{\beta}) \, g_3\,.
\end{split}
\end{equation}
Here $\mathfrak{v}$ is a covector on $\mathbf{S}^2$ and $g_i$ for $i=1,2,3$ are functions of $r$ alone. The norm of these fluctuations reads:
\begin{equation}\label{eq:rotnorm}
   \lads^{-2} \int \d^4 x \,\sqrt{g} \left(
\tilde{h}^{\mu \nu}\,h_{\mu \nu} + a_\mu \, a^\mu
    \right) = \int_{r_+}^\infty \d r \left(g_1^2 + g_2^2 - g_3^2
    \right).
\end{equation}
Hence, we see that $g_i$ must be sufficiently regular around $r=r_+$ and vanish sufficiently fast around $r=\infty$ for the perturbation to be an element of the Hilbert space we wish to consider. 

The Maxwell perturbations are automatically divergent-free. Moreover, they also satisfy $F_{\mu \nu} a^\nu = 0$. Therefore, the only non-trivial gauge constraint is
\begin{equation}
    0 =  \nabla^\mu \tilde{h}_{\mu \nu} = \nabla^\mu h_{\mu \nu}.
\end{equation}
In other words, we simply need the metric perturbation to respect the harmonic gauge. This (one) equation may be immediately solved to obtain
\begin{equation}
    g_1 = \beta  \, \frac{(r\,f' + 4\,f) \,g_2 + 2 \,r\,f  \,g_2' }{4 \pi  \,n \,r}.
\end{equation}
With this in mind, the eigenvalue problem becomes a system of two second-order ODEs for $g_2$ and $g_3$. 

Passing to the $z$-coordinate~\eqref{eq:rzmap}, and examining the fall-offs near the boundary and horizon,  should yield insights about field redefinitions. Assuming, 
\begin{equation}
g_2 \sim (1-z)^{s_2} \,, \qquad  g_3 \sim (1-z)^{s_3} \,,
\end{equation}
we find the following four decays near the horizon $z=1$, 
\begin{equation}
\begin{aligned}
s_2 &= 
    \frac{3}{2} \pm \sqrt{ \frac{9}{4} - 4\, \lambda } \,, \qquad 
s_2 &=
    \frac{3}{2} \pm \sqrt{\frac{1}{4} - \frac{1}{2}\, \lambda} \,, \\ 
s_3 &= 
    \frac{5}{2} \pm \sqrt{ \frac{9}{4} - 4\, \lambda }\,, \qquad 
s_3 &= 
    \frac{1}{2} \pm \sqrt{\frac{1}{4} - \frac{1}{2}\, \lambda} \,,
\end{aligned}
\end{equation}
which can be identified with  graviton and photon asymptotics. The choice of positive signs ensures that we get to impose Dirichlet boundary conditions at infinity. Smoothness at the horizon leads to 
\begin{equation}
g_2 \sim (1+z)^{\abs{n}-1} \,, \qquad 
g_3 \sim (1+z)^{\abs{n}}.
\end{equation}
At the end of the day, we can motivate the following reparameterization: 
\begin{equation}
\begin{split}
g_2(z) &= (1+z)^{\abs{n}-1}\, (1-z) \, V_2(z) \,, \\  
g_3(z) &=   (1+z)^{\abs{n}}\, (1-z)^\frac{1}{2}\, V_3(z) \,,
\end{split}
\end{equation}
supplemented with the boundary conditions $V_2(1) = V_3(1) =0$, which imposes Dirichlet BCs as long as
\begin{equation}
    \lambda < \frac{7}{16}\,.
\end{equation}
Due to the factors of $(1-z)^{\abs{n}-1}$, the perturbations will fail to be smooth for $n=0$. Hence, we shall consider only $n \ge 1$ (without loss of generality, we may take $n$ to be positive).

\begin{figure}[ht]
\begin{subfigure}[b]{\linewidth}
\centering
\subcaptionbox{The eigenvalue spectrum at low temperatures. \label{fig:rnrotzoom}}[\linewidth]{
\includegraphics[width=0.7\textwidth]{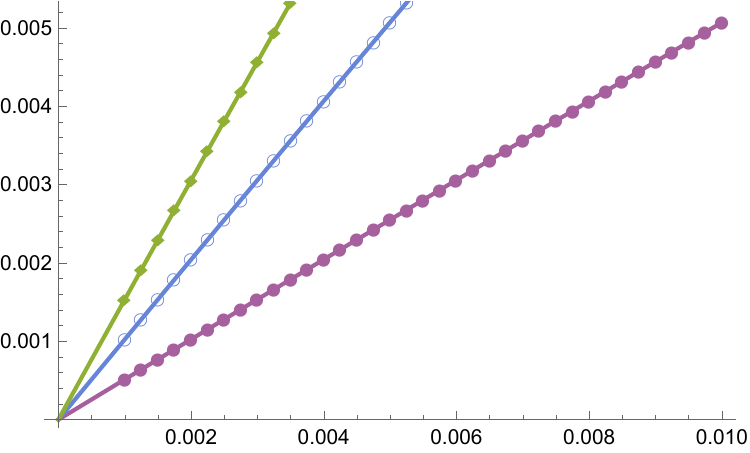}
\begin{picture}(0.3,0.4)(0,0)
\put(-320,175){\makebox(0,0){$\lambda_n$}}
\put(0,25){\makebox(0,0){$\lads\,\kappa$}}
\put(-240,135){\makebox(0,0){$n=3$}}
\put(-150,160){\makebox(0,0){$n=2$}}
\put(-130,125){\makebox(0,0){$n=1$}}
\end{picture}}
\end{subfigure}
\vspace{2cm}
\begin{subfigure}[b]{0.45\linewidth}
\centering
\subcaptionbox{The eigenvalue spectrum over an extended range.\label{fig:rnrotfull}}[\linewidth]{
\includegraphics[width=0.95\textwidth]{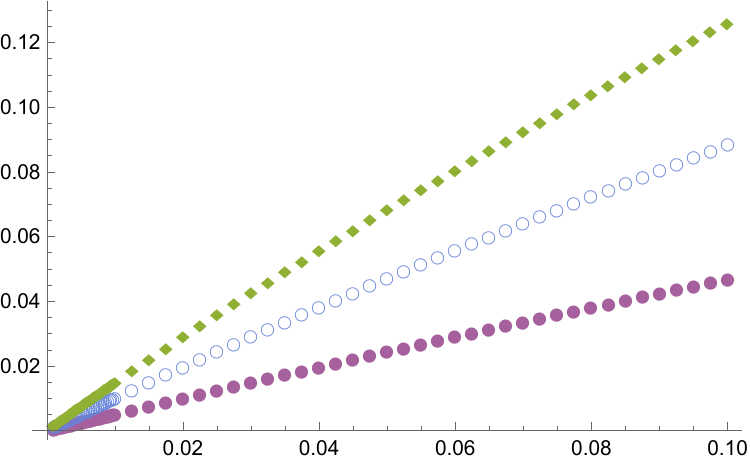}
\begin{picture}(0.3,0.4)(0,0)
\put(-200,100){\makebox(0,0){$\scriptstyle{\lambda_n}$}}
\put(0,15){\makebox(0,0){$\scriptstyle{\lads\,\kappa}$}}
\end{picture}}
\end{subfigure}
\hspace{1cm}
\begin{subfigure}[b]{0.45\linewidth}
\subcaptionbox{Rescaled eigenvalues demonstrating scaling.\label{fig:rnrotcollapse}}[\linewidth]{
\includegraphics[width=0.95\textwidth]{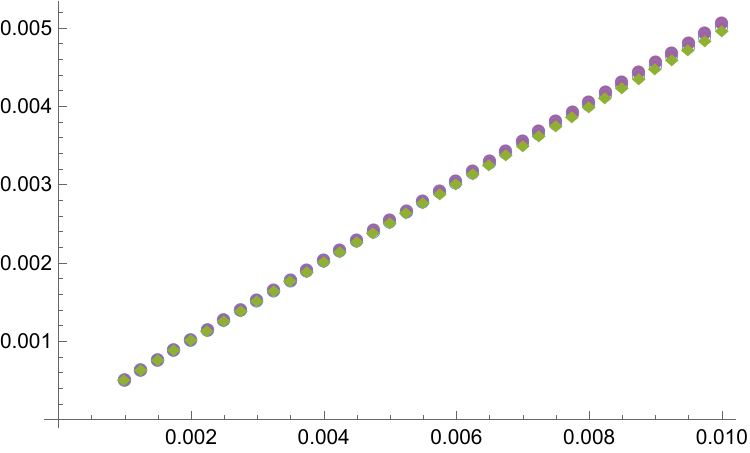}
\begin{picture}(0.3,0.4)(0,0)
\end{picture}}
\put(-205,100){\makebox(0,0){$\scriptstyle{\frac{\lambda_n}{n}}$}}
\put(-20,15){\makebox(0,0){$\scriptstyle{\lads\,\kappa}$}}
\end{subfigure}
\vspace{-1.5cm}
\caption{The uplift of the   rotational field zero modes to the full Reissner-Nordstr\"om-\AdS{4} black hole geometry ($\mu=10$). We show the data for very low temperatures in the top panel~\cref{fig:rnrotzoom}, demonstrating linear behavior with temperature. In the bottom left panel~\cref{fig:rnrotfull}, we show the behavior over a greater range of temperatures, making the departures from linear growth manifest. Finally, in the bottom right panel~\cref{fig:rnrotcollapse}, we plot the eigenvalues rescaled by the overtone number $\frac{1}{n}\,\lambda_n$ to demonstrate that $\lambda_n \sim n\,T$.  } 
\label{fig:rn_rotation}
\end{figure}
Tracking once again the lowest-lying eigenvalue\textbf{}, using the numerical scheme described in~\cref{sec:HypSchw}, for different $n$ we obtain our rotational modes. Sample results of this numerical analysis for the lowest three Matsubara modes are illustrated in~\cref{fig:rn_rotation} for $\mu =10$.
The numerical data for the eigenvalues for $n=1,2,3$ fit well to a quadratic polynomial for low values of surface gravity. For example, for $\mu = 10$ we may look at $\kappa\, \lads \in [10^{-3},10^{-2}]$. Letting, 
\begin{equation}
 \lambda_n = a_n + b_n \, \kappa\, \lads + c_n \, (\kappa\, \lads)^2   \,,   
\end{equation}
we find   
\begin{equation}
\begin{aligned}
b_1 &= 0.51382 \,, & \qquad c_1 &= - 0.669075 \,, & \qquad a_1 \sim \order{10^{-6}}\,, \\ 
b_2 &= 1.02451 \,, & \qquad c_2 &= - 2.286737  \,, & \qquad a_2 \sim \order{10^{-6}} \,. \\ 
b_3 &= 1.53324 \,, & \qquad c_3 &=  - 4.363498  \,, & \qquad a_3 \sim \order{10^{-5}} \,.
\end{aligned}
\end{equation}
This is indeed consistent with these modes becoming zero modes as we attain extremality, i.e., at  $\kappa =0$. In particular, as expected, at very low temperatures, we have a set of modes whose eigenvalues scales linearly with temperature. Furthermore, 
\begin{equation}
    \frac{b_2}{b_1} = \frac{1.02451}{0.51382} \approx 2 - 0.00450\,,\qquad \frac{b_3}{b_2} = \frac{1.53324}{1.02451} \approx \frac{3}{2} - 0.00344\,.
\end{equation}
This illustrates that the eigenvalues have a slope that scales linearly with the overtone number $n$, consistent with them being the extension of the Schwarzian mode onto the full geometry. The upshot of this numerical results is that one has 
\begin{equation}
    \lim_{T\to 0} \frac{\lambda_n \lads^2}{16\pi G_N} \approx \textbf{} \alpha\, n\, \frac{T}{T_q} + \order{T^2}\,,\qquad \abs{n} \geq 1\,,
\end{equation}
in excellent agreement with our expectation based on the near-horizon analysis.

\begin{figure}[ht]
\begin{subfigure}{0.45\linewidth}
\centering
\subcaptionbox{Norm against the proper distance from horizon.\label{fig:rnrothnorm}}[\linewidth]{
\includegraphics[width=0.9\textwidth]{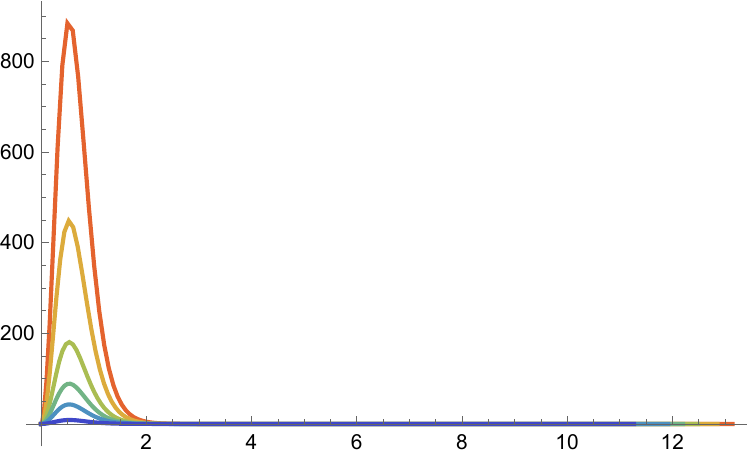}
\begin{picture}(0.3,0.4)(0,0)
\put(-190,100){\makebox(0,0){$\scriptstyle{\norm{X}^2}$}}
\put(0,15){\makebox(0,0){$\scriptstyle{\frac{\rho}{\lads}}$}}
\end{picture}}
\end{subfigure}
\hspace{0.6cm}
\begin{subfigure}{0.45\linewidth}
\subcaptionbox{Norm density versus proper distance from horizon\label{fig:rnrothnormdensity}}
[\linewidth]{
\includegraphics[width=0.9\textwidth]{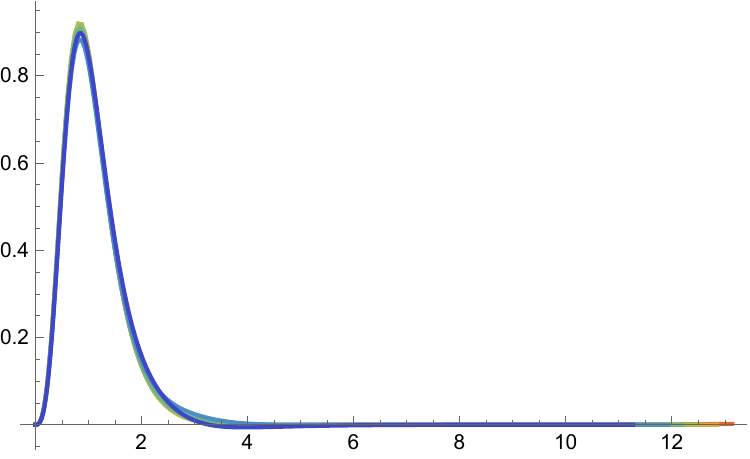}
\begin{picture}(0.3,0.4)(0,0)
\put(-200,100){\makebox(0,0){$\scriptstyle{\sqrt{f}\,\norm{X}^2}$}}
\put(0,15){\makebox(0,0){$\scriptstyle{\frac{\rho}{\lads}}$}}
\end{picture}}
\end{subfigure}
\caption{
The behavior of the norm of the perturbations, with $n=2$, as a function of proper distance $\rho \sim \lads \, \log(r/r_+)$ from the horizon. As with earlier results, this is for a Reissner-Nordstr\"om-\AdS{4} black hole geometry with  $\mu=10$. 
The norm is defined as the integrand of~\eqref{eq:rotnorm}. 
The data is shown for $\kappa \lads \in \{0.01, 0.005,0.002,0.001,0.0005,0.0002, 0.0001\}$.  On the left panel~\cref{fig:rnrothnorm} we show the bare norm which increases in amplitude as we go to lower temperature, similar to the feature seen in the BTZ case~\cref{fig:h_squared_btz}. However, the norm density, which is normalized to integrate to unity, displays data collapse as depicted in~\cref{fig:rnrothnormdensity}. 
} 
\label{fig:rn_rot-hsq}
\end{figure}
We can also find eigenfunctions, see~\cref{fig:rn_rot-hsq}.  Moreover, for the considered temperatures, these fluctuations are localized near the horizon, thus confirming the fact that they are connected to the rotational modes in the near-horizon region.

\subsubsection{Flat space limit} \label{sec:flatlimit}

Thus far, we have discussed only black holes with AdS asymptotics. The presence of the timelike conformal boundary makes the analysis significantly easier because, as we have seen, infinity is a regular singular point of the eigenvalue equations at hand. Instead of trying to solve a significantly harder problem in the asymptotically flat context, we can try to take the limit with $\lads \to \infty$. Of course, out of the discussed examples, only Reissner-Nordstr\"om admits this limit.\footnote{A Ricci-flat black hole with a hyperbolic horizon would fail to be asymptotically flat, and does not seem to be of particular physical interest.} However, there are a few subtleties involved in the process:
\begin{itemize}[wide,left=0pt]
\item Extremal Reissner-Nordstr\"om-AdS has $\mu >1$ whereas asymptotically flat solution must have $\mu = 1$. This follows from~\eqref{eq:rnTmu} when we take  $\lads \to \infty$ (equivalently, take $r_+ \simeq r_- \ll \lads$). For concreteness, we will do in such a way that the black hole's charge at extremality remains constant. We will see important consequences of the constraint $\mu = 1$ in~\cref{sec:BTZ_rot_mode}.
\item The quadratic fluctuation operators (e.g., the Lichnerowicz operator and its charged analogue) being dimensionful, the eigenvalues have mass dimension $+2$. Hitherto, we took this into account by using $\lads$ to supply the dimensions, by defining
\begin{equation}
    \hat{L} X = \frac{\lambda}{\lads^2} X\,.
\end{equation}
In order to take the limit, we will instead use the black hole size to set the dimensions. Hence, the analog of the parametrization~\eqref{eq:fit_Sch} should be rather (ignoring for simplicity the constant  and quadratic deviation terms characterized by $a_n$ and $c_n$)
\begin{equation}
    \lambda_n \, \frac{r_0^2}{\lads^2} = \tilde{b}_n\, (\kappa\, r_0) \,, \qquad  \tilde{b}_n \equiv b_n \frac{r_0}{\lads}\,,
\end{equation}
where $r_0$ is the black hole radius at extremality. We expect $\tilde{b}_n$ to have a sensible  limit as we take $\lads \to \infty$.
\item The final complication we need to deal with is the following. In our analysis, we have focused on eigenmodes of the quadratic fluctuation operator that are in $L^2$, which placed an upper bound on the dimensionless eigenvalue $\lambda_n$, cf.~\eqref{eq:lambdabd}. As we make the black hole smaller in AdS units, we wish to ensure that this is respected. In addition, we are interested in ensuring that we are in the regime where the eigenvalues scale linearly with temperature. Both of these are satisfied only when the black hole is sufficiently cold, which demands 
\begin{equation}
    \kappa \,r_0 = \order{\frac{r_0^2}{\lads^2}} \,.
\end{equation}
This makes the numerical analysis much more challenging. We were nevertheless able to attain low enough temperatures to see the effect we describe below with $\frac{r_0}{\lads} \sim \order{10^{-1}}$.
\end{itemize}

\begin{figure}[ht]
\begin{subfigure}[b]{0.45\linewidth}
\centering
\subcaptionbox{Divergence of $\tilde{b}_2$ as $\mu \to 1$.\label{fig:b2tvsmu}}[\linewidth]{
\includegraphics[width=0.95\textwidth]{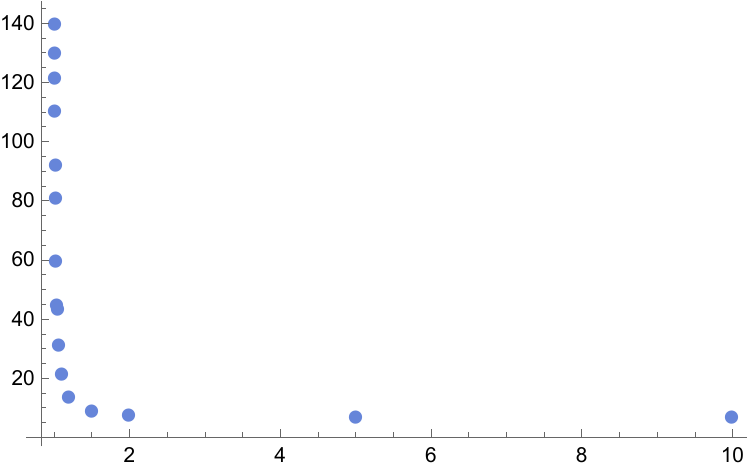}
\begin{picture}(0.3,0.4)(0,0)
\put(-200,120){\makebox(0,0){$\scriptstyle{\tilde{b}_2}$}}
\put(0,20){\makebox(0,0){$\scriptstyle{\mu}$}}
\end{picture}}
\end{subfigure}
\hspace{1cm}
\begin{subfigure}[b]{0.45\linewidth}
\subcaptionbox{ The rate of divergence: $\tilde{b}_2 \sim (\mu-1)^{-1}$.\label{fig:b2tvsmum1inv}}[\linewidth]{
\includegraphics[width=0.95\textwidth]{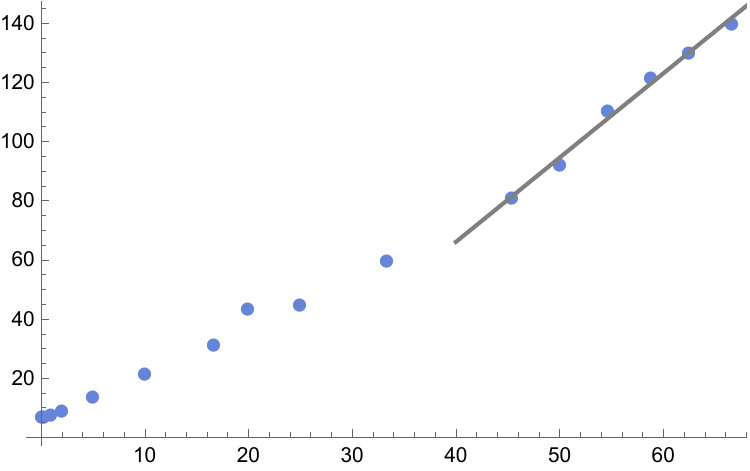}
\begin{picture}(0.3,0.4)(0,0)
\end{picture}}
\put(-200,120){\makebox(0,0){$\scriptstyle{\tilde{b}_{2}}$}}
\put(-10,20){\makebox(0,0){$\scriptstyle{\frac{1}{\mu-1}}$}}
\end{subfigure}
\caption{The behavior of the $\mathrm{U}(1)$ modes in a Reissner-Nordstr\"om black hole as a function of the (dimensionless) chemical potential. We use the chemical potential as a proxy to take the flat space limit, since $\mu \to 1$ as $r_+ \simeq r_- \ll \lads$. In the limit, we see that the linear fit coefficient $\tilde{b}_2$ diverges as $(\mu-1)^{-1}$ strongly suggesting that these modes are absent in the asymptotically flat black hole spacetime. }
\label{fig:rn-u1-blowup}
\end{figure}

Having explained all these subtleties, we are ready to investigate the limits themselves. Note that since (with reasonable precision) $b_n = \frac{n}{2} b_2$, we may restrict ourselves to studying $\tilde{b}_2$. Our numerical analysis for the $L^2$ normalizable modes reveals the following:
\begin{itemize}[wide,left=0pt]
\item For the Schwarzian  and rotational family of eigenmodes, the rescaled fit parameter $\tilde{b}_2$ remains constant as we take $r_0 \ll \lads$.  This suggests that modes robustly survive the limit $\lads \to \infty $, and should therefore contribute to the quantum effects around the asymptotically flat Reissner-Nordstr\"om black hole.
\item On the other hand for the $\mathrm{U}(1)$ modes we find something surprising; $\tilde{b}_2$ blows up as we go with $\lads \to \infty$ as depicted in~\cref{fig:rn-u1-blowup}. In particular, we find that 
$\tilde{b}_2 \to (\mu-1)^{-1} $ as $\mu \to 1$. Recall that we are using the chemical potential (which is dimensionless in our convention) to characterize the limit where the cosmological scale diverges. This strongly suggests that these modes should not contribute in the $\lads \to \infty$ limit. We will reconcile this conclusion with the near-horizon analysis in~\cref{sec:BTZ_rot_mode}, and propose a criterion to identify when this can happen.
\end{itemize}

\section{Role of rotational modes}\label{sec:rot}

The Schwarzian mode is universally present in all near-extremal black hole geometries. As we noted in~\cref{sec:intro}, one also expects isometries of the background geometry to generate a set of zero modes. We indeed saw their avatars in the full geometry in~\cref{sec:schw} for the neutral hyperbolic \AdS{4} and the charged Reissner-Nordstr\"om-\AdS{4} black holes. We now turn to the question of understanding these zero modes from the near-horizon perspective. 

The reason for doing so is to address a puzzle raised in~\cite{Rakic:2023vhv}. This paper examined the rotational zero mode in the near-horizon geometry of the extremal Kerr black hole using the harmonic gauge fixing condition for gravitational perturbations. They demonstrated that in this gauge, the putative zero mode associated with the rotational isometry was not regular at the poles of the deformed $\mathbf{S}^2$. The lack of smoothness suggested a potential absence of such rotational zero modes in axisymmetric spacetimes.\footnote{
In the cases we have analyzed  the issue arises for axisymmetric solutions (viz., for abelian rotations). For example, in the spherically symmetric case, one can dimensionally reduce in the near-horizon region to JT gravity with the isometries realized as non-abelian gauge fields.}
This work did not examine other geometries closely, but noted two facts regarding the absence of such rotational zero modes: a) the result was consistent with the expectation of low-temperature behavior of the extremal BTZ geometry, and b) if true, and moreover if one insisted on working in harmonic gauge, this result could have serious consequences for supersymmetric black holes such as the BMPV solution. 

We now turn to a detailed analysis of the situation by working in the near-horizon geometry. As a first step, we examine axisymmetric solutions supported by matter or cosmological constant. For the Kerr \AdS{4} black hole, we demonstrate the presence of a rotational zero mode, satisfying the harmonic gauge condition, as required. We also check that something similar occurs in the presence of background matter fields by examining the Kerr-Newman solution. These two examples suffice to reveal the problem in the Kerr geometry -- the harmonic gauge condition breaks down! We explain why this is the case, and argue that there is indeed a rotational zero mode in the Kerr geometry, thereby resolving the puzzle raised in~\cite{Rakic:2023vhv}. For completeness, we then also demonstrate that the supersymmetric BMPV solution does indeed have all the zero modes expected from the analysis of ~\cite{Sen:2012cj}. The content of this section addresses fact b) from the previous paragraph. We will also clarify the physics behind fact a) in section \ref{sec:BTZ_rot_mode}.

\subsection{Rotational modes in Kerr \texorpdfstring{\AdS{4}}{AdS4}}

We now turn to our first example, the Kerr \AdS{4} black hole. We will work in the near-horizon geometry, which is obtained by zooming into the \AdS{2} throat region, cf.~\cite{Hartman:2008pb}. The line element is given by (nb: we use $x = \cos\theta$ instead of the polar angle)
\begin{equation}
ds^2 = 
    F(x) \left( \left(y^2-1\right) \d \tau^2 + \frac{\d y^2}{y^2-1} + G(x)\frac{\d x^2}{1-x^2}\right)+H(x) (\d \phi+\i  \Omega \, (y-1)\,\d \tau )^2\,.
\end{equation}
The functions appearing in the metric are 
\begin{equation}
 F(x) = \frac{a^2 \, x^2+r_+^2}{V} \,, \quad 
G(x) = \frac{V}{1-\frac{a^2\, x^2}{\lads^2}} \,, \quad 
H(x) = (1-x^2) \,\frac{\lads^2 \left(a^2+r_+^2\right)^2 \left(\lads^2-a^2 \,x^2\right)}{\left(\lads^2-a^2\right)^2 \left(a^2 x^2+r_+^2\right)} \,,
\end{equation}
with 
\begin{equation}
\Omega=\frac{2\, a \,r_+ \left(1-\frac{a^2}{\lads^2}\right)}{V \left(a^2+r_+^2\right)} \,, \qquad 
a= \frac{r_+ \sqrt{\lads^2+3 r_+^2}}{\sqrt{\lads^2-r_+^2}}\,, 
\qquad 
V=\frac{a^2+\lads^2+ 6\, r_+^2}{\lads^2} \,.
\end{equation}

The idea as in~\cite{Rakic:2023vhv} is to seek a zero-mode generated by a large diffeomorphism along the Killing field $\partial_\phi$. We will need to add a compensating vector field to ensure that the resulting perturbation lies in harmonic gauge. Therefore, consider the following vector field
\begin{equation}\label{eq:KerrAdSrotxi}
\xi^\mu \partial_\mu = 
H(\tau,y)\,  \partial_\phi + \alpha(x) \,\nabla^\mu H \, \partial_\mu \,, \qquad 
H(\tau, y) = e^{\i \,n \,\tau} 
    \left(\frac{y-1}{y+1}\right)^{\frac{\abs{n}}{2}}\,.
\end{equation}
This choice is suggested by the structure of the gauge field zero modes that arise in the \AdS{2} throat, were we to ignore the warping along the $x$ direction, and dimensionally reduce to a 2d gravity theory. In particular, since $H(\tau,y) \to e^{\i\,n\,\tau}$ as $y\to \infty$, this vector field is not normalizable. It therefore generates a large diffeomorphism (and not just a gauge transformation). The second piece involving the yet to be determined function $\alpha(x)$ is a compensating (normalizable) diffeomorphism, which is included here to bring the resulting perturbation into harmonic gauge.

\begin{figure}[ht]
\centering
\includegraphics[width=.7\textwidth]{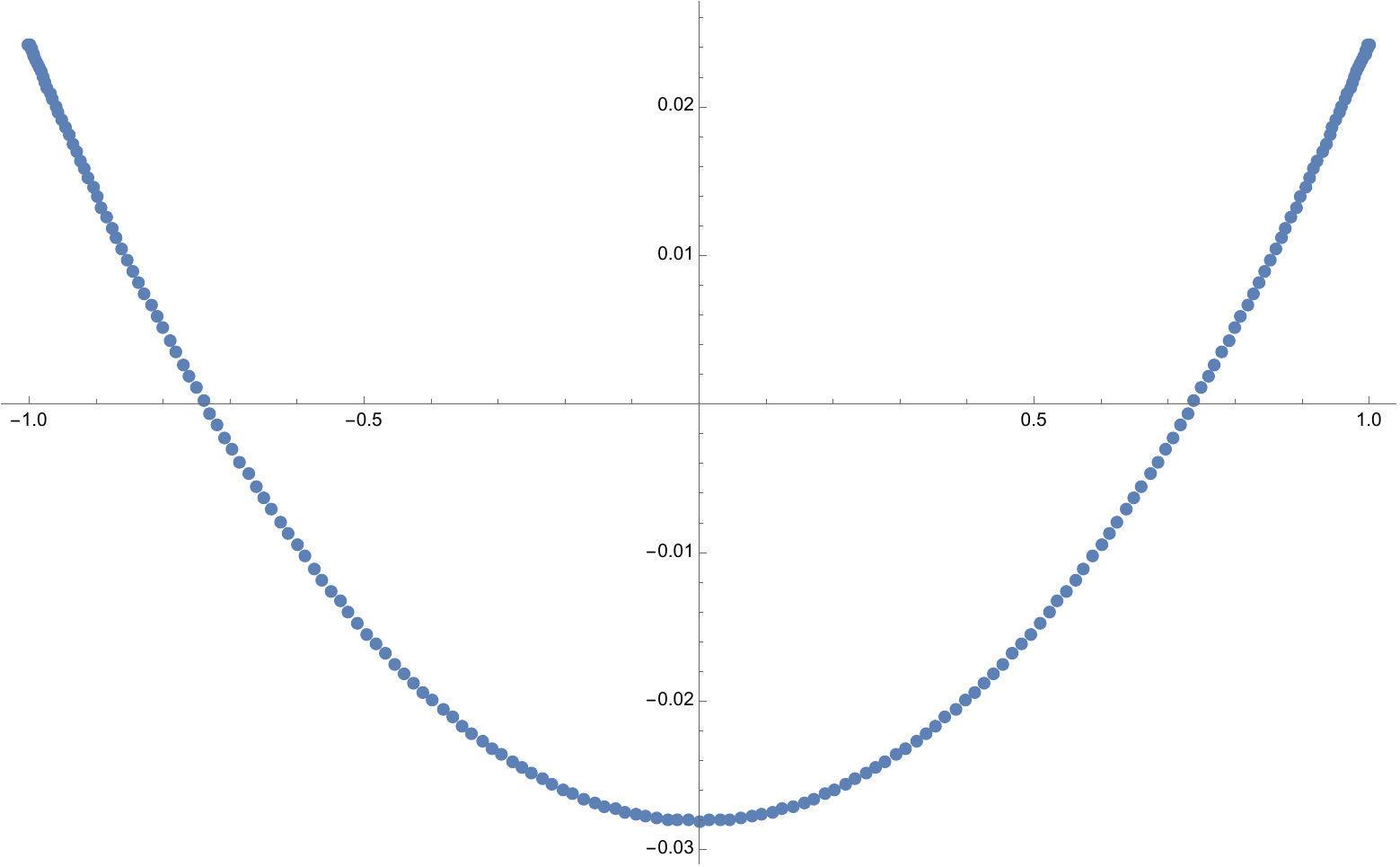}
\setlength{\unitlength}{0.1\columnwidth}
\begin{picture}(0.3,0.4)(0,0)
\put(-3.65,4.5){\makebox(0,0){$\alpha(x)/\ell_{\textrm{AdS}}$}}
\put(-0.2,2.5){\makebox(0,0){$x$}}
\end{picture}
\caption{The solution for the profile of the function $\alpha(x)$ in rotational zero mode $\xi$ defined in~\eqref{eq:KerrAdSrotxi} at $r_+ = 0.2\,\lads$. Note importantly, that the function is well-behaved at the poles of the $\mathbf{S}^2$, $x=\pm 1$.}
\label{fig:KerrAdS}
\end{figure}
The resulting metric perturbation $h_{\mu\nu} = 2\, \nabla_{(\mu}\xi_{\nu)}$ is automatically trace-free $h\indices{_\mu^\mu} =0$. Thus, to ensure it preserves the harmonic gauge, we only need to require it is divergence-free $\nabla^\mu h_{\mu\nu} = 0$ as well. This leads to a simple (but long) equation for $\alpha$. It is a second order, linear inhomogeneous ODE. As such, it is straightforward to solve this equation numerically. We depict an illustrative solution for a particular choice of $r_+$ in~\cref{fig:KerrAdS}. Perhaps surprisingly, this equation does not depend on $n$. One can also check directly that the metric perturbation thus generated lies in the kernel of the spin-2 Lichnerowicz operator~\eqref{eq:Lichnerowicz_AdS}.

The upshot of this calculation is that there is a smooth solution as long as $\Lambda = -\frac{3}{\lads^2} \neq 0$.  We will shortly see that this suggests that something singular occurs in the Ricci-flat limit, which will be relevant for the Kerr solution.

\subsection{Rotational modes in Kerr-Newman}

Next we turn to the Kerr-Newman geometry as a solution in Einstein-Maxwell theory. The authors of ~\cite{Rakic:2023vhv} examined the near-horizon geometry for zero modes, and in particular constructed the explicit form of the Schwarzian mode. The line-element of the near-horizon spacetime takes the form 
\begin{equation}\label{eq:KNnearhor}
\d s^2 
= \Gamma (x) \left(\left(y^2-1\right)\d \tau^2+\frac{\d y^2}{y^2-1}+\frac{\d x^2}{1-x^2}+\left(1-x^2\right) \gamma(x)^2 \left(\d\phi +\i\, \Omega\, (y-1)\, \d \tau \right)^2 \right) ,
\end{equation}
and is supported by a Maxwell field, which in turn is 
\begin{equation}
A = \i f(x)\, \Omega\, \d \tau + \left(
f(x) - \frac{e}{\Omega}
        \right) \d \phi\,.
\end{equation}
The functions appearing above are 
\begin{equation}
    \Gamma(x) = r_+^2 + a^2\, x^2\,, \qquad 
    \gamma(x) = \frac{r_+^2 + a^2}{r_+^2 + a^2 \,x^2}\,, \qquad f(x) = Q \frac{r_+^2 + a^2}{2\,r_+\, a} \frac{r_+^2 - a^2 x^2}{r_+^2 + a^2\, x^2} \,.
\end{equation}
The physical parameters, $\Omega$,  $e$, and $r_+$ are themselves 
\begin{equation}
    \Omega = \frac{2\, a\, r_+}{r_+^2 +a^2} \,, \qquad 
    e = \frac{Q^3}{r_+^2 + a^2} \,, \qquad 
    r_+= \sqrt{a^2 + Q^2}\,.
\end{equation}
Note that the gauge field has a finite expression as $a\to 0$, despite the factors of $\Omega^{-1}$ the expression. 

As in the vacuum case, we work with the harmonic gauge for the metric and Lorentz gauge for Maxwell perturbations. In~\cite{Rakic:2023vhv} the Schwarzian mode was exhibited as an element in the kernel of the Lichnerowicz operator, which for Einstein-Maxwell theory is given by\footnote{We define the operator by the kinetic term for the gravitons parameterized as $h^{\mu\nu} \, (\Delta_{L}^{\rm EM} h)_{\mu\nu}$ in harmonic gauge, which is slightly different from the presentation in~\eqref{eq:Lopmeasure}. We are also employing the harmonic and Lorenz gauge choices instead of the one used in~\cref{sec:RNads} in the near-horizon geometry.} 
\begin{equation}\label{eq:LichEM}
\begin{split}
(\Delta_{L}^{\rm EM} h)_{\mu\nu} 
&= 
  ( \Delta_L h)_{\mu\nu} -\frac{1}{2} \,g_{\mu\nu} \,g^{\alpha\beta}\,(\Delta_L h)_{\alpha\beta} \\
&\quad  
    + \frac{1}{4} \left[ 
    - F^2 \left(h_{\mu\nu} - \frac{g_{\mu\nu}}{2}\,h \right) - 4\,h\, F_{\mu\rho}F\indices{_\nu^\rho} + 4\,F_{\mu\rho} \, F_{\nu\sigma}\, h^{\rho \sigma} + 8\, F_{\mu\rho} \,F\indices{_\sigma^\rho} \, h\indices{_\nu^\sigma} 
    \right] .
\end{split}          
\end{equation}
Furthermore, they also pointed out that a putative rotational zero mode, which one would expect to be generated by the diffeomorphism along the Killing field $\partial_\phi$, fails to be in the kernel of~\eqref{eq:LichEM}. 

We will now examine the latter statement closely, and verify that there is indeed a rotational zero mode. We will consider an ansatz identical to what we chose in Kerr \AdS{4}. Imagine  
generating a large diffeomorphism of the near-horizon geometry~\eqref{eq:KNnearhor} along the vector field $\xi$ introduced in the Kerr \AdS{4} context in~\eqref{eq:KerrAdSrotxi}, reproduced below for convenience
\begin{equation}\label{eq:rotKerrxi}
\xi = H(\tau,y)\,  \partial_\phi + \alpha(x)\, \nabla^\mu H \, \partial_\mu \,,  \qquad 
    H(\tau, y) = e^{\i\, n \,\tau} \left( \frac{y-1}{y+1}\right)^\frac{\abs{n}}{2} \,.
\end{equation}
The component along $\partial_\phi$ is the part that was considered in~\cite{Rakic:2023vhv}. 

\begin{figure}[ht]
\centering
\includegraphics[width=.7\textwidth]{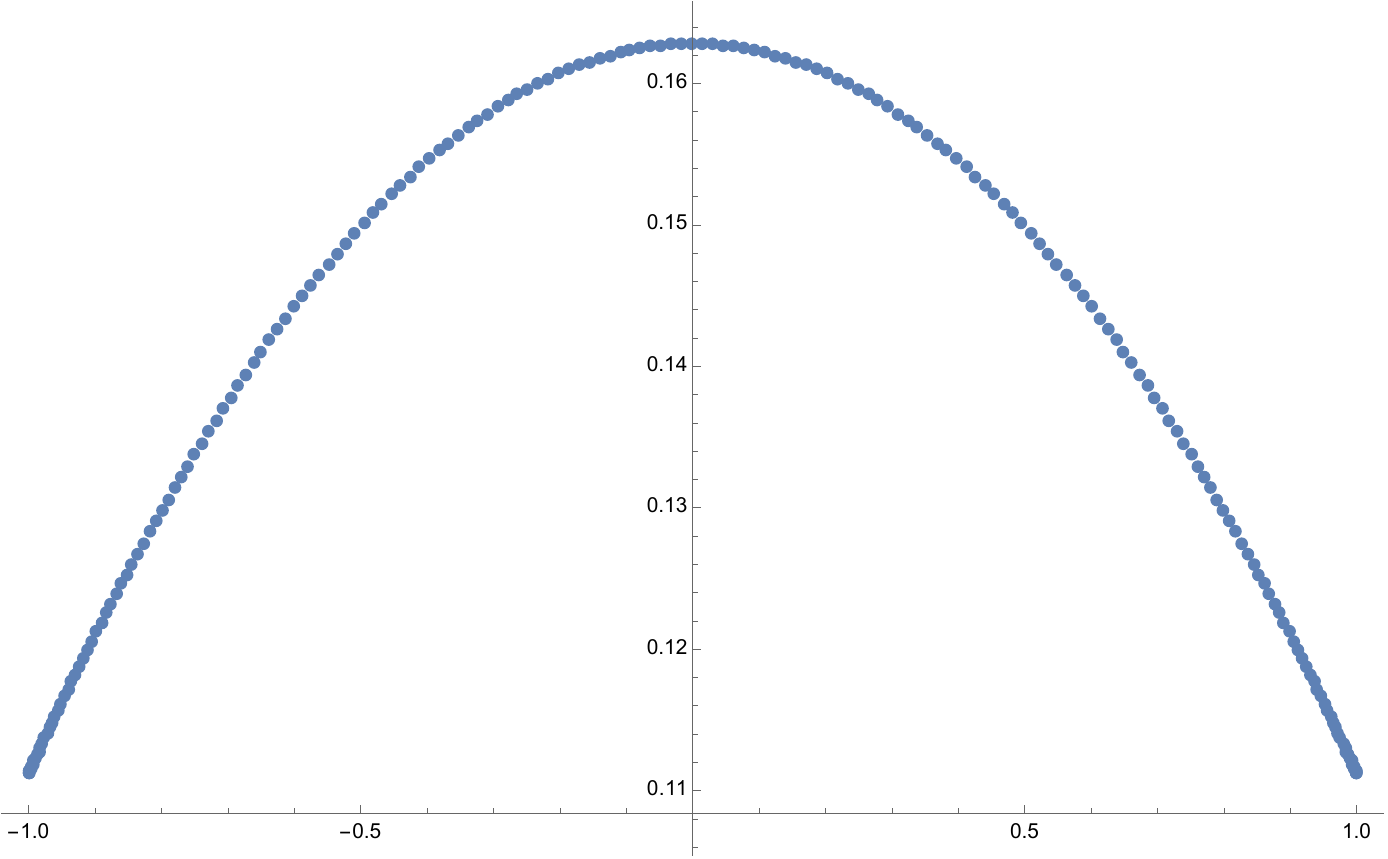}
\setlength{\unitlength}{0.1\columnwidth}
\begin{picture}(0.3,0.4)(0,0)
\put(-3.8,4.5){\makebox(0,0){$\alpha(x)/r_+^2$}}
\put(-0.1,0.4){\makebox(0,0){$x$}}
\end{picture}
\caption{The solution for the profile of the function $\alpha(x)$ in rotational zero mode $\xi$ defined in~\eqref{eq:alphaKerrode} at $a = 0.2\,r_+$ (thus $Q = 0.98\, r_+$). As in~\cref{fig:KerrAdS} we note that the function is well-behaved at the poles of the $\mathbf{S}^2$, $x=\pm 1$.}
\label{fig:KerrNew}
\end{figure}

Since we want the perturbation generated by $\xi$ to be a physical zero mode, it must satisfy the harmonic and Lorenz gauge conditions. It is immediate to see that the latter cannot be satisfied as written, since $\xi$ also Lie drags the gauge field. Per se, this is not a problem because we may accompany the diffeomorphism along $\xi$ by a gauge transformation by a scalar $\chi$. Thus, the Maxwell perturbation is
\begin{equation}\label{eq:KerrdelA}
    \delta A= \d \chi + \mathcal{L}_\xi A.
\end{equation}
We wish to choose $\chi$ in such a way that this expression is divergence-free, i.e., $\nabla^\mu \delta A_\mu =0$. The structure of $\xi$ leads to the following choice:
\begin{equation}\label{eq:chiKerrsol}
\chi = - \i \, e \frac{\partial_\tau H}{1+y} \,\frac{\alpha}{\Gamma}\,.
\end{equation}
Then, we are left with the task of imposing the harmonic gauge condition on $\mathcal{L}_\xi g$. This leads to an inhomogeneous second order ODE for $\alpha(x)$   
\begin{equation}\label{eq:alphaKerrode}
\Gamma\dv{x}((x^2-1)\, \gamma\, \dv{\alpha}{x}) +
\left[2\,\Gamma\, \gamma +\Omega^2\, (x^2-1)\, \gamma^3\, \Gamma- \dv{x}((x^2-1)\, \gamma\, \Gamma')  \right]\alpha 
    =(1-x^2)\, \Omega\, \gamma^3\, \Gamma^2 \,. 
\end{equation}
This equation is easy to solve numerically, and the result is exhibited in~\cref{fig:KerrNew}. 
In particular, we find a unique smooth solution as long as $Q \neq 0$. In the  $Q\to 0$ limit, however, the coefficient in front of $\alpha$ in~\eqref{eq:alphaKerrode} vanishes. It then follows that the homogeneous equation has a solution $\alpha(x) = \textrm{constant}$. For that reason, a smooth solution to the inhomogeneous equation of interest ceases to exist.

We have thus established a near-horizon large diffeomorphism that respects the gauge conditions chosen for the metric and the gauge field. It then follows that the perturbation has vanishing on-shell action. We verified this statement by explicit calculation of the on-shell Einstein-Maxwell action. Notice, however, that owing to the mixing with the gauge field, as evidenced by the gauge transformation along $\chi$, the resulting metric perturbation, $h_{\mu\nu} = 2 \,\nabla_{(\mu}\xi_{\nu)}$, will not be annihilated by~\eqref{eq:LichEM}. In~\cite{Rakic:2023vhv} the authors did not analyze the structure of the compensating diffeomorphism and gauge transformation, but only noted that the diffeomorphism along $H(\tau,y)\, \partial_\phi$ is not annihilated by the Lichnerowicz operator, which, while true, is insufficient to conclude anything about the zero modes of the kinetic operators.

\subsection{Rotational modes in the near-horizon of extremal Kerr spacetime}

As presented above, there are various rotating black holes that admit rotational zero modes in their near-horizon region. On the other hand, it was observed in \cite{Rakic:2023vhv} that this is not the case for the Kerr black hole in the harmonic gauge.   We will argue that this is a failure of the harmonic gauge on the NHEK background rather than any physical phenomenon.

There are two ways in which a gauge choice may fail: either there exist perturbations that cannot be transformed into the chosen form, or there are non-zero perturbations satisfying all the gauge conditions that may not be transformed to zero. For the harmonic gauge, these two notions are essentially equivalent. Indeed, let us consider a perturbation $h_{\mu \nu}$ and let
\begin{equation}
    \mathsf{GF}_\mu = \nabla^\nu \left( h_{\mu \nu} - \frac{1}{2}\, g_{\mu \nu} \,h \right) .
\end{equation}
We will assume that $h_{\mu \nu}$ is smooth and decays sufficiently fast at infinity to ensure that it is normalizable with respect to the ultralocal measure. It follows that $\mathsf{GF}_\mu$ is also normalizable. 

Now we note that $h_{\mu \nu}$ is in the harmonic gauge if and only if $\mathsf{GF}_\mu = 0$. If this is not the case, we may modify it by a gauge transformation
\begin{equation}
    h_{\mu \nu} \mapsto h_{\mu \nu} + \nabla_\mu \zeta_\nu + \nabla_\nu \zeta_\mu \,,
\end{equation}
for a normalizable vector field $\zeta^\mu$. To ensure that the new perturbation is in the harmonic gauge, $\zeta^\mu$ must satisfy the following equation:
\begin{equation}\label{eq:harmGFop}
    \left(\delta\indices{^\mu_\nu} \, \nabla^2 + 
    R\indices{_\nu^\mu} \right) \zeta_\mu = -\mathsf{GF}_\nu \,.
\end{equation}
This is an elliptic problem for the compensating diffeomorphism generated by $\zeta^\mu$. It has a solution for all $\mathsf{GF}_\mu$ only if the cokernel of the operator defined by the left-hand side vanishes. But this operator is self-adjoint, and so the condition is equivalent to the requirement that its kernel be zero.\footnote{As such, due to $\i$ factors in the (complex Euclidean) metric, the Laplacian does not have to be self-adjoint. However, it is easy to see that is when restricted to axisymmetric perturbations, it does, which suffices for the argument.} We thus see that for the harmonic gauge, the two possible modes of failure are equivalent.

The only thing that is left is to prove that the operator $\delta\indices{^\mu_\nu} \, \nabla^2 +     R\indices{_\nu^\mu}$ on the near-horizon extreme Kerr (NHEK) geometry indeed has a non-zero kernel. Of course, since the NHEK is Ricci flat, the operator reduces to the Laplacian, $ \delta\indices{^\mu_\nu} \, \nabla^2$.

Let us start by inspecting a hyperbolic disk:
\begin{equation}
    \d s^2_{\textrm{AdS}_2} = (y^2-1)\,
    \d\tau^2 + \frac{\d y^2}{y^2-1}\,.
\end{equation}
The spectral theory on this background is well-known~\cite{Camporesi:1994ga}. In particular, the (co)vector Laplacian has one series of normalized eigenstates, namely \footnote{This should be familiar from the structure of gauge field zero modes encountered in the near-horizon \AdS{2} geometry.}
\begin{equation}
\zeta = \d H \,, \qquad 
    H(\tau, y)  = e^{\i n \,\tau} \left(\frac{y-1}{y+1}
    \right)^{\frac{\abs{n}}{2}} \,.
\end{equation}

It is an easy exercise to check that $\zeta$ pull-backed to the NHEK background, which has the line element \cite{Bardeen:1999px}
\begin{equation}
    \d s^2_{\textrm{NHEK}} = \left(1+x^2\right) \left( \left(y^2-1\right) \, \d \tau^2 +\frac{\d y^2}{y^2-1} +\frac{\d x^2}{1-x^2}\right) + 4\frac{1-x^2}{1+x^2}  \left(\d\phi+\i\, (y-1) \,\d \tau\right)^2 \,, 
\end{equation}
satisfies  
\begin{equation}\label{eq:kernel}
\nabla^2 \zeta_\mu = \nabla_\mu \nabla^2 H + 
R\indices{_\mu^\nu} \; \nabla_\nu H = 0 \,.
\end{equation}
Thus, as promised, $\nabla^2\, \delta_\mu^{\ \nu}$ has a non-zero kernel and the harmonic gauge cannot be imposed. That resolves the puzzle encountered in~\cite{Rakic:2023vhv}.

Let us add a couple more remarks to close out this discussion. Firstly, we can consider a more general near-horizon geometry which is a warped bundle with a \AdS{2} factor, such as, for example, any $d$-dimensional black hole with $\mathrm{U}(1)^{d-3}$ symmetry group, cf~\cite{Kunduri:2007vf}. Then, $\zeta$ is still well-defined as a pull-back. Moreover, $\nabla^2 H = 0$ and the first equality of~\eqref{eq:kernel} still holds. It follows that in general $\zeta$ satisfies,
\begin{equation}
    \left(\delta\indices{^\mu_\nu}\, \nabla^2 - R\indices{^\mu_\nu} \right) \zeta_\mu = 0.
\end{equation}
One may recognize that this is the operator that describes Maxwell fluctuations (around an uncharged saddle). That shows that $\zeta$ is universally a zero mode for $U(1)$ theory. Moreover, on any Ricci-flat background, $\zeta$ is in the kernel of the Laplacian and the harmonic gauge is invalid. Thus, not only do we have an issue for the Kerr spacetime, but  similar problems will also persist for the higher-dimensional Myers-Perry solutions. Secondly, the one-loop determinant for the ghost associated with the diffeomorphism is exactly $\det \left(-\delta\indices{^\mu_\nu}\, \nabla^2 - R\indices{^\mu_\nu} \right)$. As we have seen, this determinant vanishes on the Ricci-flat backgrounds, thus providing yet another sign of the failure of the harmonic gauge. Finally, a compactification of the theory to two-dimensional JT gravity, coupled to a rotational ${\rm U}(1)$ gauge field and matter, suggests that the rotational modes are present. An explicit calculation can be made in the two-dimensional harmonic gauge combined with the Lorenz gauge for the gauge field (this is \emph{not} the dimensional reduction of the 4d harmonic gauge condition for Kerr!). This observation would be more convincing if there were an improvement of the 4d harmonic gauge that led to the same answers.

\subsection{Rotational zero modes in BMPV}

In~\cite{Rakic:2023vhv} it was noted that the authors were unable to determine any rotational zero modes in the near-horizon geometry of the five-dimensional BMPV black hole. They pointed out this would be a serious  problem because these modes needed to be present to match the macroscopic gravitational computations with the microscopic result obtained from string theory~\cite{Sen:2012cj}. In light of the above, it behooves us to re-examine the solution, and fill this gap.

The near-horizon geometry of the BMPV black hole reads (cf.~\cite{Gupta:2021roy})
\begin{equation}
    \d s^2 =\left(y^2-1\right) \d \tau^2 +\frac{\d y^2}{y^2-1} +\d \psi^2 +\sin^2 \psi\, \d\chi^2+
    \cos^2 \omega \left(\i\, (y-1) \,\tan \omega\, \d \tau+\cos \psi \,\d\chi +\d\phi \right)^2,
\end{equation}
where $\psi \in [0,\pi], \chi \in [0,2\pi], \phi \in [0,4\pi]$. The parameter $\omega \in  [0,\frac{\pi}{2}]$ measures the angular momentum in units of the mass. The solution has $\mathrm{SU}(2)\times \mathrm{U}(1)$ rotational isometry. From the gravitational fluctuations one expects 7 zero modes; 3 from the Schwarzian set of modes, 3 associated with  the $\mathrm{SU}(2)$ rotations, and 1 associated with the $\mathrm{U}(1)$ isometry. The former six are easy to find, as already mentioned in~\cite{Rakic:2023vhv}. We will focus on the $\mathrm{U}(1)$ rotational zero mode. Notice that this metric corresponds to the very near horizon, near \AdS{2}, regime of the BMPV black hole, not the larger \AdS{3} throat.

Let us borrow the ansatz from the previous subsections and choose a vector field, 
\begin{equation}
    \xi = H(\tau,y) \,\partial_\phi + \alpha\, \nabla^\mu H \,\partial_\mu\,.
\end{equation}
Here $H(\tau,y)$ is the same function encountered in~\eqref{eq:KerrAdSrotxi}.  The perturbation generated by $\xi$ is $\mathcal{L}_\xi g$. It is straightforward to check that it is traceless. Moreover, its divergence reads
\begin{equation}
    \nabla^\mu \left( \nabla_\mu \xi_\nu + \nabla_\nu \xi_\mu \right) = 
    -\frac{1}{2} \left( 3\,b + b\, \cos 2\omega + \sin 2 \omega \right) \nabla_\nu H.
\end{equation}
Hence, we see that $\xi$ with 
\begin{equation}
    \alpha = - \frac{\sin 2 \omega}{3 + \cos 2 \omega}
\end{equation}
is indeed a zero mode. For the non-rotating solution, where $\omega =0$, this is indeed a rotational zero mode associated with the  $\mathbf{S}^3$. In this case, we have a set of $\mathrm{SO}(4)$ zero modes, of which this is a part. For  $\omega\neq 0$, however, the $\mathrm{U}(1)$ zero mode identified above does not mix with the other rotations. This is in perfect agreement with the predictions of \cite{Sen:2012cj}.

\section{Low-temperature thermodynamics of the BTZ black hole}\label{sec:BTZ_rot_mode}

Here, we continue the discussion from the previous section, and focus on the case of the BTZ spacetime, for reasons that will become clear shortly.  

Let us recall the puzzle regarding the low-temperature thermodynamics of BTZ, alluded to in the previous section. To sharpen the puzzle, it is useful to import some results reviewed in \cref{Sec:BTZ_full}. The one-loop determinant for gravitons around the BTZ geometry can be computed from first principles, in an ensemble of fixed temperature and angular velocity, either using the heat-kernel method~\cite{Giombi:2008vd} or with a formal argument using the asymptotic symmetries~\cite{Maloney:2007ud}. In the low-temperature limit, up to a possible shift in the extremal limit, leads to~\cite{Ghosh:2019rcj}
\begin{equation}
Z_{\rm graviton}^{\rm 1-loop} \sim T^{\frac{3}{2}}\,.
\end{equation}
At low temperatures, the prefactor of $T^{3/2}$ is consistent with the presence of Schwarzian modes but not with the rotational ones. The latter are expected, in an ensemble of fixed angular velocity, to produce another factor of $T^{1/2}$ around an individual saddle.\footnote{To obtain full one-loop determinant, one should sum over integer shifts of the chemical potential for rotations at infinity, which is the subset of the ${\rm SL}(2,\mathbb{Z})$ transformed black hole. The scaling with temperature is the same, both with and without performing the sum.} Contrary to the expectation in \cite{Rakic:2023vhv}, the BMPV analysis from the previous section suggests that indeed there should be a rotational zero mode in the near-horizon region of the BTZ geometry, leading to an extra $T^{1/2}$ factor. Why is the one-loop determinant around an individual near-extremal BTZ geometry $T^{3/2}$ and not $T^2$? 

We shall first show that the extremal BTZ geometry does indeed have a rotational zero-mode in the near-horizon region. Consider the  BTZ solution~\eqref{eq:btzmet}. In the near-extremal limit, we work at fixed $J$ and expand around the extremal value $r_+ = r_-$ with the scaling 
\begin{equation}\label{eq:rpmT}
r_+ = r_0 +\frac{\Ts}{4} + \frac{\Ts^2}{32\,r_0} +\order{\Ts^3} \,, \qquad 
r_- = r_0 -\frac{\Ts}{4} - \frac{3\,\Ts^2}{32\,r_0} +\order{\Ts^3}\,.
\end{equation}
Plugging~\eqref{eq:rpmT} into the BTZ line element~\eqref{eq:btzmet} together with
\begin{equation}\label{eq:btzzoom}
\varphi \to \varphi- \i\, \tE \,,\qquad \textbf{}
r = r_0 + \frac{\Ts}{4}\, y\,, 
\qquad \tE \to \frac{\tau}{\Ts}\,,
\end{equation}
we zoom into the extremal near-horizon metric, which is a fibration of a circle over \AdS{2},
\begin{equation}\label{eq:btznh}
\begin{split}
\d s^2 &= 
\frac{1}{4} \left( (y^2 -1)\,\d\tau^2 + \frac{\d y^2}{y^2 -1} + 4\,r_0^2 \left( \d\varphi-\i\, \frac{y}{2\,r_0}\,\d \tau \right)^2 \right) .
\end{split}
\end{equation}

Before tackling the rotational mode, let us confirm analytically that the off-shell modes identified in~\cref{Sec:BTZ_full} are indeed the Schwarzian modes. 
These modes should be in the kernel of the Lichnerowicz operator in the extremal near-horizon geometry~\eqref{eq:btznh}. By direct computation, one finds 
\begin{equation}\label{eq:zeromodes}
\begin{split}
\d s_\text{Sch}^2 
&= 
	\left( \frac{y-1}{y+1} \right)^\frac{\abs{n}}{2}
    \left[ \left(-\d \tau^2 +  \frac{\d y^2}{(y^2-1)^2} \right)
    \cos(n\tau)-  2 \,\i\,\, \frac{\d\tau \, \d y}{y^2-1}  \sin(n\tau) 
   \right] \\ 
\end{split}
\end{equation}	
Here $n\geq 2$ and the amplitude of the modes can be normalized by demanding the modes be suitably orthonormal. Now consider the off-shell modes $h^{(n)}_{\mu\nu}$ obtained in~\eqref{eq:hbtzb}, and apply the scaling~\eqref{eq:btzzoom}. It is not hard to show that the line element reduces to 
\begin{equation}
    h^{(n)}_{\mu\nu} \,\d x^\mu\, \d x^\nu = 8\,r_0^4\, \Ts^2 \, \d s_\text{Sch}^2  + \order{\Ts^3}\,.
\end{equation}
This adds to the list of checks that all we find in the full geometry are the uplift of the Schwarzian modes.

Let us now return to the question of the rotational mode. We can borrow the ansatz from the analysis in~\cref{sec:rot} and choose a vector field
\begin{equation}
    \xi = H(\tau,y) \,\partial_\phi + \alpha \,\nabla^a H \partial_a,
\end{equation}
where the function $H(\tau,y)$ is the same as the one defined in \eqref{eq:KerrAdSrotxi}. The perturbation generated by $\xi$ is $\mathcal{L}_\xi g$. It is straightforward to check that it is traceless. Moreover, its divergence reads
\begin{equation}
    -\nabla^\mu \left( \nabla_\mu \xi_\nu + \nabla_\nu \xi_\mu \right) = \left(2 r_0 + 4\alpha\right) \,\nabla_\nu H.
\end{equation}
Hence, we see that $\xi$ with $\alpha = -\frac{r_0}{2}$ is indeed a zero mode, since it is a pure (large) diffeomorphism satisfying the gauge condition. 

At this point, it is appropriate to comment on the choice of ensemble. The rotational zero-modes written above, as well as the ones in the previous section, naturally arise in a fixed charge ensemble. They affect the normalizable mode, but keep the non-normalizable mode in the gauge field (which only in 2d controls the charge) unchanged. These modes also contribute to the grand canonical ensemble in the following way. While in higher dimensions the grand canonical ensemble corresponds to fixing the holonomy of a gauge field, that is not the case in the near-horizon geometry. It is easy to see this from the 2d perspective. Consider a gauge field ${\sf A}_\mu$ in two dimensions. It was shown in~\cite{Ghosh:2019rcj,Iliesiu:2020qvm,Iliesiu:2022onk} that a fixed-holonomy boundary condition in the full geometry translates to a mixed boundary condition in the throat with
\beq\label{eq:mixedbdycond}
\delta( {\sf A}_\mu - c \, n^\nu {\sf F}_{\mu\nu} ) = 0,
\eeq
where $c$ is a coefficient determined by the details of how the throat is glued to the exterior, and is proportional to the quantity $1/T_q^{\mathrm{U}(1)}$ introduced below. This means that a fluctuation mode that affect the normalizable piece of ${\sf A}$ is consistent with the 2d grand canonical boundary condition only if one supplements it with a shift in the charge that keeps \eqref{eq:mixedbdycond} fixed. In all cases relevant to this paper, the effect from the small change in the charge is subleading and can be neglected in the evaluation of the action. This manifests itself in two dimensions in the fact that the Yang-Mills coupling is always suppressed by factors of $G_N$, reducing the Yang-Mills theory to an effective $BF$ model with mixed boundary conditions. This implies that the modes written in the previous paragraph should also contribute to the grand canonical ensemble.

Assuming these rotational modes can be regularized by turning on the temperature, we would find out that it produces an additional factor of $T^{1/2}$ to the total graviton one-loop determinant. This stands in stark contrast both with the heat-kernel method and our results on the spectrum of the Lichnerowicz operator from~\cref{Sec:BTZ_full}. But the  conclusion that there should be an additional $T^{1/2}$ factor is too naive for two reasons:
\begin{itemize}[wide,left=0pt]
    \item First, it is not guaranteed that a zero-mode that exists in the near-horizon region will extend to the full (in this case asymptotically $\mathrm{AdS}_3$) geometry. Indeed, while this might not be completely evident from our analysis in~\cref{Sec:BTZ_full}, the Schwarzian modes found in that section are the only ones that exist in the full geometry with an action that vanishes linearly with the temperature. This conclusion can be extracted from the thorough analysis of the eigenfunctions of the Lichnerowicz operator in~\cite{Datta:2011za,Castro:2017mfj}. This is not surprising; for example, it is not guaranteed that a mode in the throat, when extended to the full geometry, remains normalizable. The fact that no rotational modes are found in the analysis of~\cite{Datta:2011za,Castro:2017mfj} suggests this is the case. In fact, the main purpose of this paper is to verify in non-trivial examples that the Schwarzian modes always successfully extend to the full geometry.  
    \item The rotational zero-modes would produce a factor of $T^{1/2}$ only if their eigenvalue with a fine-temperature regulator is linear with the temperature. But what is the coefficient of proportionality? The Schwarzian modes eigenvalues are always proportional to $T/T_q$ setting the temperature scale at which they become strongly coupled. This scale can be identified from the classical thermodynamics, see for example \cite{Sachdev:2019bjn}, through the relation
    \beq
T_q = 4\pi^2\frac{1}{\left. \frac{\partial S}{\partial T}\right|_{T=0}}.
\eeq
   In other words, the Schwarzian coupling is the coefficient in the linear-in-temperature term in the classical entropy, viz., it is determined by the specific heat at extremality. This temperature scale is of order $\order{G_N}$ in the semiclassical limit. Similar scales can be defined for any gauge mode $T_q^{\mathrm{U}(1)}$ below which the gauge sector nearly zero-modes become strongly coupled. As explained in~\cite{Sachdev:2019bjn} (or in~\cite[Section 3.4]{Iliesiu:2020qvm}) this scale can again be extracted from classical thermodynamics 
\begin{equation}\label{eq:chargesus}
T_q^{\mathrm{U}(1)} = \frac{1}{K}\,, \qquad K= \left. \frac{\partial Q}{\partial \mu}\right|_{T=0}.
\end{equation}
This applies to any gauge mode and can be generalized also to non-Abelian ones. The coefficient $K$ is the charge susceptibility (but is referred to as the compressibility of the system in the condensed matter literature). Similarly, the rotational modes  also are governed by the angular momentum  susceptibility
\begin{equation}\label{eq:rosus}
   T_q^{\mathrm{rot}} = \frac{1}{K}\,, \qquad K= \left. \frac{\partial J}{\partial \Omega}\right|_{T=0} \,.
\end{equation}
This separation between Schwarzian and gauge-mode scales is usually unimportant since all these quantities are finite and of the same order of magnitude $\order{G_N}$, but something special happens for BTZ. An extremal rotating BTZ black hole satisfies $\Omega(J,T=0)=1$ regardless of the angular momentum. This means that the angular momentum susceptibility diverges! This implies that the finite-temperature action of the would-be $\mathrm{U}(1)$ modes diverges, essentially removing them from the spectrum. This is not independent of the previous bullet point; the finite-temperature action of the throat zero-modes arises precisely from the way the throat is glued to the exterior spacetime. Both perspectives imply that such gluing is problematic for the rotational modes. 
\end{itemize}

Taking these two observations into account, the situation becomes clear; the rotational modes of BTZ that one finds in the near-extremal solution near the horizon are not physical. They do not exist in the full geometry.  Further evidence for this assertion is provided by the fact that the other one-loop effect, viz., $\log S_0$ correction to the black hole entropy, is reproduced from the near-horizon analysis by the Schwarzian modes alone.  

A general lesson can be drawn from this specific case. We propose that whenever the zero temperature susceptibility associated to a gauge symmetry or background isometry (extracted from the classical thermodynamical analysis) is singular, this symmetry does not lead to any physical light mode at low temperatures.\footnote{For black holes with non-compact spatial horizon cross-section, like the planar \AdS{} or hyperbolic \AdS{} black holes, the Schwarzian contribution is likewise suppressed because the specific heat is proportional to the spatial volume.  }

A singular limit of the susceptibility arising from gravitational solutions is not uncommon. Besides the BTZ black hole, an example is provided by the prototypical case of the asymptotically flat Reissner-Nordstr\"om black hole. In the case of a flat space black hole, in the conventions of~\cref{sec:RNads}, the extremal chemical potential is simply $\mu=1$ irrespective of the charge.\footnote{This might not be surprising. If the $\mathrm{U}(1)$ gauge field arises from a five-dimensional Kaluza-Klein reduction, as is usually the case in string theory, the near-horizon geometry is $\mathrm{AdS}_2 \times \mathbf{S}^1 \times \mathbf{S}^2$ with the first two factors being precisely the same as the BTZ throat.} This leads to a singular charge susceptibility $K\to\infty$. This singularity is regulated by the presence of a cosmological constant, leading to
\begin{equation}
\mu = \sqrt{ \frac{1}{2} + \sqrt{\frac{1}{4}+\frac{3\,Q^2}{\lads^2}}} 
\quad \Rightarrow \quad K = \frac{\lads^2}{3\,\abs{Q}} + \cdots\,. 
\end{equation}
Our general guess, together with the expression above for $K$ in the large $\lads$ limit, predicts the rate at which the $\mathrm{U}(1)$ eigenvalues should diverge in the flat space limit. In particular, we expect $\lambda \sim (\mu-1)^{-1}$ as we approach the asymptotically flat solution. This is indeed what we see from our numerical analysis, as already explained in~\cref{sec:flatlimit}, see~\cref{fig:rn-u1-blowup}. Based on this numerical analysis, we would like to argue that gauge fields which source the background charge do not contribute to quantum effects for the asymptotically flat Reissner-Nordstr\"om spacetime.

\section{Discussion}

In this paper, we constructed off-shell modes in the full higher-dimensional asymptotically flat (or \AdS{}) metric that behave in the near-extremal limit precisely like the Schwarzian modes present near the horizon in the \AdS{2} description. Our results confirm that the modes identified in previous work on quantum corrections to the near-extremal black hole spectrum and dynamics, relying on the analysis of quantum effects in the throat using two-dimensional gravity techniques, are indeed physical and relevant in higher dimensions. We end with some open questions which we hope to address in the future. 

\begin{itemize}[wide,left=0pt]
\item  To carry out our analysis we had to rely on numerics to find the Schwarzian modes in the full geometry. It would be desirable to find the profiles analytically. This is hopeless to achieve for the full spectrum of metric fluctuations, but the Schwarzian modes have special properties that might make this possible. 
\item It would be interesting to find the extension of the Schwarzian modes for the Kerr black hole in \AdS{4}, or even in flat space  to the full spacetime. This would have to be tackled after using a gauge-fixing term which does not suffer the drawbacks of the harmonic gauge in Ricci-flat spacetimes. This would be an interesting exercise to undertake, although one also needs to address the role that superradiance plays near extremality. 
\item  Likewise, it would be interesting to extend some of these results to supergravity. It was recently uncovered how to perform the gravitational computation of the index in the full geometry, where one works at a finite temperature and identifies a complex Euclidean saddle associated with a particular value of chemical potential, cf.~\cite{Cabo-Bizet:2018ehj,Cassani:2019mms,Bobev:2019zmz, Bobev:2020pjk,Larsen:2021wnu,Iliesiu:2021are,BenettiGenolini:2023rkq,Anupam:2023yns,Chen:2023mbc,Boruch:2023gfn}. The index is, of course, temperature independent and thus are no logarithmic corrections of the kind discussed herein. However, at low temperatures, the Schwarzian modes play a non-trivial role in correcting the density of states.
\item Since we focused on evaluating the determinant by working in Euclidean signature, it sufficed for us to extract off-shell eigenmodes of the quadratic fluctuation operator. The determinant can, however, be expressed as a product over physical on-shell quasinormal (and their conjugate, anti-quasinormal) modes~\cite{Denef:2009kn}. These are on-shell in Lorentz signature. It would be interesting to understand how to interpret the analysis herein in terms of these quasinormal modes~\cite{Jia:2024tbd}.
\item Finally, analogs of the Schwarzian modes appear in the evaluation of the wavefunction of the universe with a positive cosmological constant and with spatial $\mathbf{S}^1 \times \mathbf{S}^2$ topology \cite{Maldacena:2019cbz}. When the size of the circle is much larger than the size of the sphere, there is a long period of $\mathrm{dS}_2$ inflation and similar light modes were shown to arise. In that case, the role of quantum effects was to make the wavefunction normalizable. It would be interesting to find those modes in the full inflating geometry. 
\end{itemize}

\section*{Acknowledgements}

It is a pleasure to thank Jan Boruch, Gary Horowitz, Guanda Lin, Sameer Murthy, Jorge Santos and especially Ashoke Sen for useful discussions. M.K.~and D.M.~were supported by NSF grants PHY-2107939 and PHY-2408110, and by funds from the University of California. I.R.~was supported by U.S. Department of Energy grant DE-SC0020360 under the HEP- QIS QuantISED program. M.R.~was supported by U.S.~Department of Energy grant DE-SC0009999 and funds from the University of California. G.J.T.~was supported by the University of Washington and the DOE award DE-SC0024363.
 
D.M., I.R., and M.R.~would like to thank KITP for hospitality during the program, “What is string theory? Weaving perspectives together”, which was supported by the grant NSF PHY-2309135 to the Kavli Institute for Theoretical Physics (KITP). M.R.~and G.J.T.~would  like to thank the Aspen Center for Physics, which is supported by National Science Foundation grant PHY-2210452 where this work was completed.


\providecommand{\href}[2]{#2}\begingroup\raggedright\endgroup

\end{document}